\documentclass[aps,prd,showpacs,floatfix,superscriptaddress,11pt]{revtex4-1}
\usepackage{xr}
\usepackage{amsmath,amssymb,physics}
\usepackage[toc,page]{appendix}
\usepackage{verbatim,graphicx,mathtools}
\usepackage{txfonts}
\usepackage{color}
\usepackage[all]{xy}
\usepackage{color}
\usepackage[all]{xy}
\usepackage{subfigure}
\usepackage{xcolor}
\usepackage{tikz}
\usepackage{tkz-euclide}
\usepackage{comment}
\usepackage{hyperref}

\newcommand{\eref}[1]{(\ref{#1})}
\newcommand{\nn}{\nonumber}
\newcommand{\be}{\begin{eqnarray}}
\newcommand{\ee}{\end{eqnarray}}

\newcommand{\bmat}{\left ( \begin{array}{cc} }
	\newcommand{\emat}{\end{array} \right ) }

\def\Tr{\textrm{Tr}}

\newcounter{jvct}

\newcounter{yjcc}

\newcounter{drc}

\newcommand{\beq}{\begin{equation}}
\newcommand{\beqs}{\begin{equation*}}
\newcommand{\eeq}{\end{equation}}
\newcommand{\eeqs}{\end{equation*}}

\allowdisplaybreaks

\begin{document}

\begin{flushright}
KIAS-P20042
\end{flushright}

\title{Sparse Sachdev-Ye-Kitaev model, quantum chaos and gravity duals}
\author{Antonio M. Garc\'\i a-Garc\'\i a}
\email{amgg@sjtu.edu.cn}
\affiliation{Shanghai Center for Complex Physics,
	School of Physics and Astronomy, Shanghai Jiao Tong
	University, Shanghai 200240, China}

\author{Yiyang Jia}
\email{yiyang.jia@stonybrook.edu}
\affiliation{Department of Physics and Astronomy, Stony Brook University, Stony Brook, New York 11794, USA}
\author{Dario Rosa}
\email{Dario85@kias.re.kr}
\affiliation{School of Physics, Korea Institute for Advanced Study,
	85 Hoegiro Dongdaemun-gu, Seoul 02455, Republic of Korea}
\author{Jacobus J. M. Verbaarschot}

\email{jacobus.verbaarschot@stonybrook.edu}
\affiliation{Department of Physics and Astronomy, Stony Brook University, Stony Brook, New York 11794, USA}

\begin{abstract}
 We study a sparse Sachdev-Ye-Kitaev (SYK) model with $N$ Majoranas where only $\sim k N$ independent matrix elements are non-zero. We
  identify a minimum $k \gtrsim 1$ for quantum chaos to occur by a level statistics analysis.
  The spectral density in this region, and for a larger $k$, is still given by the Schwarzian prediction of the dense SYK model, though with renormalized parameters.
  Similar results are obtained for a beyond linear scaling with $N$ of the number of non-zero matrix elements.
  This is a strong indication that this is the minimum connectivity for the sparse SYK model to still have a quantum gravity dual.
We also find an intriguing exact relation between the leading correction to moments of the spectral density due to sparsity and the leading $1/d$ correction of Parisi's U(1) lattice gauge theory in a $d$ dimensional hypercube.
In the $k \to 1$ limit, different disorder realizations of the sparse SYK model show
emergent random matrix statistics that for fixed $N$ can be in any
universality class of the ten-fold way. The agreement with random matrix statistics is restricted to short range correlations, no more than a few level spacings,
in particular in the tail of the spectrum.
In addition, emergent discrete global
symmetries in most of the disorder realizations for $k$ slightly below one
give rise to $2^m$-fold degenerate spectra, with $m$ being a positive integer. For $k =3/4$, we observe a large number of such emergent global symmetries with a maximum $2^8$-fold degenerate spectra for $N = 26$.

\end{abstract}\maketitle
\newpage
\tableofcontents
\section{{Introduction}}
Models of interacting fermions with infinite range interactions in zero spatial dimension \cite{french1970,french1971,bohigas1971a,mon1975} were introduced about fifty years ago to describe qualitative aspects of nuclear dynamics.
Later, they were broadly employed \cite{benet2001} to model quantum chaotic dynamics in a many-body context and also certain aspects of quantum magnetism \cite{sachdev1993}.

More recently \cite{kitaev2015,maldacena2016,jensen2016,sachdev2010},
a variant of these models based on $N$ Majoranas \cite{kitaev2015}, the so called Sachdev-Ye-Kitaev (SYK) model, has attracted a lot of attention as a toy model for holography,  and for its potential to reveal novel insights in the dynamics of strongly interacting quantum matter. In the low temperature (strong coupling) limit, the SYK model
shares the same pattern of soft breaking of conformal symmetry \cite{maldacena2016a} by finite temperature and quantum $(1/N)$ effects as that of Jackiw-Teitelboim (JT) gravity \cite{teitelboim1983,jackiw1985}, a two-dimensional gravity theory with a dilaton in Anti-de Sitter space with non-trivial boundary conditions. This symmetry breaking pattern dictates  low temperature thermodynamic
properties \cite{Georges_2001,kitaev2015,maldacena2016,jevicki2016,garcia2016} such as a linear specific heat, and an exponential growth of low energy excitations. These are all expected features in field theories with a black hole gravity dual.

Another distinctive feature of these systems is quantum chaos \cite{maldacena2015}. Quantum chaos reveals itself in level statistics described by random matrix theory \cite{bohigas1984} and also in
 the exponential growth at the scrambling time of quantum corrections measured by certain out-of-time-order correlation functions, with a growth rate controlled by the Lyapunov exponent. Kitaev \cite{kitaev2015} found
that this feature occurs in the SYK model and that, in the strong-coupling, low temperature limit, the Lyapunov exponent saturates a previously proposed universal bound on chaos \cite{maldacena2015}.
Regarding level statistics, both the  SYK model \cite{garcia2016,garcia2017,cotler2016,you2016,Jia:2019orl} and JT gravity \cite{garcia2020,saad2019} are well described by random matrix theory \cite{wigner1951,dyson1962a,dyson1962b,dyson1962c,dyson1962d} which indicates that the system is quantum  chaotic at all times scales.

A natural question to ask is how the above features, which determine the existence of a quantum black hole dual, are robust to deformations of the SYK model.
Typically, generalizations of the SYK model to higher spatial dimensions \cite{berkooz2017} or involving more Majoranas than the usual four-body interaction \cite{li2017} share similar features. However, the addition of an integrable two-body interaction \cite{garcia2018b,Nosaka:2018iat,Nosaka:2019tcx} prevents the saturation of the Lyapunov exponent. Moreover, in a certain range of parameters, the system is not quantum chaotic as spectral correlations are well described by Poisson statistics, typical of an integrable system.

Another example of a generalized SYK model in which quantum chaos may not occur is that of
a two-site coupled SYK model that in the low temperature limit is dual \cite{maldacena2018} to an eternal traversable wormhole.  It was shown in Ref.~\cite{garcia2019b} that the traversable wormhole phase is not quantum chaotic. Quantum chaotic features are only observed for higher temperatures where the gravity dual undergoes a thermodynamic transition  to a quantum two-black-hole background.

Another plausible deformation of the SYK  model is to relax the requirement of infinite range interactions. Indeed, in the context of condensed matter physics \cite{altshuler1997}, interacting quantum dots describing realistic electronic interactions are qualitatively similar to the SYK model with complex fermions but with a Fock space geometry living on a Cayley tree rather than on a complete graph. The effect in the SYK model of a sharp-cut off in Fock space distances \cite{garcia2019} induces a metal-insulator transition. However, not much is known about the requirements on the range or the form of the interactions
that guarantees the existence of a gravity dual.
Progress on this problem would not only bring a more detailed understanding
on the conditions for a field theory to have a gravity dual but also it might
be useful to identify systems to test experimentally holography predictions.

Here we study the properties of a sparse SYK model where some of the
couplings are randomly set to zero with a probability $1-p$, where $p \sim k/N^\alpha$, $\alpha >0$ and $k$ is a positive real number. This model was first articulated in a talk given by Brian Swingle \cite{Swingle:2019}.
Our main aim is to characterize the maximum sparseness for which both the spectral density and spectral correlations are consistent with that of a gravity dual.
Namely, the spectral density is still described by the Schwarzian prediction of the dense SYK model \cite{kitaev2015,maldacena2016} and level statistics are still quantum chaotic \cite{garcia2016,cotler2016} and therefore well modeled by random matrix theory. For that purpose, we have computed analytically the spectral density, and the partition function, by an explicit calculation of the moments of the Hamiltonian. We show that it still has a Schwarzian form, and therefore it is likely related to a gravity dual, provided that $\alpha \leq 3$ and, for $\alpha = 3$,  $k \sim 1$ or larger. A study of spectral correlations confirms agreement with random matrix theory in this region of parameters which indicates the dynamics is still quantum chaotic at late time scales. 
For $k=1$ extra symmetries and chiral symmetries  emerge for some disorder realizations, and level statistics of
the three Wigner-Dyson ensembles and the three chiral ensembles are observed
for an ensemble of 26 Majorana fermions.
For $k<1$ we find a large number of emergent discrete symmetries as well as chiral symmetries leading to exact degeneracies in powers of 2.

We note that formally the SYK Hamiltonian is defined over random hypergraphs.
As we shall see, there are not many mathematically rigorous results for
generic random hypergraphs as a function of the degree of sparseness.
The situation is different in the simpler case of random sparse graphs,
usually termed Erdos-Renyi graphs \cite{erdos1960} which can be cast as $L\times L$ matrices.
There is a rather rigorous characterization \cite{huang2015,erdos2012,huang2020,bauerschmidt2017} of the  bulk spectral properties: level statistics consistent with the prediction of random matrix theory will occur if the fraction of nonzero matrix elements satisfies
$p \geq L^{\epsilon} /L$ with $\epsilon > 0 $. In this region, the spectral density is given by the semicircle law \cite{erdos2013,rodgers1988}. These results are fully consistent with numerical \cite{evangelou1992} and analytical results \cite{fyodorov1991,mirlin1997} in the physics literature. Close to the edge of the spectrum, the spectral region related to the gravity dual, it was demonstrated rigorously  \cite{erdos2012} that, for $p \geq L^{\epsilon}/L^{2/3}$,  spectral correlations are described by RMT. As far as we know, it is unclear whether this bound is optimal.

These findings cannot be directly applied to the sparse SYK model as its Hamiltonian is not represented by a graph but by a more complex random sparse hypergraph for which not many explicit results for the density or spectral correlations are available. An exception \cite{erdos2014} is the spectral density of a SYK-like model that can be cast as a $\sqrt N$-hypergraph.
We refer to \cite{dumitriu2019} and references therein for recent mathematical results about the conditions to observe the semicircle law in random hypergraphs. We are not aware of any level statistics characterization of random hypergraph in the mathematical literature. In the physics literature, we refer to Ref.~\cite{verbaarschot1984,Bagrets2017} for an analytical calculation of the two-level correlation function in a fermionic model with infinite range interactions.

The paper is organized as follows: in Section \ref{sec:sykmodel}, we introduce the sparse SYK model, the mechanism to tune the sparseness and the regularity condition that makes the connectivity on the hypergraph uniform for each disorder realization. Section \ref{sec:anal} is devoted to an analytical evaluation of the spectral density as a function of the degree of sparseness. We notice a striking equivalence between leading corrections to the moments of the density, due to the sparsity of the model, and leading $1/d$ corrections of the same quantity in the Parisi's $U(1)$ gauge model on a $d$-dimensional hypercube \cite{Parisi:1994jg}. Based on exact analytical results for low order moments, we propose two approximate analytical expressions for moments of any order. For one of them, we write down a closed analytical expression for the spectral density. In Section \ref{sec:specnum}, these predictions are compared to numerical results resulting from the exact diagonalization of the sparse SYK Hamiltonian. In Section \ref{sec:levelsta}, we turn to the study of the conditions for the existence of quantum chaos by an analysis of spectral correlations in both the bulk and the edge of the spectrum as a function of the degree of sparseness. Section \ref{sec:emersym} is focused on the description of emergent global symmetries that only occur in the limit of strong sparsity. For a fixed number of Majoranas, these additional symmetries, that depend on the disorder realization,  lead to spectral degeneracies and spectral correlations described by random matrix ensembles of different universality classes including those with chiral symmetry. Finally, in Section \ref{sec:conclusions}, we summarize the main results of the paper and list some problems for future research. The numerical implementation of the regularity condition
is discussed in Appendix \ref{app:regularity}. In Appendix \ref{app:N26GOEGSE},
we discuss two examples of emergent symmetries.

 \section{Sparse SYK Model}\label{sec:sykmodel}
We investigate the following Hamiltonian representing $N$ strongly interacting Majorana fermions \cite{kitaev2015} with sparse $q$-body infinite range interactions. For $q=4$,

\begin{equation}\label{hami}
H \, = \, \sum_{1\leq i<j<k<l\leq N} x_{ijkl} J_{ijkl} \, \gamma_i \, \gamma_j \, \gamma_k \, \gamma_l \, ,
\end{equation}
where we have used Euclidean Dirac matrices $\gamma_i$ to represent Majorana fermions. Dirac matrices satisfy the anti-commutation relations
\begin{eqnarray}
\{ \gamma_i, \gamma_j \} =2 \delta_{ij}.\label{clif}
\end{eqnarray}
They are the same as the anti-commutation relations of Majorana fermions up to a factor of two, which will be absorbed in the definition of the variance of $J_{ijkl}$ in Eq. \eqref{eqn:Jdefq4}.
 The sparseness is implemented by the random variable $x_{ijkl}$: $x_{ijkl}=1$ with probability $p$ and $x_{ijkl}=0$ with probability $1-p$. We may think of the interactions to be defined on random hypergraphs: $i= 1, 2, \ldots, N$ labels the $N$ vertices. A hyperedge connecting the $i, j, k, l$ vertices is present if $x_{ijkl}=1$. The expected number of the hyperedges (and hence the number of terms in the Hamiltonian) is $p\binom{N}{4}$.  The hypergraph gets sparser as $p$ gets closer to zero, and $p=1$ gives the maximal number of hyperedges, $\binom{N}{4}$, which results in the conventional ``dense'' SYK model.
 The couplings $J_{ijkl}$ is a Gaussian random variable with the distribution
\begin{equation}\label{eqn:Jdefq4}
P(J_{ijkl}) \, = \, \sqrt{\frac{2^{3}N^{3}p}{3! \pi J^2}} \exp\left( - \, \frac{ 2^{3} N^{3}pJ_{ijkl}^2}{3! J^2} \right) \, ,
\end{equation}
where $J$ sets the scale of the distribution. We will set $J = 1$ for later numerical calculations.
We focus on a probability $p$ that scales as $p \sim N^{-3}$ for which it is convenient to define an order-one quantity $k$:
\begin{equation}
k = \frac {p}N \binom{N}{4}.
\end{equation}
We shall study other scalings $p \sim N^{-\alpha}$, but unless stated explicitly,
we set $\alpha =3 $.

Although the numerical results of this paper will be restricted to $q=4$, certain analytical results will also be available for other integer values of $q > 0$. For general $q$, we write the Hamiltonian as
 \be\label{eqn:HdefGeneralq}
H =\sum_\alpha x_\alpha J_\alpha \Gamma_\alpha,
\ee
where $\alpha$ is a multi-index with $q$ elements:
\be
\alpha=\{i_1,i_2,\ldots,i_q\}, \ 1\leq i_1<i_2<\cdots<i_q\leq N,
\ee
so that $\alpha$ can take $\binom{N}{q}$ different values. The random coupling $J_\alpha$ follows a Gaussian distribution

\begin{equation}\label{eqn:JdefGeneralq}
P(J_{\alpha}) \, = \, \sqrt{\frac{2^{q-1}N^{q-1}p}{(q-1)! \pi J^2}} \exp\left( - \, \frac{ 2^{q-1} N^{q-1}pJ_{\alpha}^2}{(q-1)! J^2} \right) \, ,
\end{equation}
and $\Gamma_\alpha$ is the product of $q$ Dirac matrices indexed by $\alpha$:
\begin{equation}
\Gamma_\alpha = i^\frac{q(q-1)}{2} \gamma_{i_1}\gamma_{i_2}\cdots\gamma_{i_q}.
\end{equation}
The hyperedge variable $x_\alpha$ is defined analogously, the probability $p$ now scales as $N^{1-q}$ and
\begin{equation}
k = \frac{p}{N} \binom{N}{q}.
\end{equation}
When $k$ is small, some of the random hypergraphs are disconnected. The disconnectedness makes the Hamiltonian split into a sum of sub-Hamiltonians defined on independent tensor subspaces. In this case, the spectral statistics become a superposition of statistics from different sectors possibly belonging to different symmetry classes. To mitigate
this complication, we can consider regular hypergraphs only, that is, we can impose that every vertex has the same degree $kq$, namely each vertex is contained in the same number $kq$ of hyperedges. This regularity condition is imposed as a set of constraints on $x_\alpha$:
\begin{equation}\label{eqn:regCondition}
\sum_{\alpha \ni m}x_\alpha = kq, \quad\ \text{for any } m=1, 2, \ldots, N,
\end{equation}
where $\sum_{\alpha \ni m}$ means that, among the $\binom N q$ choices of $\alpha$, we sum over those $\alpha$ that contain a given integer $m$. The value of $k$ must be chosen so that $kq$ is a positive integer. This regularity condition implies that every realization of the Hamiltonian contains \textit{exactly} $kN = p \binom N q$ number of independent non-zero terms, as opposed to the case without the regularity condition where $kN$ is only the number of nonzero terms on average. For ordinary random graphs, where each edge connects two vertices, graphs are almost surely connected provided they are regular and its vertex degree is larger than $2$ \cite{gilbert1959,erdos1960,chartrand1966}. For random regular hypergraphs, since a hyperedge connects more than two vertices, we expect connectivity to be more easily achievable and hence the vertex degree need not be as large. This is indeed true for any vertex degree $kq>1$ for which any random regular hypergraphs will be almost surely connected \cite{chan2020}. However, as we will see, even with regularity condition, in the very sparse regime $1/q<k\leq 1$, spectral statistics can still be a superposition of independent spectra because there are new emergent global symmetries.
\section{Spectral density: Analytical results}
\label{sec:anal}
We evaluate analytically the spectral density by an explicit computation of the moments of the sparse SYK Hamiltonian,
\be
M_{2l}=  2^{-N/2}\langle\Tr  H^{2l}\rangle. \label{moment}
\ee
The spectral density can be expressed as
\be
\label{rhog}
\rho(E) &=&
2^{-N/2}\frac 1{2\pi} \int_{-\infty}^\infty dt e^{- iE t} \left \langle \Tr e^{iH t} \right \rangle = \frac 1{2\pi} \int_{-\infty}^\infty dt e^{- iE t}
\sum_l \frac 1{(2l)!} (it)^{2l} M_{2l}.
\ee
Since we have a Gaussian distribution of $J_\alpha$ (in the notation of Eq. \eqref{eqn:HdefGeneralq}), the calculation of the average requires to consider all possible
Wick contractions. In the end, we will also need to average over the random variable $x_\alpha$.

\begin{figure}
\begin{center}
\begin{tikzpicture}
\draw[fill=black] (-2,0) circle (1pt);
\draw[fill=black] (-1,0) circle (1pt);

\node at (-2.1,-0.25) {$\alpha_1$};
\node at (-0.8,-0.25) {$\alpha_2$};
\draw (-2,0)--(-1,0);

\draw[fill=black] (3,0) circle (1pt);
\draw[fill=black] (4,0) circle (1pt);
\draw[fill=black] (3.5,1.732/2) circle (1pt);

\node at (3.6,1.732/2+0.2) {$\alpha_1$};
\node at (2.9,-0.25) {$\alpha_2$};
\node at (4.2,-0.25) {$\alpha_3$};

\draw (3.5,1.732/2) --(3,0)--(4,0)--(3.5,1.732/2);

\end{tikzpicture}
\end{center}
\caption{Two examples of intersections graphs $G$ for the dense SYK model. The left graph represents $\binom{N}{q}^{-2}\sum_{\alpha_1 \alpha_2} (-1)^{c_{\alpha_1\alpha_2}}$ which is the $\eta$ defined in Eq. \eqref{eqn:etadef}; the right graph represents $\binom{N}{q}^{-3}\sum_{\alpha_1 \alpha_2 \alpha_3}(-1)^{c_{\alpha_1\alpha_2}+c_{\alpha_1\alpha_3}+c_{\alpha_2\alpha_3}}$ which is the $T_6$ defined in Eq. \eqref{eqn:T6def}.}\label{fig:interGraphDenseSYK}
\end{figure}
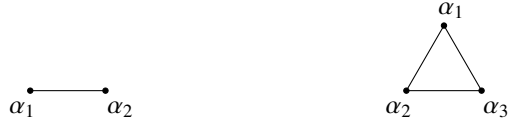

After averaging over $J_\alpha$, the result depends on whether pairs of two factors  $\Gamma_\alpha$ are adjacent or not. In the former case
we can use that
\be\label{eqn:GammaSqr}
\Gamma_\alpha^2 =1,
\ee
while in the latter case, the $\Gamma_\alpha$'s can be made adjacent by using
\cite{garcia2016},
\be\label{eqn:GammaCommu}
\Gamma_\alpha \Gamma_\beta - (-1)^{c_{\alpha\beta}}  \Gamma_\beta \Gamma_\alpha =0
\ee
where $c_{\alpha\beta}=|\alpha\cap\beta|$ is the number of indices that $\alpha$ and $\beta$ have in common. An exact calculation of a generic trace requires us to keep track of correlations with other factors $\Gamma$. This is in general a challenging combinatorial problem but some low-order moments have been evaluated exactly \cite{garcia2016,garcia2018c} for the dense SYK model. The simplest Wick contraction in which Eq. \eqref{eqn:GammaCommu} plays a role is
\begin{equation}
2^{-N/2} \sum_{\alpha,\ \beta}\Tr \Gamma_\alpha\Gamma_\beta\Gamma_\alpha\Gamma_\beta =\sum_{\alpha,\ \beta} (-1)^{c_{\alpha\beta}},
\end{equation}
out of which we will define an order-one quantity
\begin{equation}\label{eqn:etadef}
\eta := \binom{N}{q}^{-2}\sum_{\alpha,\ \beta} (-1)^{c_{\alpha\beta}}= {N \choose q}^{-1}  \sum_{c_{\alpha\beta}=0}^q  (-1)^{c_{\alpha\beta}} {q \choose c_{\alpha\beta}} {N-q \choose q-c_{\alpha\beta}}.
\end{equation}
In general, a Wick contraction that contributes to $M_{2l}$ is a trace of a product of $2l$ matrices $\Gamma$, whose subscripts form $l$ pairs that are summed over.  Repeatedly using Eqs. \eqref{eqn:GammaSqr} and \eqref{eqn:GammaCommu}, we can move all the pairs of $\Gamma$'s with the same subscripts next to each other and produce a purely combinatorial expression of the form
\begin{equation}\label{eqn:contractionCrossing}
\binom{N}{q}^{-l}\sum_{\alpha_1, \alpha_2, \ldots, \alpha_p} (-1)^{\sum_{\text{crossings}}c_{\alpha_i\alpha_j}},
\end{equation}
where $c_{\alpha_i\alpha_j}=|\alpha_i\cap\alpha_j|$ and $\sum_{\text{crossings}}$ includes all those pairs of $i, j$ for which $\alpha_i$ and $\alpha_j$ form a ``crossing'' configuration in the trace:
\begin{equation}
2^{-N/2}\Tr\left(\ldots\Gamma_{\alpha_i}\ldots\Gamma_{\alpha_j}\ldots\Gamma_{\alpha_i}\ldots\Gamma_{\alpha_j}\ldots\right).
\end{equation}
The binomial factor in front of Eq. \eqref{eqn:contractionCrossing} is to normalize the sum to an order-one quantity, which can be alternatively understood as normalizing the moments $M_{2l}$ to \textit{reduced moments} $M_{2l}/M_{2}^l$ as we shall see soon. There is an \textit{intersection graph} representation \cite{garcia2018c} of the quantity defined in Eq. \eqref{eqn:contractionCrossing}:
\begin{enumerate}
\item Draw $l$ vertices labeled by $\alpha_1,\alpha_2,\ldots,\alpha_p$.
\item If any $c_{\alpha_i\alpha_j}$ is in the $\sum_{\text{crossings}}c_{\alpha_i\alpha_j}$, connect the vertices $\alpha_i$ and $\alpha_j$ by an edge.
\end{enumerate}
We can now rewrite Eq. \eqref{eqn:contractionCrossing} in terms of intersection graphs: a Wick contraction represented by an intersection graph $G$ that contributes to the $2l$-th reduced moment has a value of \footnote{If $q$ is odd, we need an extra factor of $(-1)^{qE_G}$ where $E_G$ is the total number of edges in $G$.}
\begin{equation}\label{eqn:intersGraphValueDense}
\eta_G=\binom{N}{q}^{-l}\sum_{\alpha_1,\alpha_2,\ldots,\alpha_l} (-1)^{c(G)},
\end{equation}
where $c(G)=\sum_{(\alpha_i\alpha_j)\in G}c_{\alpha_i\alpha_j}$, $(\alpha_i\alpha_j)$ is an edge of the graph $G$ connecting the vertices $\alpha_i$ and $\alpha_j$ and $c_{\alpha_i\alpha_j}$ is the number of common indices in $\alpha_i$ and $\alpha_j$. We give two examples of intersection graphs and what they represent in Fig.~\ref{fig:interGraphDenseSYK}.

 In this notation, the reduced moments for the dense SYK model can be written as
\begin{equation}
M_{2l, \rm SYK}/M_{\rm SYK,2}^l = \sum_{G}\eta_{G},
\end{equation}
where $G$ are all the $l$-vertex intersection graphs representing Wick contractions. An important approximation to the dense SYK model moments is the so-called \textit{Q-Hermite approximation} \cite{cotler2016,erdos2014,garcia2017,garcia2018c}:
\begin{equation}\label{eqn:etaQHapprox}
\eta_G \approx \eta^{E(G)},
\end{equation}
where $\eta$ is defined in Eq. \eqref{eqn:etadef} and $E(G)$ is the number of edges in $G$. Under this approximation, the dense SYK moments are
\begin{equation}\label{eqn:riordanTouchard}
M_{2l, \rm SYK}/M_{\rm SYK,2}^l \approx \sum_{G}\eta^{E(G)}= \frac 1{(1-\eta)^l} \sum_{i=-l}^l (-1)^i
{2l\choose i+l}\eta^{i(i-1)/2},
\end{equation}
where the second equality is the Riordan-Touchard formula \cite{riordan1975,touchard1952}. The moments of the Riordan-Touchard formula are exactly the moments of the spectral density for the Q-Hermite polynomials \cite{ismail1987}:
\be\label{eqn:QHden}
\rho_{\rm QH}(E) = c_N\sqrt{1-(E/E_0)^2} \prod_{m=1}^\infty
\left [1 - 4 \frac {E^2}{E_0^2}
\left ( \frac 1{2+\eta^m+\eta^{-m}}\right )\right],
\ee
where
\be \label{eqn:groundStateE}
E_0=-\sqrt{ \frac {4 \sigma^2}{1-\eta}}
\ee
is the ground state energy. This is the reason why we called Eq. \eqref{eqn:etaQHapprox} the Q-Hermite approximation for the (dense) SYK moments.

 When it comes to averaging over $x_\alpha$, the regularity condition matters. Without regularity condition, a generic averaging $\langle x_{\alpha_1}x_{\alpha_2}\cdots x_{\alpha_l}\rangle$ can be worked out by simply noting that $x_\alpha^2 = x_\alpha$, $\langle x_\alpha\rangle = p$ and that two $x$-variables are statistically independent if they have different subscripts. With regularity condition, two $x$-variables can be correlated even if they have different subscripts (essentially because, given the regularity constraint, the various $x$-variables are not extracted independently from each other), which makes the combinatorial  problem much more difficult. For this reason, in this paper analytical results are only available for the model without regularity condition, and the regular model will only be studied numerically. However, since the regularity condition in Eq. \eqref{eqn:regCondition} implements $N$ constraints on $\binom{N}{q}$ otherwise independent variables, we expect that the regularity condition only modifies the moments by contributions of order $1/N^{q-1}$ which are subleading with respect ot those considered below.

Without regularity condition, the second moment is given by,
\be
M_2=2^{-N/2}\Tr H^2 &=& {N\choose q} \langle J_{\alpha}^2\rangle \langle x_{\alpha}^2 \rangle = {N\choose q}  \times \frac{(q-1)!J^2}{2^qN^{q-1}p} \times p={N\choose q}\frac{(q-1)!J^2}{2^qN^{q-1}},
  \ee
  which is the same as for the dense SYK model. To calculate the fourth moment we need to be careful with the average over $x_\alpha$ variables because
\begin{equation}
\langle x_\alpha x_\beta\rangle=
\begin{cases}
p \qquad&\text{if $\alpha= \beta$,}\\
p^2 &\text{if $\alpha\neq \beta$.}
\end{cases}
\end{equation}
We write down the result in terms of the reduced fourth moment without regularity condition,
\be\label{fourm}
M_4/M_2^2  &=&
\frac 1{p^2 {N\choose q}^2}\left (2p^2{N\choose q}\left ({N\choose q}-1\right )
+2p{N\choose q}    + {N\choose q} \left[ p^2 \sum_{r=1}^q (-1)^r {q\choose r}
  {N-q \choose q-r} + p{N\choose q} \right]\right )\nn\\
&=&2+ \eta+ \frac{3}{kN}\left(1-p\right)\nn\\&=&M_{4, {\rm SYK}}/M_{2,{\rm SYK}}^2 + \frac{3}{kN}\left(1-p\right),
\ee
where $M_{4,  {\rm SYK}}$ is the fourth moment of the dense SYK model  and $\eta$ is given by Eq.~\eqref{eqn:etadef}. Similarly, the result for the sixth moment without regularity condition is,
\be\label{sixm}
M_6/M_2^3 = M_{6, {\rm SYK}}/M_{2,{\rm SYK}}^3  + \frac{3}{kN} \left(1-p\right)(9 + 6 \eta )+  \frac{15}{(kN)^2} \left(1 - 3p + 2p^2 \right).
\ee
Likewise, the full expression for the eighth moment without regularity condition is,
\be\label{eightm}
M_8/M_2^4 = M_{8,\text{SYK}}/M_{2,{\rm SYK}}^4+\frac{3(1-p)}{kN}(56+86\eta+52\eta^2+16T_6)+\frac{1}{(kN)^2}(1-2 p+p^2) (144 \eta+171) \nonumber\\ +\frac{1}{(kN)^2}(1-3 p+2 p^2) (180 \eta+240)+\frac{105}{(kN)^3}(1 - 7 p + 12 p^2 - 6 p^3),
\ee
where $T_6$ is the value for the triangle intersection graph (see Fig. \ref{fig:interGraphDenseSYK}):
\be\label{eqn:T6def}
  T_6:= {N\choose q}^{-3} \frac 1{2^{N/2}}\sum_{\alpha\beta\gamma}\Tr\Gamma_\alpha\Gamma_\beta\Gamma_\gamma\Gamma_\alpha\Gamma_\beta\Gamma_\gamma,
   \ee
which in the Q-Hermite approximation Eq. \eqref{eqn:etaQHapprox} is given by $T_6\approx \eta^3$. In principle, it is possible to compute higher order moments but the final expression becomes increasingly cumbersome. It is clear from the explicit calculations so far, and from a general proof to be given soon, that the leading moment is always that corresponding to the dense SYK model. Hence, in the large $N$ limit, we already know the behavior of sparse SYK moments: they are the same as the large $N$ limit of the dense SYK model. Two commonly taken large $N$ limits are:
\begin{enumerate}
\item Fixed $q$ and $N \to \infty$: in this limit the global spectral density approaches a Gaussian.
\item Fixed $q^2/N$ and $N \to \infty$: in this limit the global spectral density approaches the density function of the Q-Hermite polynomials with $Q=\eta$ which
  for large $N$ can be approximated as
$Q\to   \exp(-2q^2/N)$ \cite{erdos2014}.
\end{enumerate}
However, we would like to understand how the large $N$ limit is approached or, in other words, we would like to understand the sparse SYK model at large but finite $N$ with $q$ fixed.
Moreover, we would also like to understand the form of the low energy excitations slightly above the ground state for the $q=4$ model, which is not captured by the two above-mentioned global limits.
The form of the spectral density in this infrared region is relevant to the type of gravitational theory the sparse SYK model might be dual to.

We now set out to study the finite $N$ behavior of the sparse SYK moments. As a start, we would like to draw the readers' attention to the subleading terms of the moments. If we apply the Q-Hermite approximation defined in Eq. \eqref{eqn:etaQHapprox}, the subleading terms in Eqs. \eqref{fourm} - \eqref{eightm} become,
\begin{equation}\label{mkn}
\begin{split}
M_4/M_2^2:&\ \frac{3}{kN},\\
M_6/M_2^3:&\ \frac{3}{kN}(9+6\eta),\\
M_8/M_2^4:&\ \frac{3}{kN}(56+86\eta+52\eta^2+16\eta^3).
\end{split}
\end{equation}
Surprisingly, these expressions are strikingly similar to certain subleading contributions to the moments of the Parisi's $U(1)$ lattice gauge theory in a hypercube \cite{Parisi:1994jg,Marinari:1995jwr,Jia:2020rfn} which we now discuss in detail.
\subsection{Relation to Parisi's $U(1)$ lattice gauge theory in large $d$ dimensions}
The Parisi's model is a $U(1)$ lattice gauge theory defined on a $d$-dimensional hypercube. The gauge links are chosen such that the magnitude of the magnetic flux through each hypercube face is $\phi$, but with random signs. The first eight reduced moments of the Parisi's model up to subleading order are \cite{Marinari:1995jwr}
\begin{equation}
\begin{split}
M_4/M_2^2=&\frac{d-1}{d}(2+\cos \phi)+ \frac{1}{d},\\
M_6/M_2^3=&\frac{(d-1)(d-2)}{d^2}(5+6\cos\phi+3\cos^2\phi+\cos^3\phi)+\frac{d-1}{d^2}(9+6\cos\phi),\\
M_8/M_2^4=&\frac{(d-1)(d-2)(d-3)}{d^3}(14+28\cos\phi+28\cos^2\phi+20\cos^3\phi+10\cos^4\phi+4\cos^5\phi+\cos^6\phi)\\
&+ \frac{(d-1)(d-2)}{d^3}(56+86\cos\phi+52\cos^2\phi+16\cos^3\phi).
\end{split}
\end{equation}
Note we will slightly abuse the terms ``leading'' and ``subleading'' for the Parisi's model: the natural parameter for the large $d$ expansion of $M_{2l}/M_2^l$ is not powers of $1/d$ but $d(d-1)\cdots(d-m+1)/d^l$ instead. Hence for the $2l$-th moment, ``leading'' means $m=l$ and ``subleading'' means $m=l-1$.
We see that if we apply Q-Hermite approximation to both the leading and the subleading moments of the sparse SYK model, and make the identification of $\eta=\cos \phi$, then the leading moments of the sparse SYK model Eq. \eqref{eqn:riordanTouchard} are exactly the same as the Parisi leading moments; the $1/kN$ coefficients of the subleading moments of the sparse SYK model Eq. \eqref{mkn} are exactly three times that of the Parisi's hypercube model. In summary we have,
\begin{equation}\label{eqn:sparseSYKtoParisi}
(\text{sparse SYK moments})_{\text{QH}} = (\text{Parisi leading coefficients}) +\frac{3}{kN}(\text{Parisi subleading coefficients} )+ \ldots,
\end{equation}
at least based on the observation of the first eight moments of both models.

We will see now why Eq. \eqref{eqn:sparseSYKtoParisi} is true  not only for the first eight but for all moments.
 In the sparse SYK model, an intersection graph $G$ represents a value of
\begin{equation}\label{eqn:intersGraphValueSparse}
\langle\eta_G\rangle_x=(kN)^{-l}\sum_{\alpha_1,\alpha_2,\ldots,\alpha_l} (-1)^{c(G)} \langle x_{\alpha_1}x_{\alpha_2}\cdots x_{\alpha_l}\rangle,
\end{equation}
where the notation $\langle\eta_G\rangle_x$ serves to distinguish it from its dense SYK counterpart $\eta_G$, and reminds us of the fact that there is an extra averaging over $x$ variables in the sparse SYK model.
 We shall distinguish the two cases in the actual drawings of the intersection graphs by annotating the vertices of the sparse SYK intersection graphs by $(\alpha_i,x_{\alpha_i})$, as opposed to by $\alpha_i$ alone for the dense SYK intersection graphs defined earlier.
 See Fig. \ref{fig:leadingPlusSubleading} for an example.
 If all the subscripts in Eq. \eqref{eqn:intersGraphValueSparse} are different, $\langle x_{\alpha_1}x_{\alpha_2}\cdots x_{\alpha_l}\rangle$ will be equal to $p^l$; if two of the subscripts become equal,  $\langle x_{\alpha_1}x_{\alpha_2}\cdots x_{\alpha_l}\rangle$ will be enhanced by a factor of $1/p$, but the restriction on the summation will suppress the sum by $\binom{N}{q}$, and the total effect is a  $1/(kN)$ suppression. It is clear then that at leading order in $1/(kN)$, Eq. \eqref{eqn:intersGraphValueSparse} is given by
\begin{equation}
\binom{N}{q}^{-l}\sum_{\alpha_1,\alpha_2,\ldots,\alpha_l} (-1)^{c(G)},
\end{equation}
coinciding with the dense SYK model value $\eta_G$. This proves that to leading order, the moments  of the sparse SYK are exactly the same as those of the dense SYK.  Hence, in the Q-Hermite approximation, they are given by the Q-Hermite moments Eq. \eqref{eqn:riordanTouchard}. In the case of the Parisi model, it is already understood that the leading contribution is given by the Q-Hermite prediction \cite{Cappelli-1998}. Therefore, to  leading order, Eq. \eqref{eqn:sparseSYKtoParisi} is proven, namely, the moments of both the dense SYK model (after Q-Hermite approximation) and the Parisi model are given by the Q-Hermite prediction. This is perhaps not too surprising, but we will see that the subleading correction in Eq. \eqref{eqn:sparseSYKtoParisi} arises in a much more subtle and surprising way.

The subleading order of Eq. \eqref{eqn:intersGraphValueSparse} can be written as
\begin{equation}\label{eqn:subleadingTerms}
(kN)^{-1}\binom{N}{q}^{-l+1}\overbrace{\left(\sum_{\alpha_1=\alpha_2,\alpha_3\ldots,\alpha_l}(-1)^{c(G)}+\sum_{\alpha_1=\alpha_3,\alpha_2,\ldots,\alpha_l}(-1)^{c(G)}+\ldots+\sum_{\alpha_1,\alpha_2,\ldots,\alpha_{l-1}=\alpha_l}(-1)^{c(G)}\right)}^{\binom{l}{2}} ,
\end{equation}
where there are $\binom{l}{2}$ sums corresponding to letting two out of the $l$ subscripts be equal. One might worry about excluding the cases where even more indices are equal, but they are of higher order and do not enter into our consideration here. We can summarize the above results as
\begin{equation}\label{eqn:leadingPlusSubleading}
\langle \eta_G\rangle_x = \eta_G+\binom{l}{2}\text{ subleading terms in Eq. \eqref{eqn:subleadingTerms}}+O\left(\frac{1}{N^2}\right).
\end{equation}
For example, Fig. \ref{fig:leadingPlusSubleading} gives an intersection graph $G$ that contributes the sixth moment, and in the form of Eq. \eqref{eqn:leadingPlusSubleading} its value can be written as
\begin{align}\label{eqn:leadingPlusSubleadingExample}
&(kN)^{-3}\sum_{\alpha_1,\alpha_2,\alpha_3} (-1)^{c_{\alpha_1\alpha_2}+c_{\alpha_2\alpha_3}} \langle x_{\alpha_1}x_{\alpha_2}x_{\alpha_3}\rangle \nn\\
=&\binom{N}{q}^{-3}\sum_{\alpha_1,\alpha_2,\alpha_3} (-1)^{c_{\alpha_1\alpha_2}+c_{\alpha_2\alpha_3}}+\frac{1}{kN} \binom{N}{q}^{-2}\left(\sum_{\alpha_2,\alpha_3}(-1)^{c_{\alpha_2\alpha_2}+c_{\alpha_2\alpha_3}}+\sum_{\alpha_1,\alpha_2}(-1)^{c_{\alpha_1\alpha_2}+c_{\alpha_1\alpha_2}}+\sum_{\alpha_1,\alpha_2}(-1)^{c_{\alpha_1\alpha_2}+c_{\alpha_2\alpha_2}}\right)\\
=&\binom{N}{q}^{-3}\sum_{\alpha_1,\alpha_2,\alpha_3} (-1)^{c_{\alpha_1\alpha_2}+c_{\alpha_2\alpha_3}}+\frac{1}{kN} \binom{N}{q}^{-2}\left(\sum_{\alpha_2,\alpha_3}(-1)^{c_{\alpha_2\alpha_3}}+\sum_{\alpha_1,\alpha_2}1+\sum_{\alpha_1,\alpha_2}(-1)^{c_{\alpha_1\alpha_2}}\right)+O(1/N^2),\nn
\end{align}
where from the second line to the third line we used $(-1)^{c_{\alpha_2\alpha_2}}=(-1)^q=1$ and $(-1)^{2c_{\alpha_1\alpha_2}}=1$.
\begin{figure}
\begin{center}
\begin{tikzpicture}
\draw[fill=black] (0,0) circle (1pt);
\draw[fill=black] (1,0) circle (1pt);
\draw[fill=black] (0.5,1.732/2) circle (1pt);

\node at (0.6,1.732/2+0.2) {$(\alpha_1, x_{\alpha_1})$};
\node at (-0.3,-0.25) {$(\alpha_2, x_{\alpha_2})$};
\node at (1.3,-0.25) {$(\alpha_3, x_{\alpha_3})$};

\draw (0.5,1.732/2) --(0,0)--(1,0);

\node at (2.2,1.732/4) {$=$};

\draw[fill=black] (3,0) circle (1pt);
\draw[fill=black] (4,0) circle (1pt);
\draw[fill=black] (3.5,1.732/2) circle (1pt);

\node at (3.6,1.732/2+0.2) {$\alpha_1$};
\node at (2.9,-0.25) {$\alpha_2$};
\node at (4.2,-0.25) {$\alpha_3$};

\draw (3.5,1.732/2) --(3,0)--(4,0);

\node at (4.5,1.732/4) {$+$};
\node at (6,1.732/4) {subleading terms};
\node at (8,1.732/4) {$+\ O\left(\frac{1}{N^2}\right)$};
\end{tikzpicture}
\end{center}
\caption{An intersection graph example of Eq. \eqref{eqn:leadingPlusSubleading}. Note the left-hand side and the first term on the right-hand side have identical graph but different labeling of the vertices, and hence the left represents $\langle \eta_G\rangle_x$ whereas the first term on the right represents $\eta_G$.}\label{fig:leadingPlusSubleading}
\end{figure}

\begin{figure}[b!]
\begin{center}
\begin{tikzpicture}
\draw[fill=black] (1,0) circle (1pt);
\draw[fill=black] (2,0) circle (1pt);
\draw[fill=black] (1.5,0.866) circle (1pt);
\draw[] (1.5,0.866)  -- (1,0) -- (2,0);

\node at (0.9,-0.2) {$\alpha_2$};
\node at (2.2,-0.2) {$\alpha_3$};
\node at (1.6,1.066) {$\alpha_1$};

\node at (2.5,0.43) {$\rightarrow$};

\draw[fill=black] (3,0) circle (1pt);
\draw[fill=black] (3.5,0.866) circle (1pt);
\draw[] (1.5,0.866)  -- (1,0) -- (2,0);

\node at (3.3,0) {$\alpha_2$};
\node at (3.65,1.066) {$\alpha_1$};
\draw[] (3.5,0.866)  -- (3,0);
\draw [] plot [smooth cycle] coordinates {(3,0) (3.2,-0.4)(2.8,-0.4)};

\node at (4.5,0.43) {$\rightarrow$};

\draw[fill=black] (5,0) circle (1pt);
\draw[fill=black] (5.5,0.866) circle (1pt);
\draw[] (5.5,0.866)  -- (5,0);

\node at (5,-0.2) {$\alpha_2$};
\node at (5.65,1.066) {$\alpha_1$};
\node at (3,-1) {merge $\alpha_2$ and $\alpha_3$};

\draw[fill=black] (7,0) circle (1pt);
\draw[fill=black] (8,0) circle (1pt);
\draw[fill=black] (7.5,0.866) circle (1pt);
\draw[] (7.5,0.866)  -- (7,0) -- (8,0);

\node at (6.9,-0.2) {$\alpha_2$};
\node at (8.2,-0.2) {$\alpha_3$};
\node at (7.65,1.066) {$\alpha_1$};

\node at (8.5,0.43) {$\rightarrow$};

\draw[fill=black] (9,0) circle (1pt);
\draw[fill=black] (9.5,0.866) circle (1pt);
\node at (9,-0.2) {$\alpha_2$};
\node at (9.65,1.066) {$\alpha_1$};

\draw [] plot [smooth cycle] coordinates {(9,0) (9.3,0.4)(9.5,0.866)(9.1,0.5)};

\node at (10.5,0.43) {$\rightarrow$};

\draw[fill=black] (11,0) circle (1pt);
\draw[fill=black] (11.5,0.866) circle (1pt);
\node at (11,-0.2) {$\alpha_2$};
\node at (11.65,1.066) {$\alpha_1$};
\node at (9,-1) {merge $\alpha_1$ and $\alpha_3$};

\end{tikzpicture}
\end{center}
\caption{The ``merge and delete'' procedures applied in the example of Eq. \eqref{eqn:leadingPlusSubleadingExample}. We show how  to obtain the subleading-moment intersection graphs from the leading-moment intersection graph in Fig.~\ref{fig:leadingPlusSubleading}.There are three subleading graphs corresponding to the merging of $\alpha_1\alpha_2$, $\alpha_1\alpha_3$ and $\alpha_2\alpha_3$, we have drawn two of them because merging $\alpha_1\alpha_2$ and merging $\alpha_2\alpha_3$ result in identical graphs. }\label{fig:mergeDelete}
\end{figure}
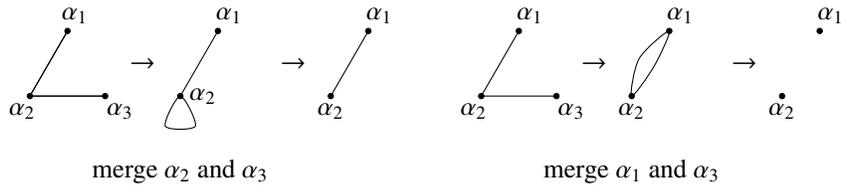
From this example, it is clear that there is a natural graphical representation of the subleading calculations:
\begin{enumerate}
\item Merge two vertices $\alpha_i, \alpha_j$ of  the intersection graph $G$, and let the merged vertices inherit the original edges.
\item There may be loops (an edge connecting a vertex back to itself, representing $(-1)^q$) and 2-multi-edges (two edges connecting the same pair of vertices, representing $(-1)^{2 c_{\alpha_i\alpha_j}}$) formed after step 1, delete all such loops and 2-multi-edges. Call the resulting graph $G_{(\alpha_i,\alpha_j)}$.
\item The subleading contribution to $\langle\eta_G\rangle_x$ is given by
\begin{equation}
\frac{1}{kN} \sum_{\{\alpha_i,\alpha_j\}\subset v(G)} \eta_{G_{(\alpha_i,\alpha_j)}},
\end{equation}
where $v(G)$ denotes the vertex set of $G$, and $\eta_{G_{(\alpha_i,\alpha_j)}}$ is the value for which the intersection graph $G_{(\alpha_i,\alpha_j)}$ would represent a dense SYK model. When the Q-Hermite approximation is applied, this gives
\begin{equation}\label{eqn:subleadingQH}
\frac{1}{kN} \sum_{\{\alpha_i,\alpha_j\}\subset v(G)} \eta^{E\left[G_{(\alpha_i,\alpha_j)}\right]},
\end{equation}
where $E\left[G_{(\alpha_i,\alpha_j)}\right]$ denotes the number of edges in the graph $G_{(\alpha_i,\alpha_j)}$.
\end{enumerate}
Fig. \ref{fig:mergeDelete} illustrates an example of the application of these rules. The above ``merge and delete'' graphical rules to calculate the subleading moments, which result in Eq. \eqref{eqn:subleadingQH}, are exactly the same as the ``averaged scheme'' defined in
\cite{Jia:2020rfn} for calculating the subleading moments of the Parisi's hypercube model, except that the averaged scheme for the Parisi's model has an extra factor of $1/3$. We can now conclude that Eq. \eqref{eqn:sparseSYKtoParisi} holds for all moments. Such coincidence does not hold to the next order in $1/(kN)$, as is evident by comparing the eighth moments of the sparse SYK and the Parisi's model at higher orders.

\subsection{The renormalized and subleading Q-Hermite approximations}

\begin{figure}[t!]
\centerline{   \includegraphics[width=8cm]{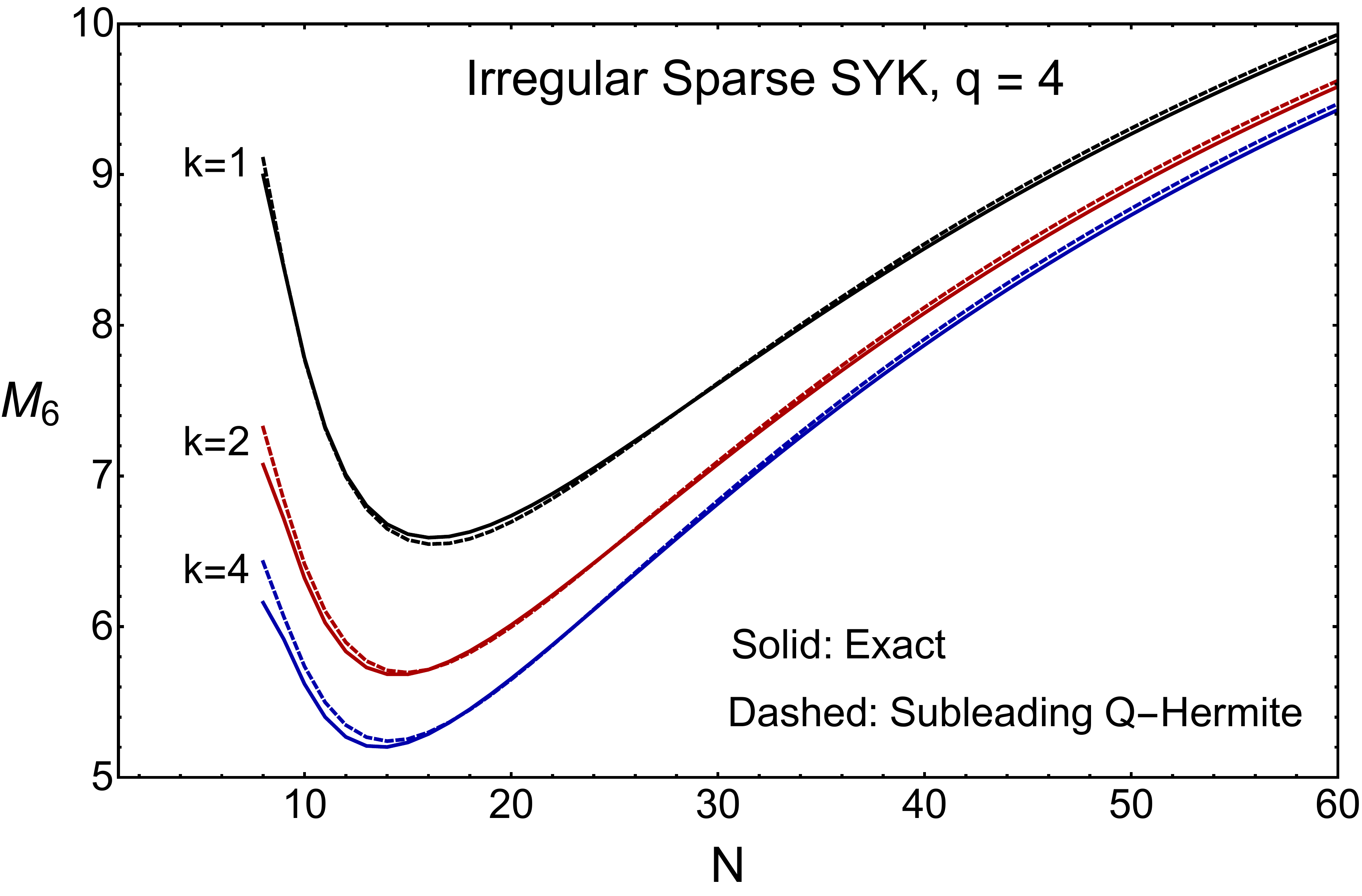}
   \includegraphics[width=8cm]{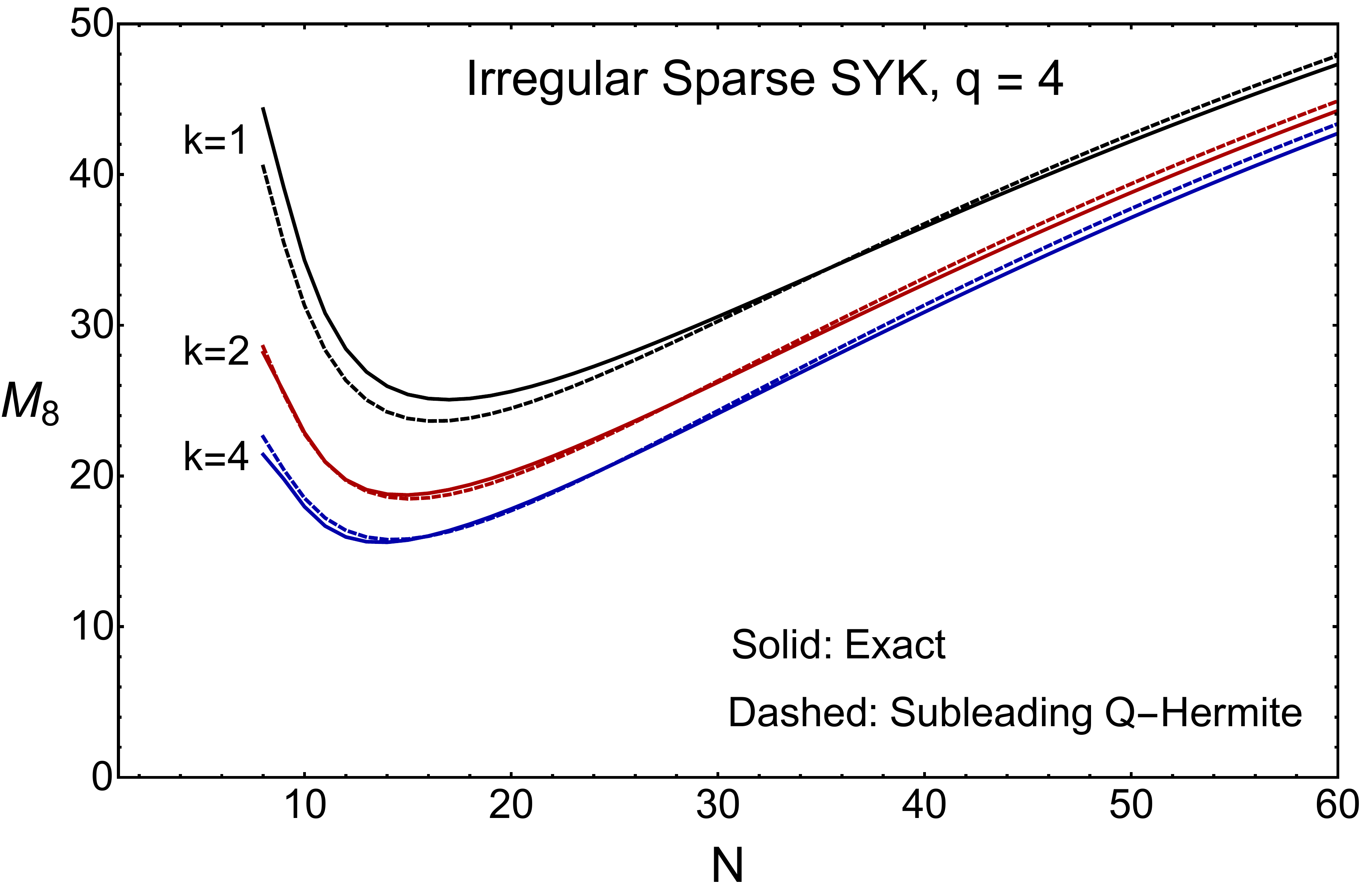}}
\caption{The accuracy of subleading Q-Hermite approximation. The sixth and eighth moments for $q=4$ are plotted with varying $N$. The solid lines represent the exact moments; the dashed lines represent the sum of leading and subleading moments, both with Q-Hermite approximation applied.  }\label{fig:subleadingQHaccuracy}
\end{figure}

The leading intersection graphs $G$ introduced in the previous section can be summed by the Riordan-Touchard
formula \cite{touchard1952,riordan1975} after applying the Q-Hermite approximation, $\eta_G\approx\eta^{E(G)}$, to both the leading and subleading terms,

\be
\begin{split}
\frac{M_{2l}}{M_2^l}\approx &   \sum_{G} \eta^{E(G)} +\frac{1}{kN} \sum_{G}\sum_{\{\alpha_i,\alpha_j\}\subset v(G)} \eta^{E\left[G_{(\alpha_i,\alpha_j)}\right]}\\
=&\frac 1{(1-\eta)^l} \sum_{i=-l}^l (-1)^i
{2l\choose i+l}\eta^{i(i-1)/2}+\frac{1}{kN} \sum_{G}\sum_{\{\alpha_i,\alpha_j\}\subset v(G)} \eta^{E\left[G_{(\alpha_i,\alpha_j)}\right]}.
\label{eqn:subleadingQHmoms}
\end{split}
\ee
We will call  Eq. \eqref{eqn:subleadingQHmoms} the \textit{subleading Q-Hermite approximation}.
One can easily check that only including the leading term results in a fairly large discrepancy with the exact result so the subleading term is an important contribution. We would like to get a grasp of how accurate the subleading Q-Hermite approximation is. In Fig. \ref{fig:subleadingQHaccuracy} we compare the exact results for the sixth and eighth moments with the sixth and eighth moments approximated this way at different values of $N$ and $k$. Rather surprisingly, this approximation works quite well, even for $k=1$ or small $N$, provided that $1/(kN) \ll 1$.
  \begin{figure}[b!]
\centerline{   \includegraphics[width=8cm]{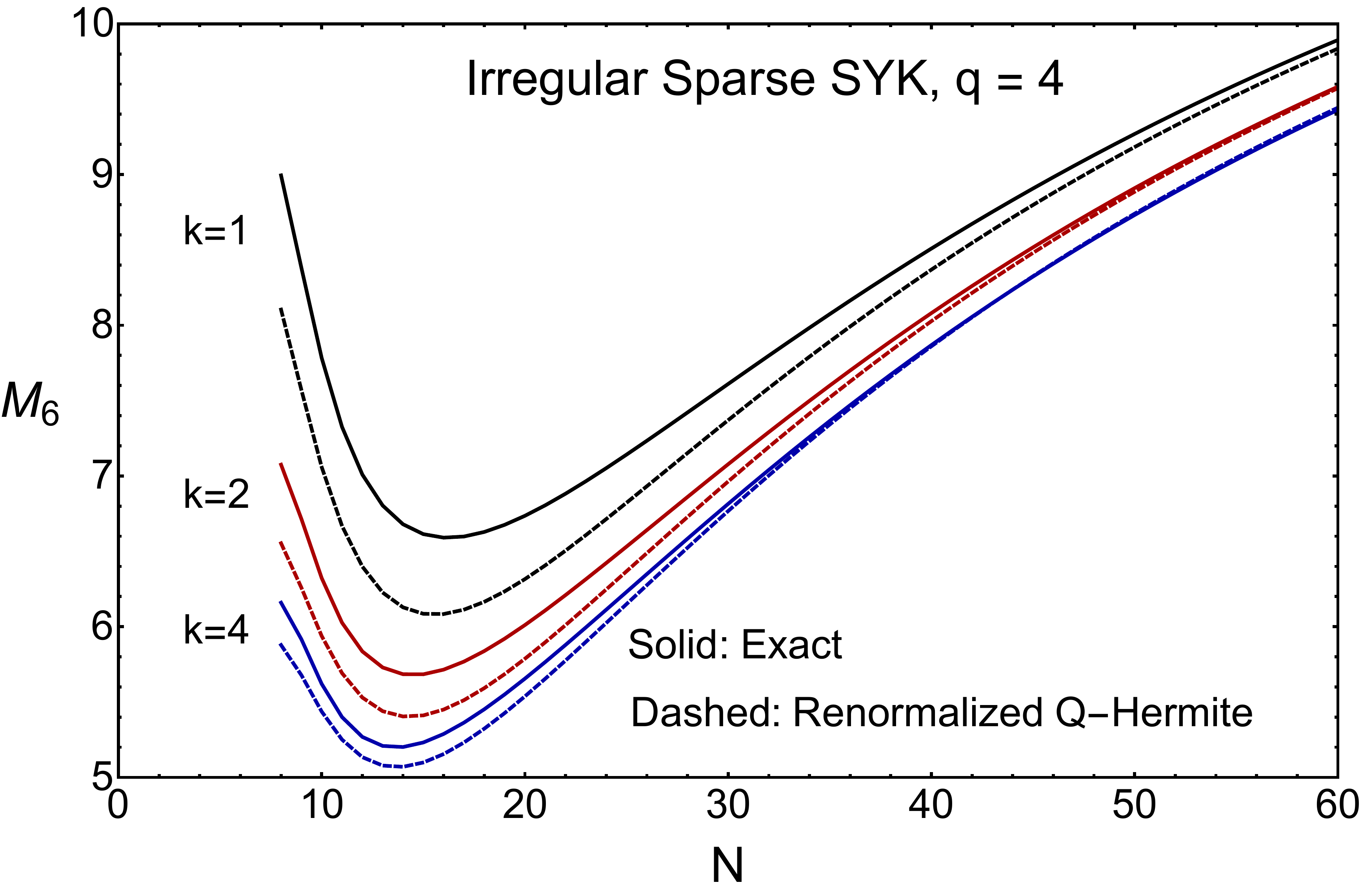}
   \includegraphics[width=8cm]{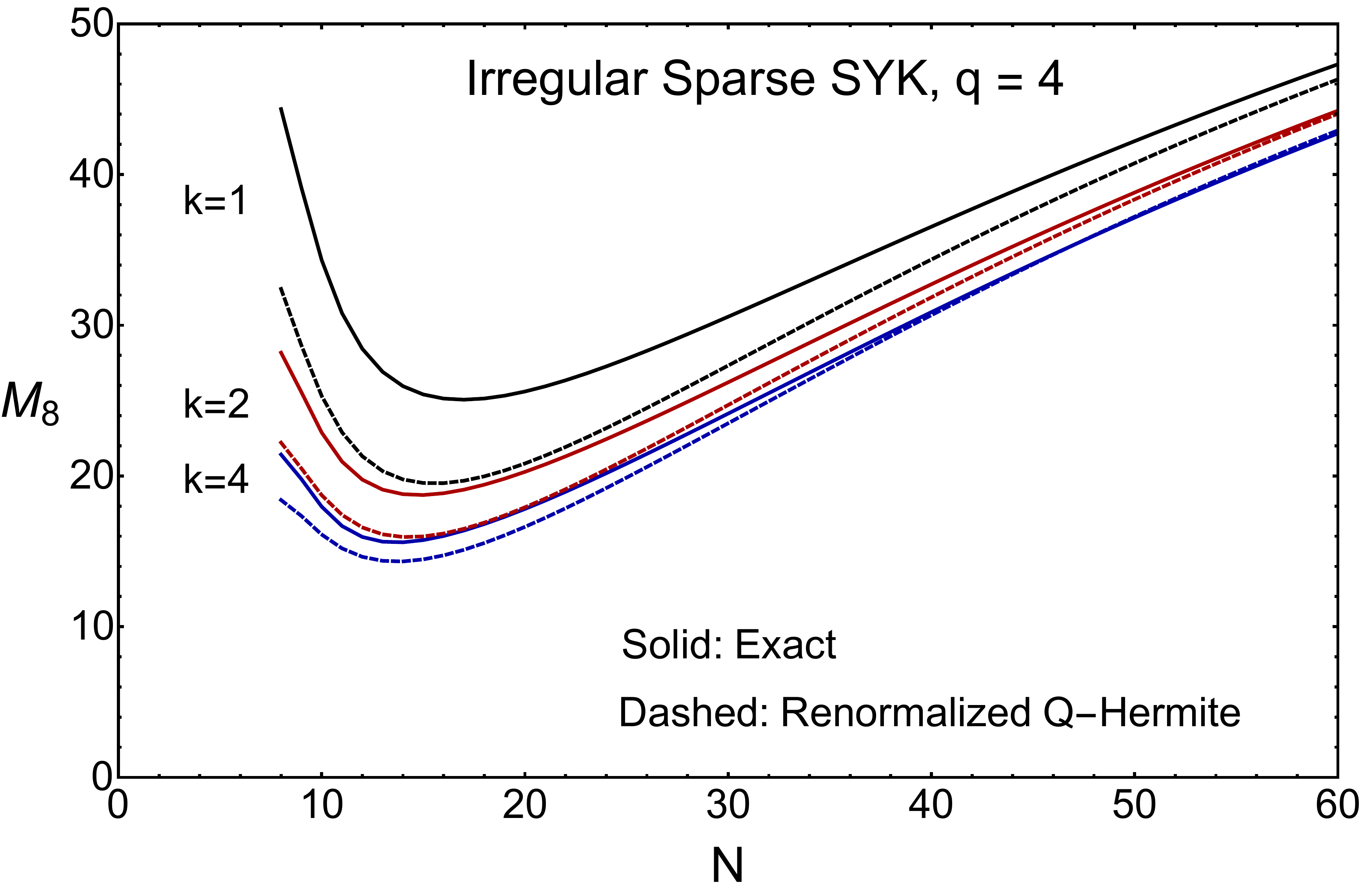}}
\caption{The accuracy of renormalized Q-Hermite approximation. The sixth and eighth moments for $q=4$ are plotted as a function of $N$. The solid lines represent the exact moments; the dashed lines represent the Q-Hermite approximation applied to the leading moments, but with a renormalized $\eta(k)=\eta +3/(kN)$.  }\label{fig:renormalizedQHaccuracy}
\end{figure}

The Riordan-Touchard formula enables us to calculate the leading term of arbitrarily high moment very efficiently, and gives the analytic expression Eq. \eqref{eqn:QHden} for the spectral density to leading order. Marinari, Parisi and Ritort \cite{Marinari:1995jwr} computed the subleading term in Eq. \eqref{eqn:subleadingQHmoms}  up to the 18th moment numerically, but we are not yet able to find a Riordan-Touchard-like formula for the subleading terms. A related difficulty is then that we are not able to write down an analytic expression for the subleading spectral density. At this point we can simply remark that it would be worthwhile to find a Riordan-Touchard-like formula since it would solve the subleading problem of two models at one stroke.
This difficulty prompts us to try a different strategy of approximating moments. We will only use the leading moment expression, but with a renormalized $\eta$, with the hope it can capture  subleading effects beyond it natural range of applicability of relatively low moments. 
More specifically,  calling this renormalized parameter $\eta(k)$, the moments are given by,
\begin{equation}
\frac{M_{2l}}{M_2^l}\approx  \sum_{G} \eta(k)^{E(G)}
=\frac 1{(1-\eta(k))^l} \sum_{i=-l}^l (-1)^i
{2l\choose i+l}\eta(k)^{i(i-1)/2}.
\label{eqn:renormalizedQHmoms}
\end{equation}
We dub this approximation the \textit{renormalized Q-Hermite approximation}.

 The lowest moment in which $\eta$ starts to make an appearance is $M_4/M_2^2$. The renormalized Q-Hermite approximation Eq. \eqref{eqn:renormalizedQHmoms} predicts $M_4/M_2^2=2+\eta(k)$ whereas the exact result Eq.  \eqref{fourm} gives $M_4/M_2^2=2+\eta+3/(kN)$ up to subleading corrections in the $1/kN$ expansion. Hence, a simple matching gives
\begin{equation}\label{eqn:renormalizedEta}
\eta(k) = \eta + \frac{3}{kN}.
\end{equation}
We remark that this renormalized Q-Hermite approximation already fails to fully capture the subleading term of the sixth moment (except at $\eta=1$). However, this approximation can be justified a posteriori: we shall see it is surprisingly close to the exact moments for certain ranges of $N$ and $k$ and to the resulting spectral density as well. 

In order to gain a more quantitative understanding of the suitability of these approximations, we compare the subleading Q-Hermite, see Fig.  \ref{fig:subleadingQHaccuracy}, and the renormalized Q-Hermite approximation, see Fig.~   \ref{fig:renormalizedQHaccuracy}, with exact results for the sixth and eighth moments of the $q=4$ model Eq.~(\ref{hami}). We have observed that:
\begin{enumerate}
\item In the very sparse limit, $k=1$, the subleading Q-Hermite approximation is the 
better approximation for $N \lessapprox 60$.
\item When $N$ is relatively small, $N \lessapprox 30$, the subleading Q-Hermite approximation is the better approximation.
\item For larger $k$, such as $k\geq 3$, the accuracy of the renormalized Q-Hermite approximation starts to catch up with that of the subleading Q-Hermite approximation, and rather surprisingly at first glance, beyond $N=40$ its accuracy exceeds that of the subleading Q-Hermite approximation. This can be understood
partly
  from the observation that for $\eta =1$ the renormalization cancels the $1/(kN)$ terms exactly in
  case of the sixth and eighth moments.
  In fact, it can be shown
  \footnote{This is because  the renormalized Q-Hermite result for the $2l$-th reduced moment is $\sum_{G} \left(\eta+3/kN)\right)^{E(G)} =\sum_{G} \eta^{E(G)}+3/kN \sum_{G} E(G)\eta^{E(G)-1} $ to subleading order. At $\eta = 1$ the subleading term of this expression becomes $3/kN \sum_{G} E(G) = \frac 1{kN} \binom{l}{2} (2l-1)!! $, where the equality follows from an edge-counting result from \cite{garcia2018c,flajolet2000}. On the other hand at $\eta = 1$ the subleading term produced by the merge and delete procedure is simply the $1/kN$ times the total number of subleading graphs, which is clearly $\binom{l}{2}$ times the total number of leading graphs, that is, $\binom{l}{2}(2p-1)!!$. So we see the two calculations give the same result at (and only at) $\eta=1$.}, that this observation for $\eta=1$ is true for all moments due to a result for edge counting of intersecting graphs. We will see in the next section that this results in a surprisingly good agreement between the renormalized Q-Hermite prediction and the numerical spectral density.

\end{enumerate}
Finally, we note that the moments in Eq. \eqref{eqn:renormalizedQHmoms} give rise to the same spectral density as in Eq. \eqref{eqn:QHden} but with $\eta$ replaced by its renormalized version $\eta(k)$:
\be
\label{eden}
\rho^{\rm ren}_{\rm QH}(E) = c_N\sqrt{1-(E/E_0(k))^2} \prod_{m=1}^\infty
\left [1 - 4 \frac {E^2}{E_0^2}
\left ( \frac 1{2+\eta(k)^m+\eta(k)^{-m}}\right )\right],
\ee
where $c_N$ is a normalization constant, and
\be \label{e0edge}
E_0(k) =-\sqrt{ \frac {4 \sigma^2}{1-\eta(k)}}
\ee
is the ground state energy.  

We compare in next section this analytical prediction with the numerical spectral density from the exact diagonalization of the sparse SYK Hamiltonian Eq.~(\ref{hami}).

\subsection{Conditions for the existence of a gravity dual}

A distinctive feature of the existence of a gravity dual in the context of the SYK model is that, for $E$ sufficiently close to the ground state $E_0$, the spectral density becomes,
\be \label{sins}
\rho_{\rm Schw}(E) \sim \sinh(\gamma \sqrt{E-E_0})
\ee
with $\gamma$ a non-universal constant directly related to $\eta$.
This is the result of the exact quantum path integral computation of the classical Schwarzian action \cite{stanford2017} which is $1/N$ exact. The classical Schwarzian captures the soft breaking from conformal to $SL(2,R)$ symmetry that characterizes both, the infrared limit of the SYK model and certain near AdS$_2$ backgrounds \cite{maldacena2016a,kitaev2015,maldacena2016}. These symmetry considerations are enough to determine the effective low energy theory that is then quantized.

The analytical moment calculation that we have carried out indicates that, for the sparse SYK model with $\alpha < 3$, corrections due to the sparsity of the Hamiltonian are subleading with respect to $1/N$ corrections which
strongly suggests that the spectral density is still given by Eq.~(\ref{sins}) and  therefore it could still have a gravity dual.
The case $\alpha = 3$ is more interesting. The leading correction due to the sparsity of the Hamiltonian is of order $1/kN$ and therefore it modifies the expansion leading to the Q-Hermite approximation in the dense SYK. However, the analytical moment calculation earlier in this section, together with the comparison of the renormalized Q-Hermite approximation with numerical results supports that Eq.~(\ref{eden}) provides a good description of the spectral density of the model for large but finite $N$ and even relatively small $k$ provided that $1/kN$ is small. In principle, this means that the expression for  the spectral density Eq.~(\ref{sins}) is still valid with $\gamma = \gamma(k)$.
This will be shown explicitly  in Fig.~\ref{fig:tail}, but
it may be argued that we had to remove by hand the strong fluctuations of $E_0$ in order to clearly
observe the edge of the spectrum which casts some doubts on the applicability of Eq.~(\ref{sins}) and indirectly on the existence of a gravity dual. We think that these concerns are unfounded. The fluctuations in $E_0$ are a direct consequence of the quantization procedure we have followed.
Instead of picking up the classical low energy effective theory and then quantizing the gravitational degrees of freedom of interest, we are quantizing the full theory without suppressing other degrees of freedom which leads to strongly enhanced fluctuations. Moreover, the collective excitations that induce fluctuations in $E_0$ are also $1/N$ suppressed so we expect them to become a smaller problem if larger $N$ could be explored numerically.
Therefore, we believe that removing the fluctuations of $E_0$, a degree of freedom of no direct interest in our analysis, is an approximation in line with that of first identifying the effective low energy classical action and then proceeding with the quantization \cite{stanford2017}. This is specially true when we have strong evidence that the renormalized Q-Hermite approach provides a very good description of the spectral density in the bulk of the spectrum.

It is a quite exciting prospect that even a strongly sparse SYK model could have a gravity dual.
If so, it may be possible to push this idea further and investigate specific conditions on the geometry of the Fock space which could be favorable to the existence of a gravity dual.
More specifically, it may be possible to establish the minimum requirements on connectivity
so that the spectral density has black-hole like features such a stretched exponential form, $\propto e^{a\sqrt{E-E_0}}$ with $a$ independent of energy, in the infrared limit.

\begin{figure}[t!]
	\centering
	\resizebox{0.49\textwidth}{!}{\includegraphics{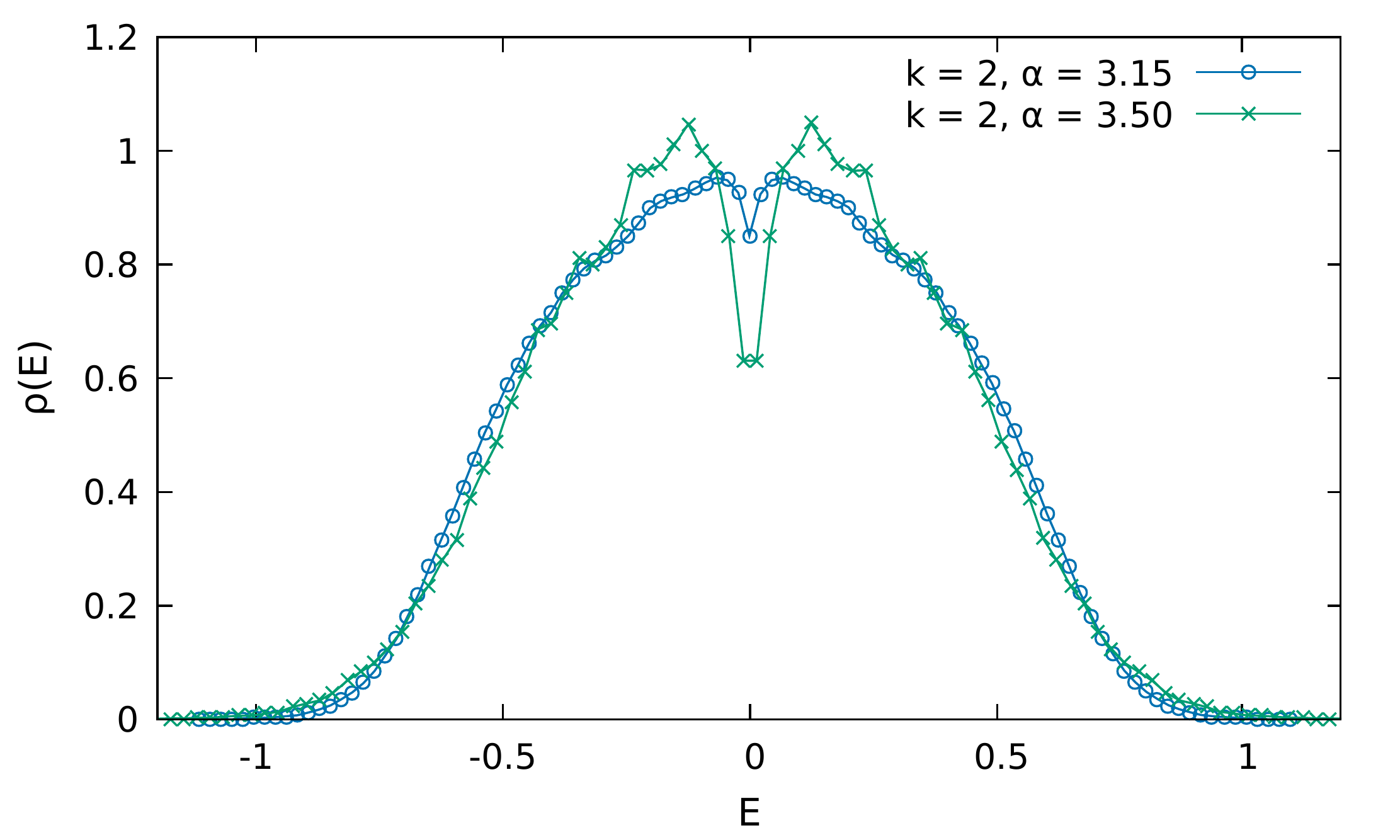}}	\resizebox{0.49\textwidth}{!}{\includegraphics{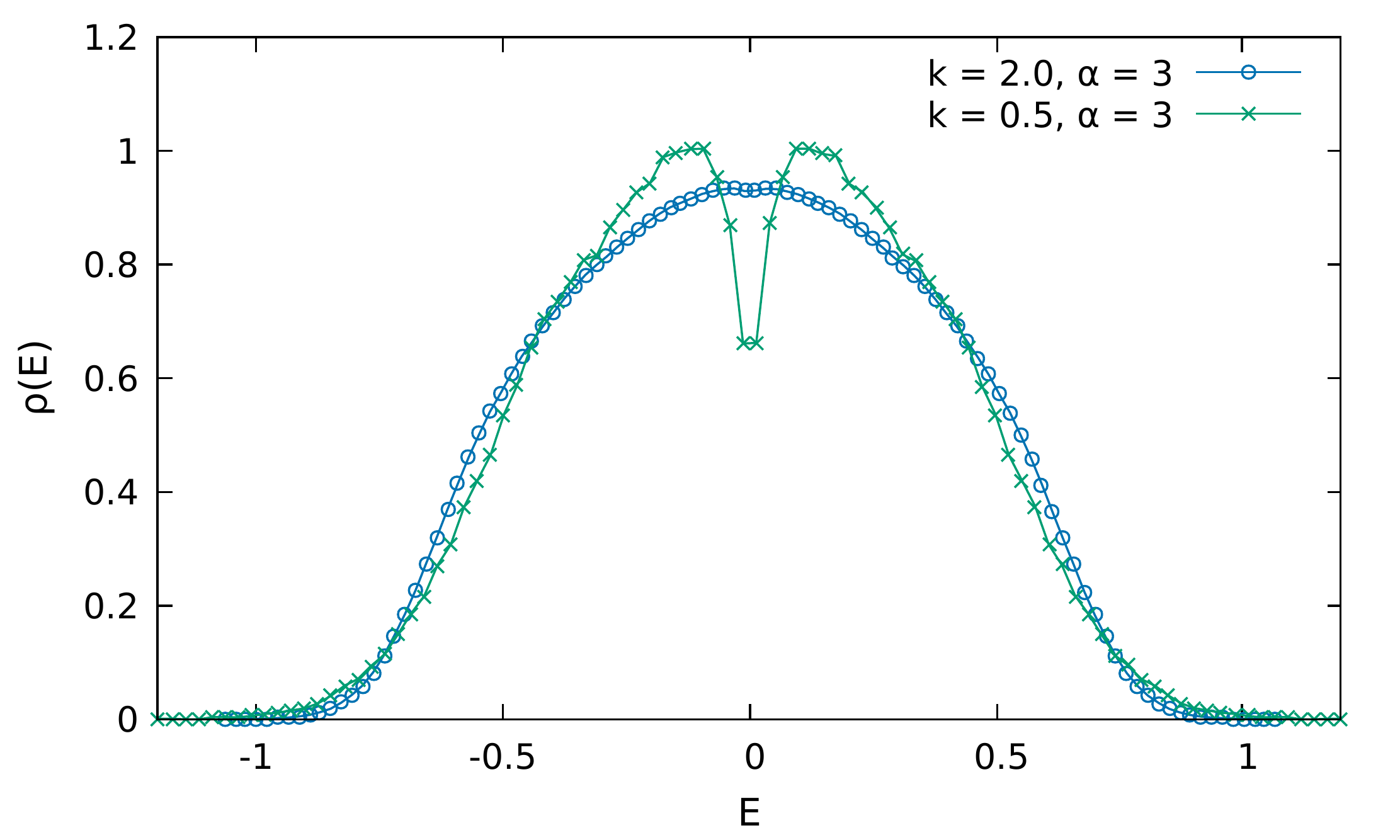}}
	\vspace{-4mm}
	\caption{Spectral density $\rho(E)$ obtained from the exact diagonalization of the Hamiltonian Eq.~(\ref{hami}) for $N = 24$ and $5000$ disorder realizations. Left: For $\alpha > 3$, a depletion of the eigenvalue density occurs for $E \sim 0$ that increases with $\alpha$. Right: For the critical scaling $\alpha = 3$, we observe similar features for $k < 1$. For $k >1$, the density is qualitatively similar to the dense SYK model.}
	\label{dendifsca}
\end{figure}

\section{Spectral Density: Numerical results}\label{sec:specnum}
We compute the eigenvalues of the Hamiltonian Eq.~(\ref{hami}) by exact diagonalization techniques. The resulting spectral density is very sensitive to  the probability $p \sim k/N^{\alpha}$. For $\alpha > 3$, and a small value of $k$, we observe, see Fig.~\ref{dendifsca},  a depletion of eigenvalues around to $E = 0$, and an increase of statistical fluctuations.

\begin{figure}[b!]
	\centering
	\resizebox{0.49\textwidth}{!}{\includegraphics{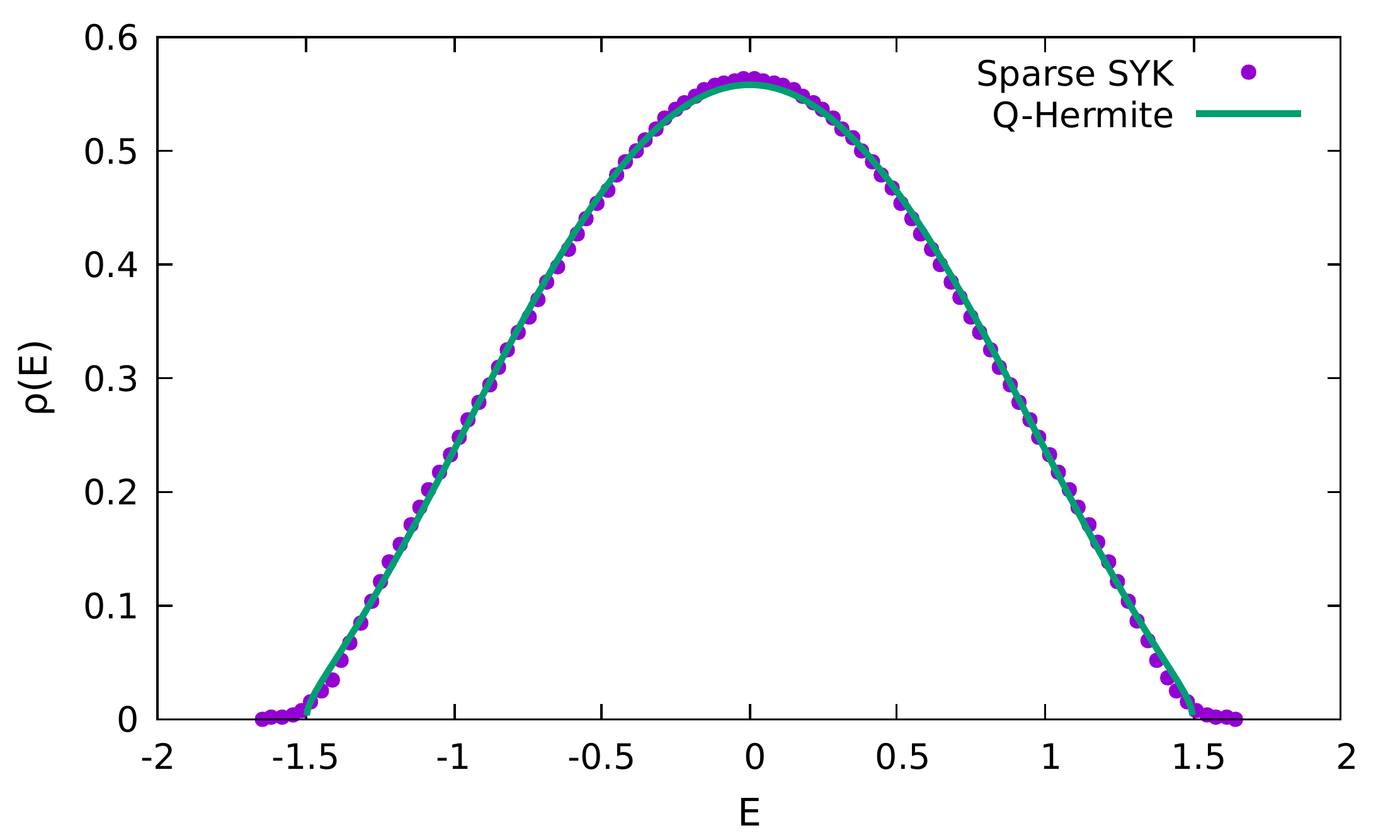}}
	\resizebox{0.49\textwidth}{!}{ \includegraphics{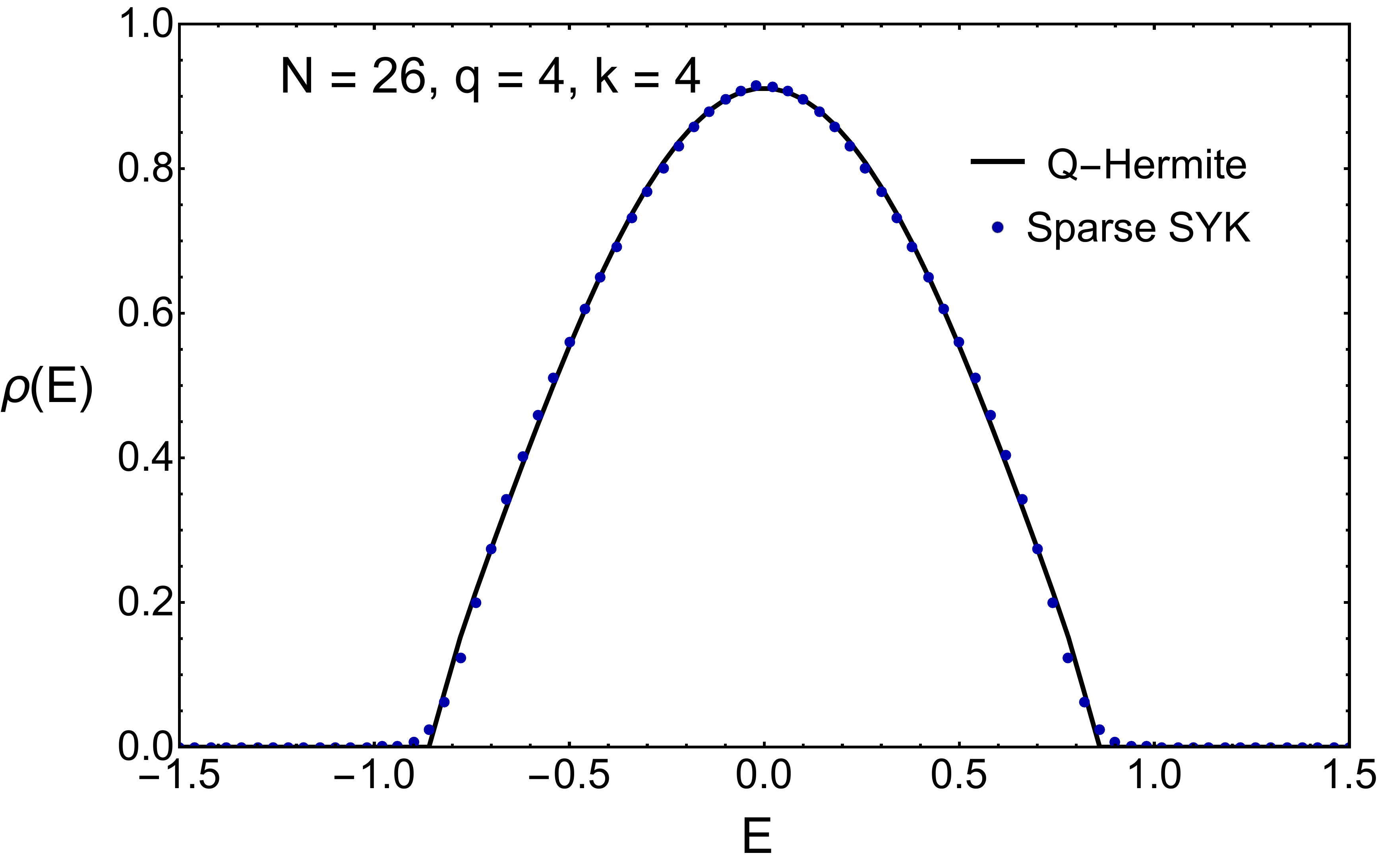}}
	\vspace{-4mm}
	\caption{Left: Spectral density $\rho(E)$ obtained from the exact diagonalization of the Hamiltonian Eq.~(\ref{hami}) for $k = 4$ and comparison with the renormalized Q-Hermite prediction Eq.~(\ref{eden})
          for $N=32$ (left) and $N=26$ (right).
	}
	\label{denana}
\end{figure}

\begin{figure}[t!]
	\centering
	\resizebox{0.49\textwidth}{!}{\includegraphics{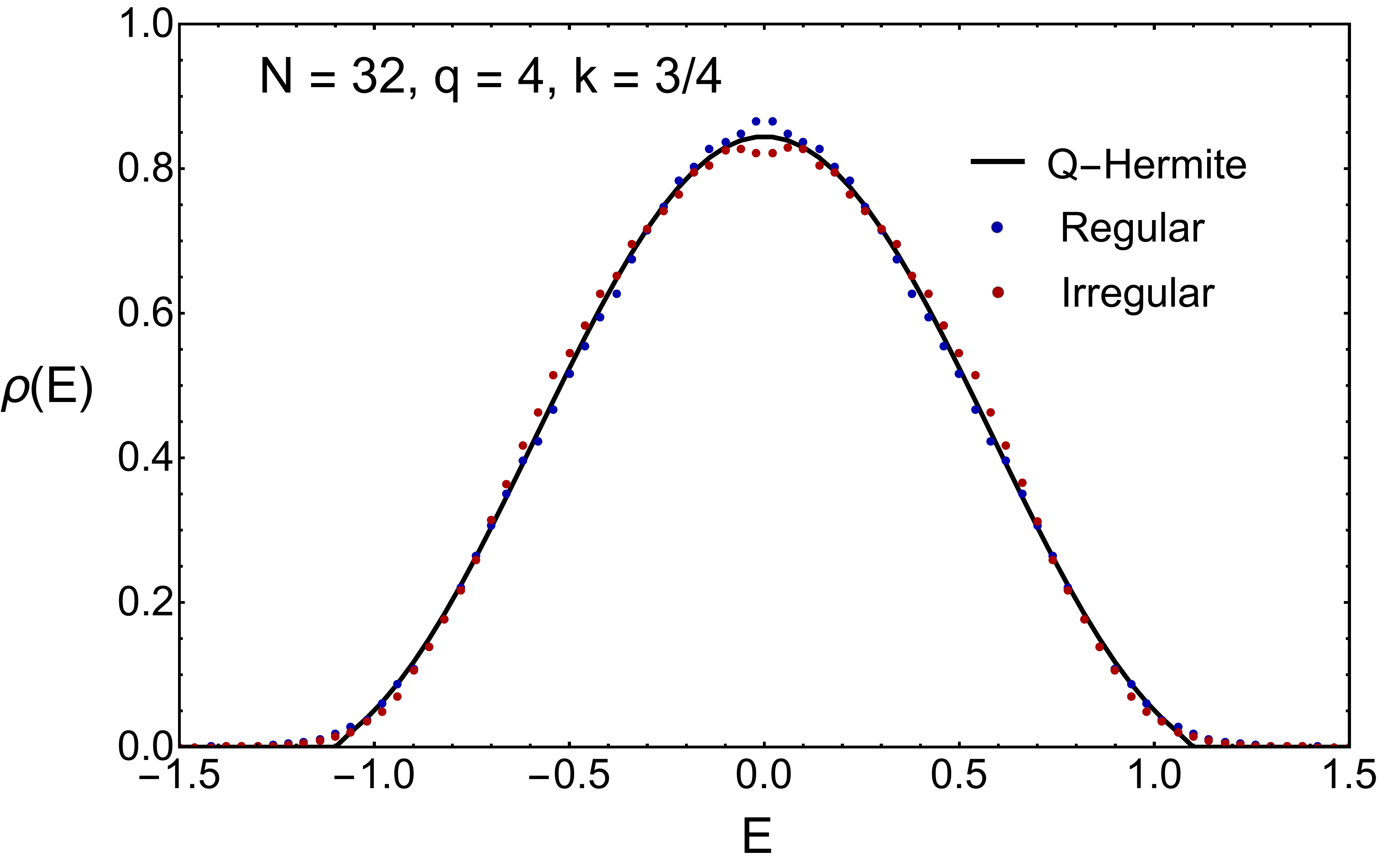}}	\resizebox{0.49\textwidth}{!} {\includegraphics{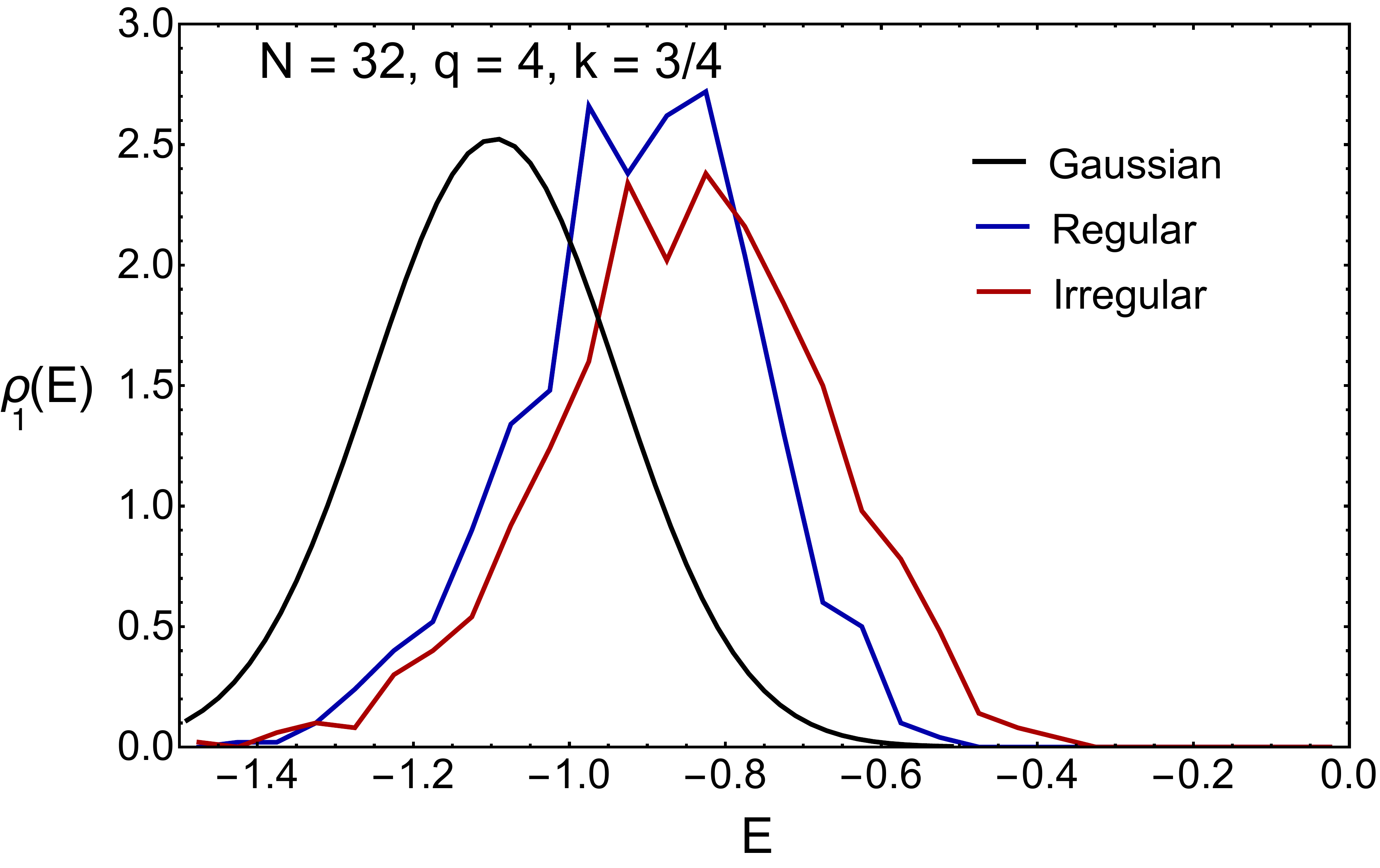}}
	\vspace{-4mm}
	\caption{Left: Spectral density $\rho(E)$ of the Hamiltonian Eq.~(\ref{hami}) for $N=32$ and
          $k=\frac 34$ both with (blue points) and without regularity condition (red points). Right: Distribution of the smallest eigenvalue of theses ensembles of 1000 configurations compared to the Q-Hermite result Eq.~(\ref{eden}).}
	\label{dencomQH}
\end{figure}

For $\alpha < 3$ and $k >1$, we expect the spectral density to be similar to that of the dense SYK model. We focus on the case $\alpha = 3$, that according to the previous analytical results, is the critical case to observe controlled deviations from the results for the dense SYK model. The first question we aim to clarify is whether the spectral density of the sparse SYK in this case is still well described by the Q-Hermite result Eq.~(\ref{eden}) so that the effect of sparsing can be included in a redefinition of $\eta$. This also means that
the low-energy excitations are well described by the Schwarzian prediction which would support the existence of a gravity dual.
In Fig.~\ref{denana}, we show the spectral density for $\alpha=3$ and $k = 4$ and
compare the result with the renormalized Q-Hermite spectral density Eq. \eqref{eden}.
Apart from deviations in the tail region,
we find excellent agreement for both $N=32$ (left)  and $N=26$ (right).
The results for $k=0.75$, where fluctuations from one realization to the next are large,
are shown in Fig. \ref{dencomQH}, left. The Q-Hermite density is again given by Eq. \eqref{eden} with the renormalized parameter $\eta(k) = \eta +3/kN$. We show results with and without regularity condition which has only a minor effect on the spectral density. The good agreement
is surprising in the very sparse regime because
the level density of each realization deviates strongly from the average result. For
example, the width of the spectrum of a realization may be a factor two larger, or the spectrum
may show
macroscopic gaps. To understand better the agreement with the Q-Hermite result we plot
in the right panel of Fig. \ref{dencomQH} the distribution of the smallest eigenvalue
with and without regularity condition. The analytical result is a Gaussian located at
the Q-Hermite prediction for the smallest eigenvalue with a width $\sigma$ determined
by $M_{22}$ (as defined in eq.~\eqref{eq:second_moment_variance}) as $\sigma = 2/kN$. The width is in agreement with the numerical results,
in particular when the regularity condition is imposed, but the average position is
well below the numerical result. How can we reconcile this with the good agreement of
the overall spectral density? Because the ensemble fluctuations of the individual
eigenvalues are much larger than the level spacing, the tail of the spectral density
is not determined by the distribution of the lowest eigenvalue, but rather by the
totality of the distribution of the excited states which are exponentially
close to the ground state.

\begin{figure}[t!]
	\centering
	\resizebox{0.49\textwidth}{!}{ \includegraphics{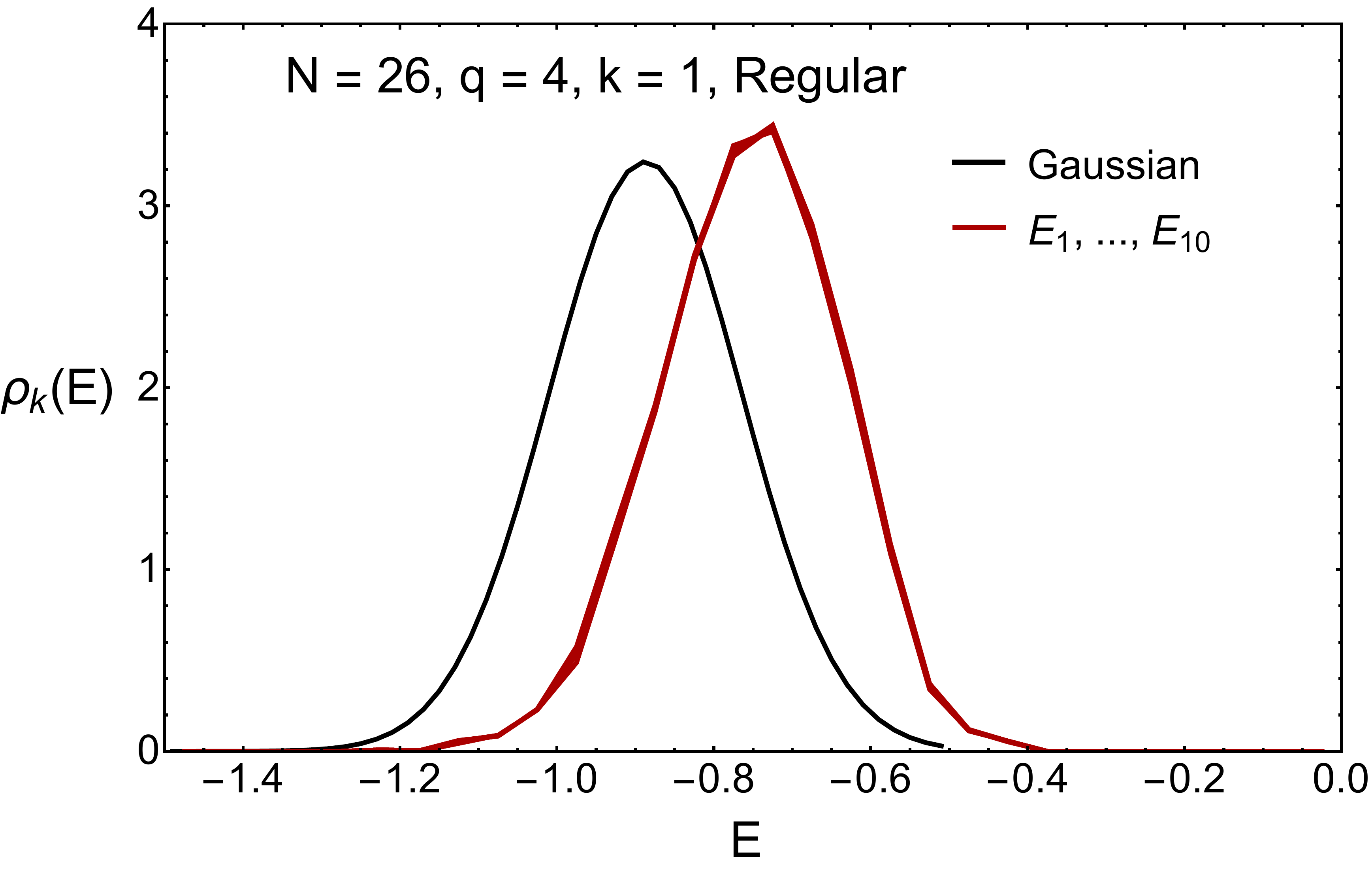}}
	\resizebox{0.49\textwidth}{!}{ \includegraphics{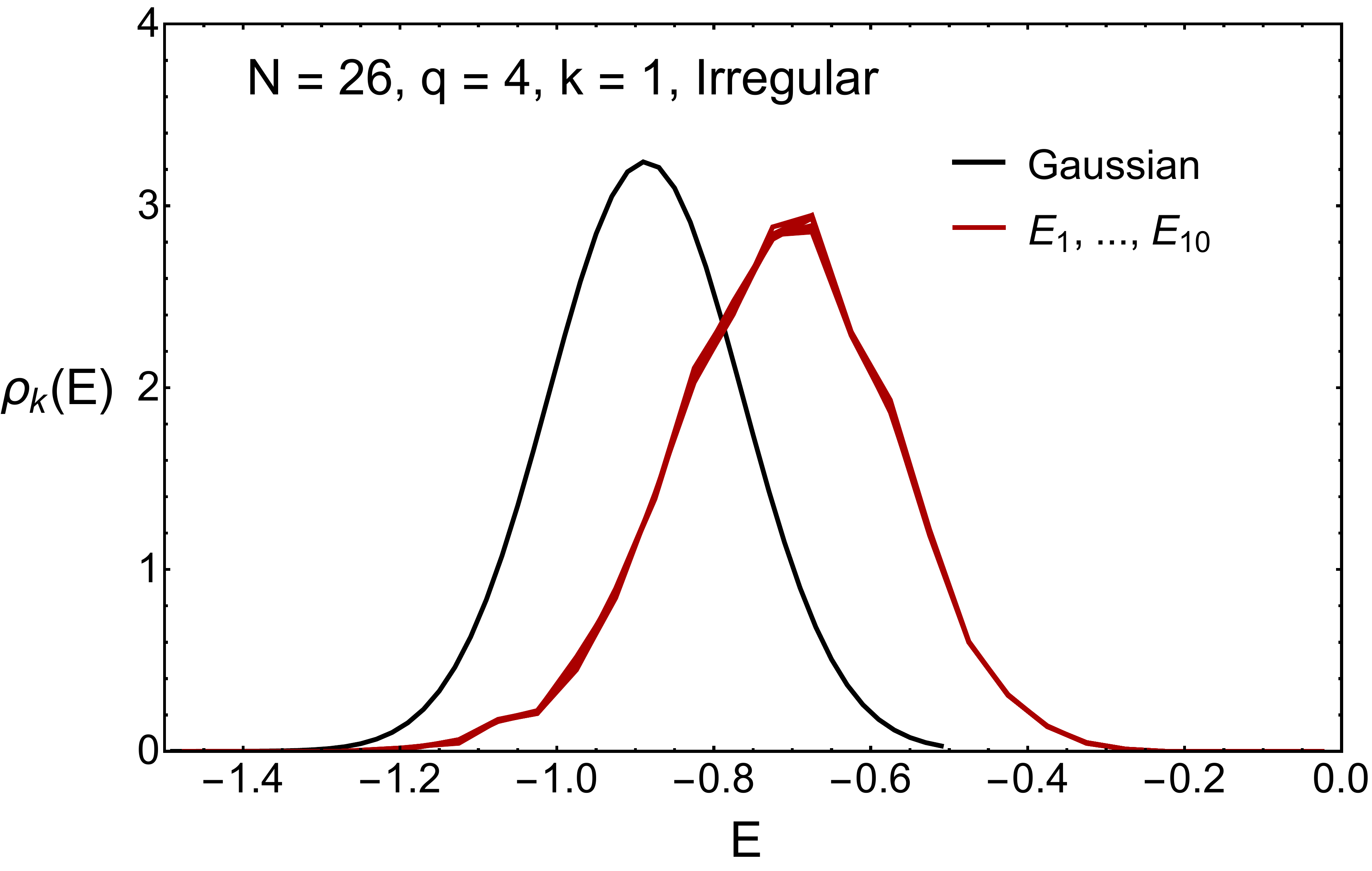}}
	\vspace{-4mm}
	\caption{Distribution of the ten smallest eigenvalues for an ensemble of
          $5000$ configurations for $N=26$ and $k=1$ imposing the regularity condition (red curves, left)
          and without imposing the regularity condition (red curves, right).
          The analytical result given by the black curve has the right width but its average
          position disagrees with the numerical results.}
	\label{fig:emin1-10}
\end{figure}

The situation gets better for larger $k$. In Fig.~\ref{fig:emin1-10} we show the
distribution of the first 10 eigenvalues (red curves) for $N=26$ and $k=1$ with (left) and without (right)  regularity
condition as well as the analytical result for the distribution
of the smallest eigenvalues (black curve). The width of the distribution is much larger than the
level spacing, and the distributions of the
first 10 eigenvalues  are almost identical. The analytical and numerical
are clearly closer, and it is clear that a large number of small eigenvalues contribute
to the tail of the spectral density.

These results are a strong indication that $\alpha = 3$ and $k \sim 1$
is the maximum degree of sparseness, or the connectivity in Fock space,
that can support the existence of a gravity dual.

Despite the good agreement, we observe visible differences in the
infrared part of the spectrum. The numerical result has a smooth tail while
the renormalized Q-Hermite density predicts an edge. The reason behind the numerical tail
is the strong fluctuations of $E_0$ for different disorder realizations. It is well
known that disorder induces collective excitations in the spectrum,
which blur the existence of spectral edges.

\begin{figure}[b!]
	\centering
	\resizebox{0.49\textwidth}{!}{ \includegraphics{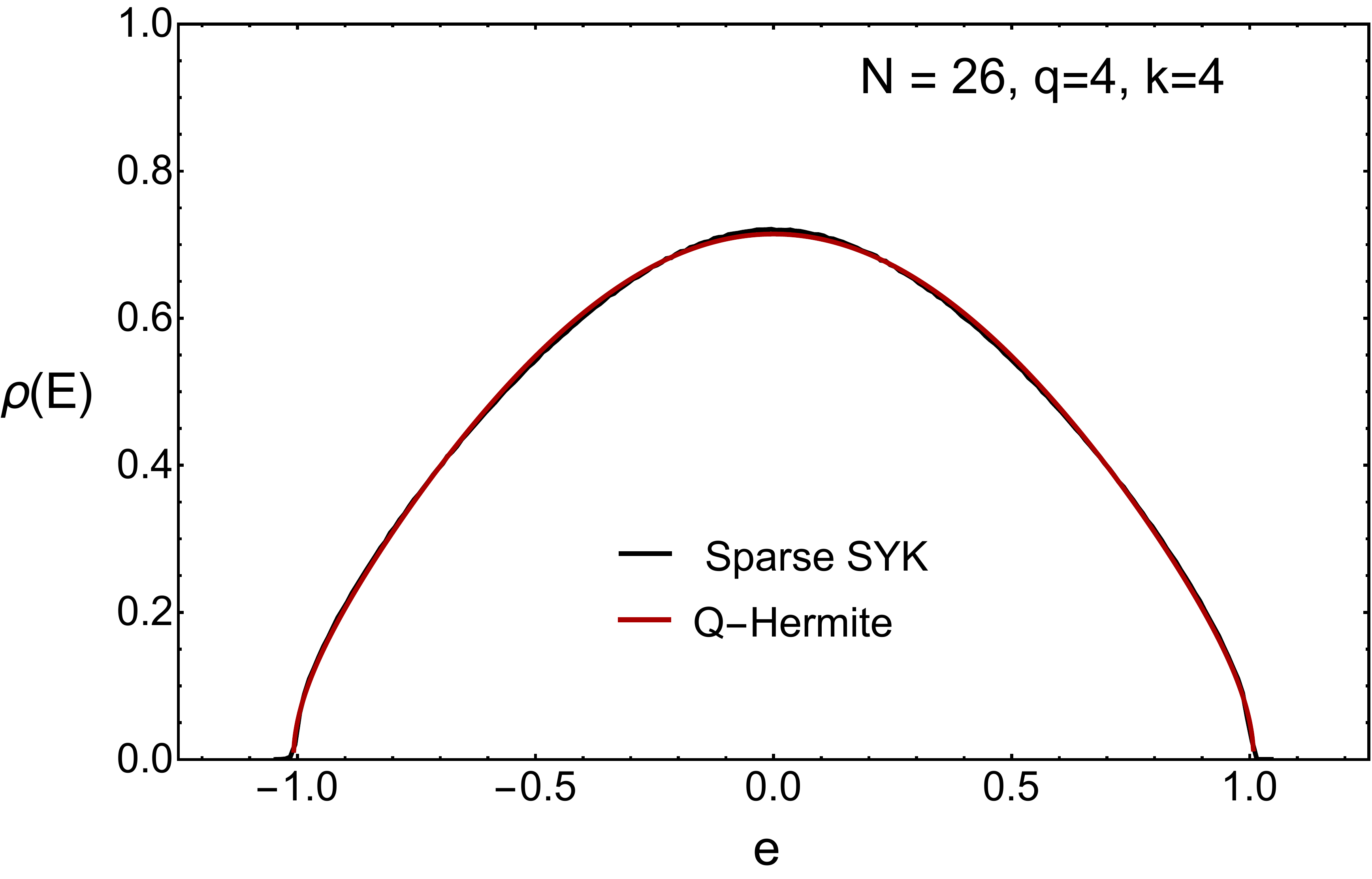}}
	\resizebox{0.49\textwidth}{!}{ \includegraphics{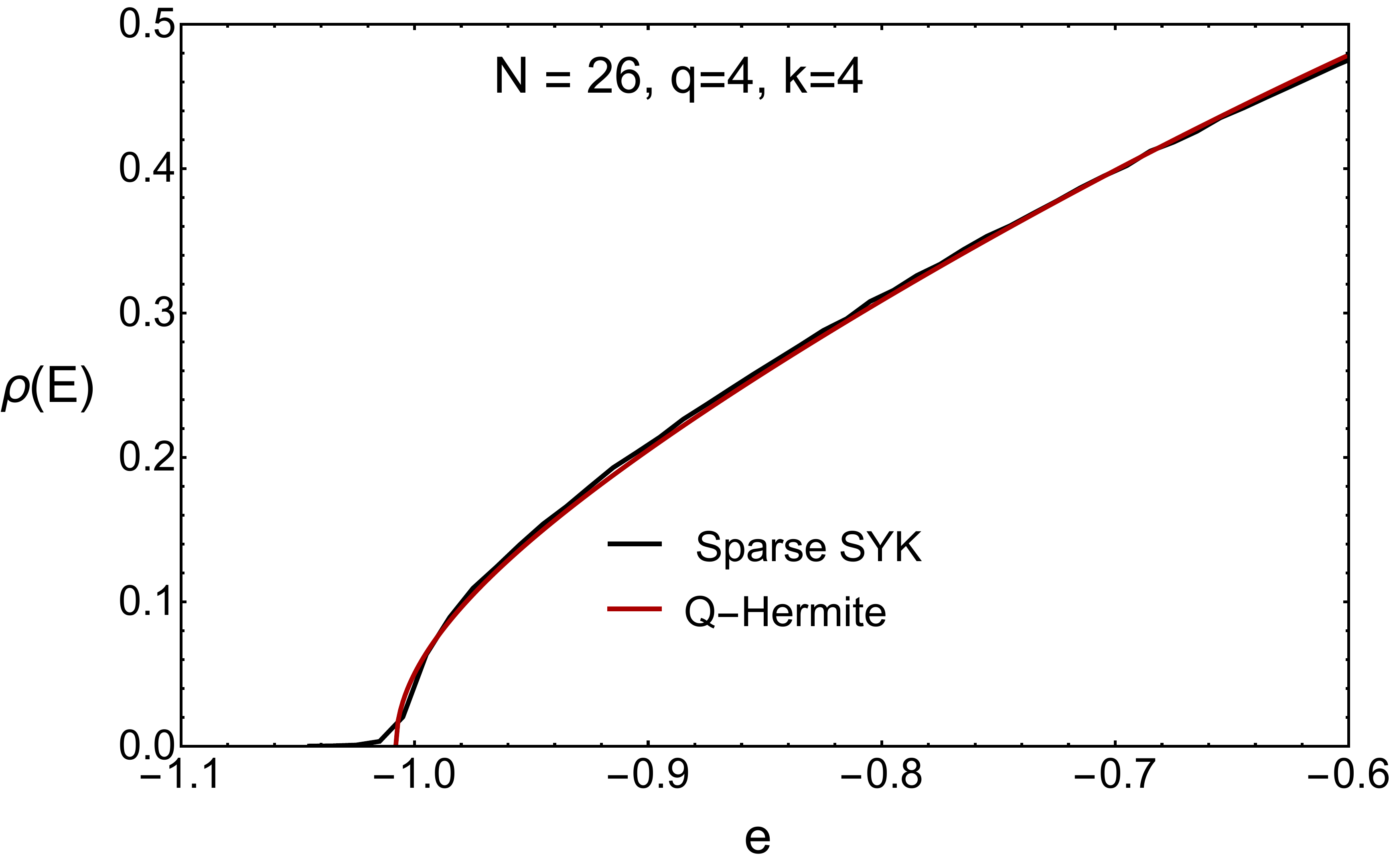}}	\vspace{-4mm}
	\caption{Spectral density $\rho(E)$ for $N = 26$ and an ensemble of $1000$ disorder realizations without imposing the regularity condition.
            The red curve line is the renormalized Q-Hermite result Eq.~(\ref{eden}) with $\eta$ a fitting parameter. The agreement is excellent even in the tail of the spectrum (right) where fluctuations are stronger. This is an indication that the sparse SYK model may have a gravity dual even for this large degree of sparseness.}
	\label{fig:tail}
\end{figure}

In order to study the tail in more detail we remove these collective excitations
by dividing all eigenvalues of each realization by its largest eigenvalue.
The spectral density of these renormalized eigenvalues
for an ensemble of 1000 realization  with $N=26$ and $k=4$ is shown in
Fig. \ref{fig:tail}. It is also shown the Q-Hermite spectral density with fitted
values for $\eta=0.129$ and $E_0=1.008$. The fitted value of $\eta$ is considerably less that the theoretical value of $0.164$ (without the $1/(kN)$ correction)
or $0.193$ (with the $1/(kN)$ correction). One might argue that $\eta$ should be given
by the value corresponding to the internal fourth moment Eq.~\eref{internal4},
but it is actually quite a bit smaller. In the right panel for Fig.~\ref{fig:tail},
we depict a magnification of the tail of the spectral density. There is an agreement
with the Q-Hermite spectral density (red curve) almost to the square root edge.

\subsection{Scale Fluctuations}\label{sec:scale}

For sparse matrices, the number of independent stochastic variables defining
the Hamiltonian is $kN$. Therefore the  relative error in an observable is $1/\sqrt{kN}$. If we decompose each  eigenvalue  into the ensemble average and a small deviation,
\be
E_i = \langle E_i \rangle +\delta E_i,
\ee
we thus have that
\be
\frac{\delta E_i}{E_i} \sim \frac 1{kN}.
\ee
This corresponds to scale fluctuations of the eigenvalues. It is natural to introduce a stochastic variable $\xi$,
\be
E_i =  \langle E_i \rangle (1+\xi)
\ee
that describes the scale fluctuations of the spectrum over different disorder realizations. The scale fluctuations follow from the variance of the second moment:
\be
\label{eq:second_moment_variance}
M_{2,2}&=&2^{-N}\langle \Tr H^2  \Tr H^2 \rangle-2^{-N} \langle \Tr H^2 \rangle^2\nn\\
&=&\left[\langle(1+\xi)^4\rangle -\langle(1+\xi)^2\rangle^2 \right ]
2^{-N}\langle\Tr H^2\rangle^2.
\ee
For the contribution  of scale fluctuations to the reduced moment we find
\be
\frac {M_{2,2}}{M_2^2}-1 = 4 \langle \xi^2 \rangle + O(\xi^4).
\ee
On the other hand, we can evaluate the above moment exactly through Wick contractions and explicit trace calculation, and the exact result is,
\be\label{m22}
\frac {M_{2,2}}{M_2^2}-1 =\frac{2}{kN}.
\ee
This results in
\be
\frac {\langle \delta E_i^2 \rangle }{\langle E_i\rangle^2}
=\langle \xi^2\rangle
= \frac 14 \left(\frac {M_{2,2} }{M_2^2}-1\right) = \frac 1{2kN}.
\label{delta-fl}
\ee
	This means that the Thouless scale is only $\sqrt N$ when $k = O(1)$.
        The $O(1/N^2)$ correction also includes the $1/N^q$ contribution from the
        dense SYK model \cite{Jia:2020rfn}.

        In Fig. \ref{fig:tail}, we have eliminated the scale fluctuations
        by normalizing the eigenvalues by the largest eigenvalue. Let us estimate
        the value of the effective value of $\eta$.
        The scale fluctuations give the following correction to the reduced
        fourth moment,
\be
\frac {M_4}{M_2^2} &=& \left . \frac {M_4}{M_2^2}\right |_{\rm int}(1 +4 \langle \xi^2\rangle)
\nn \\
&=& \left . \frac {M_4}{M_2^2}\right |_{\rm int}(1 +\frac 2{kN}),
\ee
where the subscript ``int'' (internal) refers to the fourth moment where the
contributions of the scale fluctuations have been eliminated.
This give the internal fourth reduced moment
\be
\left.\frac {M_4}{M_2^2}\right |_{\rm int} = 2+\eta - \frac 1{kN} -\frac {2\eta}{kN},
\label{internal4}
  \ee
  where the last term  is sub-leading. Indeed this gives a reduced value of
  $\eta$, but the fitted value of $\eta$ is still considerably smaller.

\begin{figure}[t!]
  	\centering
	\resizebox{0.45\textwidth}{!}{\includegraphics{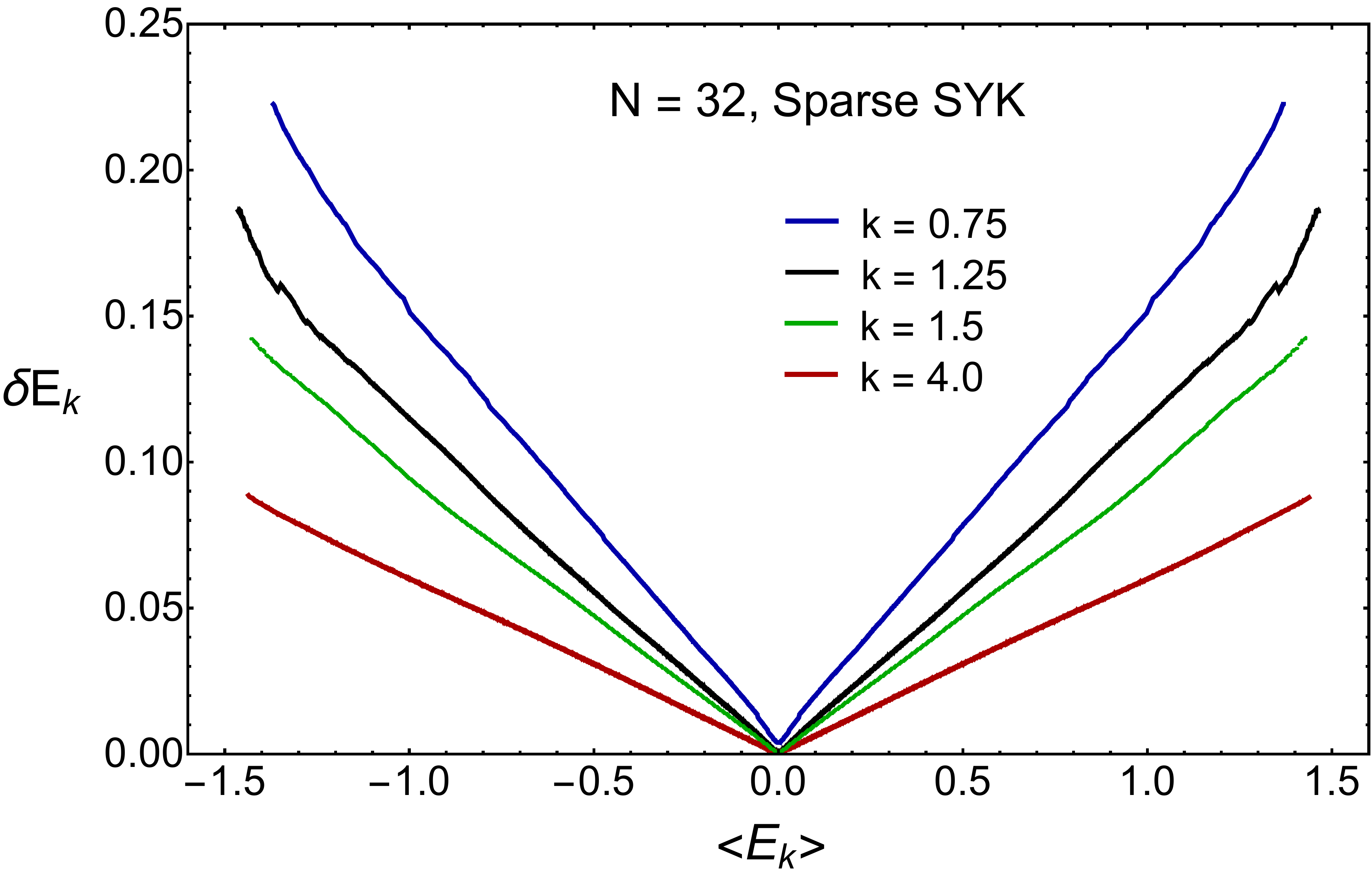}}
        \hspace{1cm}
		\resizebox{0.45\textwidth}{!}{\includegraphics{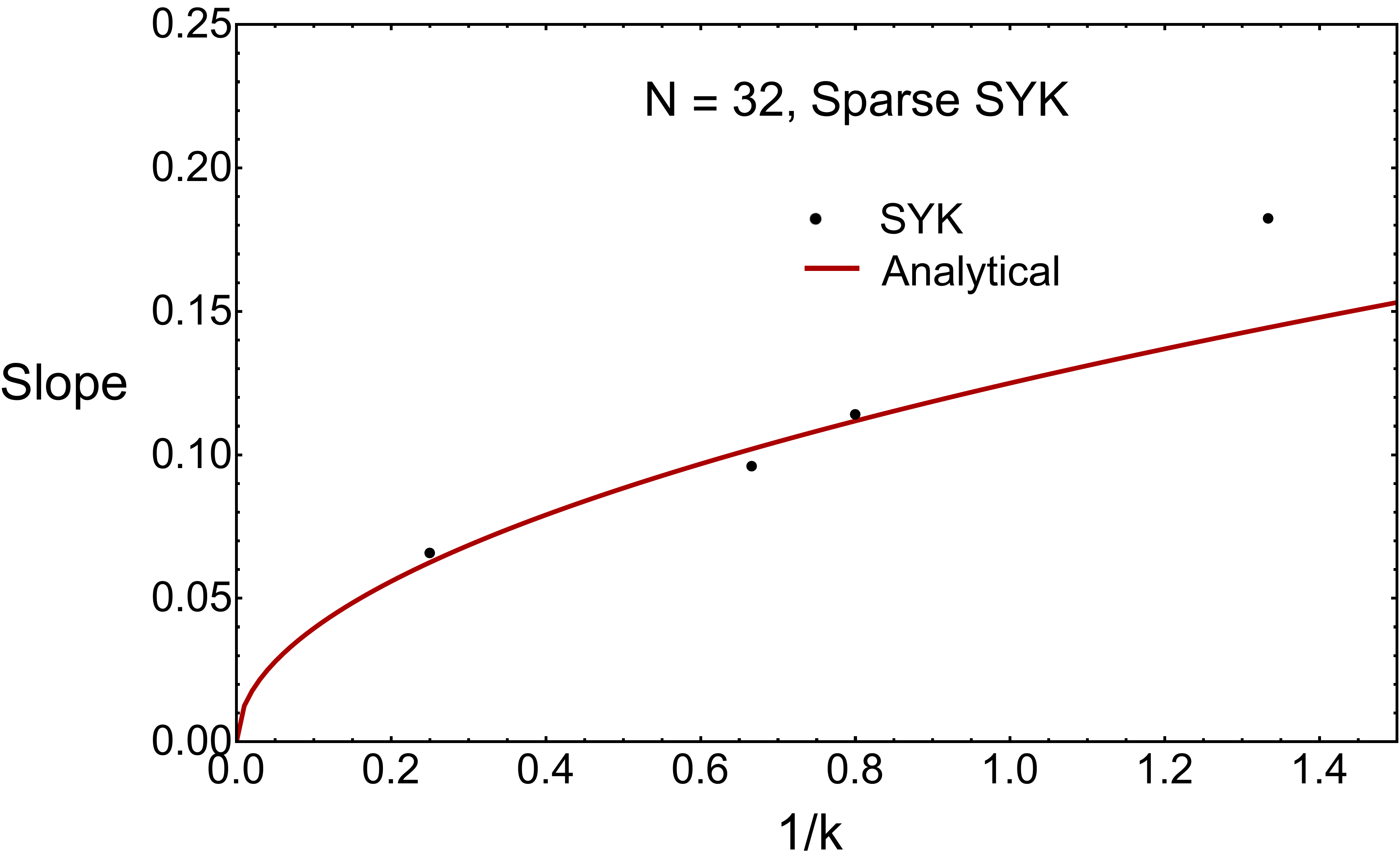}}
	\vspace{-4mm}
	\caption{Left: The root mean ensemble
		fluctuations of eigenvalues versus the eigenvalues for $N=32$, $q=4$ and various
		values of $k$ as shown in the legend of the figure. Right: The slope of
		these curves versus $1/k$ is shown in the right figure and is compared
		to the analytical result $1/\sqrt{kN}$. It is not clear why the
                point at $k=3/4$ deviates so much.}\label{delta99}
\end{figure}

It is straightforward to numerically calculate $\langle \delta E_i^2 \rangle$ for an
  ensemble of sparse SYK Hamiltonians.
  In Fig. \ref{delta99} we show $\delta E_i \equiv \langle \delta E_i^2 \rangle^{1/2}$ versus the ensemble average $\langle E_i\rangle$ of the $i$-th eigenvalue
  for $N=32$ and various values of $k$. In particular, for larger values of $k$
  there is a linear dependence on  $\langle E_i\rangle$ confirming the above
  analysis. The slope of the curves versus $1/k$ is given in the right panel
  of Fig. \ref{delta99} and compared to Eq. \eref{delta-fl}
(red solid curve). Except for the point at $k=3/4$, the agreement is excellent.

\section{Spectral statistics and quantum chaos}\label{sec:levelsta}

We now study the late time dynamics associated with the Hamiltonian Eq.~(\ref{hami}) by a level statistics analysis. Spectral correlations are a valuable probe to describe the quantum dynamics for long time scales of the order of the inverse mean level spacing. Agreement with RMT signals that the dynamics is quantum chaotic while Poisson statistics corresponds to an insulator or an integrable system \cite{guhr1998}.

The bulk of the spectrum corresponds to the high temperature phase while the low temperature/strongly coupled region is related to the lowest eigenvalues of the spectrum. In principle, only the latter is related to the existence of a gravity dual.

In order to proceed, we obtain the spectrum of the model by exact diagonalization techniques. Since the matrix representation of the Hamiltonian is extremely sparse, the use of Lanczos's algorithm allows to reach up to $N = 42$ Majoranas. As already discussed, for sufficiently small $k$, it is useful to impose the regularity condition  Eq. \eqref{eqn:regCondition}
so that all Majoranas live on a connected hypergraph.
We discuss in Appendix ~\ref{app:regularity} an efficient method for the
numerical implementation of  the regularity condition.

Except for the calculation of the form factor, the procedure of spectral unfolding is carried out by relatively low order  $< 5$ polynomials.

Since our main goal is to establish the maximum sparseness consistent with quantum chaos, we will be mostly interested in short-range spectral correlators, such as the level spacing distribution, $P(s)$, and the adjacent gap ratio. The former is defined as the probability to find two consecutive eigenvalues
$E_{i},\  E_{i+1}$ at a distance $s = (E_{i+1}-E_{i})/\Delta$ (with $\Delta$ the average local level spacing). For a fully quantum chaotic system it is given by Wigner-Dyson statistics \cite{mehta2004} which is well
approximated by the so-called Wigner surmise that depends on the universality classes \cite{guhr1998}.
For the Gaussian Orthogonal Ensemble (GOE), Gaussian Unitary Ensemble (GUE), Gaussian Symplectic Ensemble (GSE)  is given
by:
$
P_\mathrm{W,\beta}(s) = a_\beta s^\beta \exp(b_\beta s^\beta)
$
with $\beta = 1,\  2,\  4$, respectively. $a_\beta, b_\beta$ are numerical coefficients \cite{guhr1998}.

For an insulator, or a generic integrable system, it is given by Poisson statistics,
$P_\mathrm{P}(s) = e^{-s}$.
The adjacent gap ratio is defined as \cite{luitz2015,oganesyan2007,bertrand2016},
\begin{equation}
r_i = \frac{\min(\delta_i, \delta_{i+1})}{\max(\delta_i, \delta_{i+1})}
\label{eq:agr}
\end{equation}
for the ordered spectrum $E_{i-1} < E_i < E_{i+1}$  where $\delta_i = E_i - E_{i-1}$.
For a Poisson distribution, it is equal to $\left\langle r \right\rangle_\mathrm{P} \approx 0.38$ while for a random matrix ensemble it depends on the symmetry class, with
$\langle r \rangle\approx 0.53,\  0.60,\  0.67$ for the GOE, GUE, GSE \cite{atas2016}, respectively. The advantage of $\langle r \rangle$ over $P(s)$ is that it does not require us to unfold the spectrum. For that reason, we will also consider the full distribution of the adjacent gap ratio $\rho(r)$. An analytical Wigner-surmise for $\rho(r)$ is available for different random matrix ensembles \cite{atas2016},
\be
\rho_\mathrm{W,\beta}(r) = A_{\beta} \frac{{(r+r^2)}^\beta}{(1+r+r^2)^{1+3\beta/2}}
\ee
with $\beta = 1, 2, 4$ for GOE, GUE and GSE respectively. The prefactor $A_\beta$ is a numerical coefficient and $r \equiv \delta_i/\delta_{i+1}$ \cite{atas2016} (note the difference with Eq. \eqref{eq:agr}). 
We note that despite $P(s)$ and $\rho(r)$ are both short-range spectral correlators, that probe the quantum dynamics at times of the order and larger than the Heisenberg time, $\rho(r)$ is a shorter range correlator than $P(s)$. Therefore, we expect that deviations from RMT predictions will become more apparent in $P(s)$.

We start the spectral analysis with the study of spectral correlations near the center of spectrum, usually called the {\it bulk}, corresponding to the high temperature phase.

\subsection{Bulk}
\begin{figure}
	\centering
	\resizebox{0.49\textwidth}{!}{\includegraphics{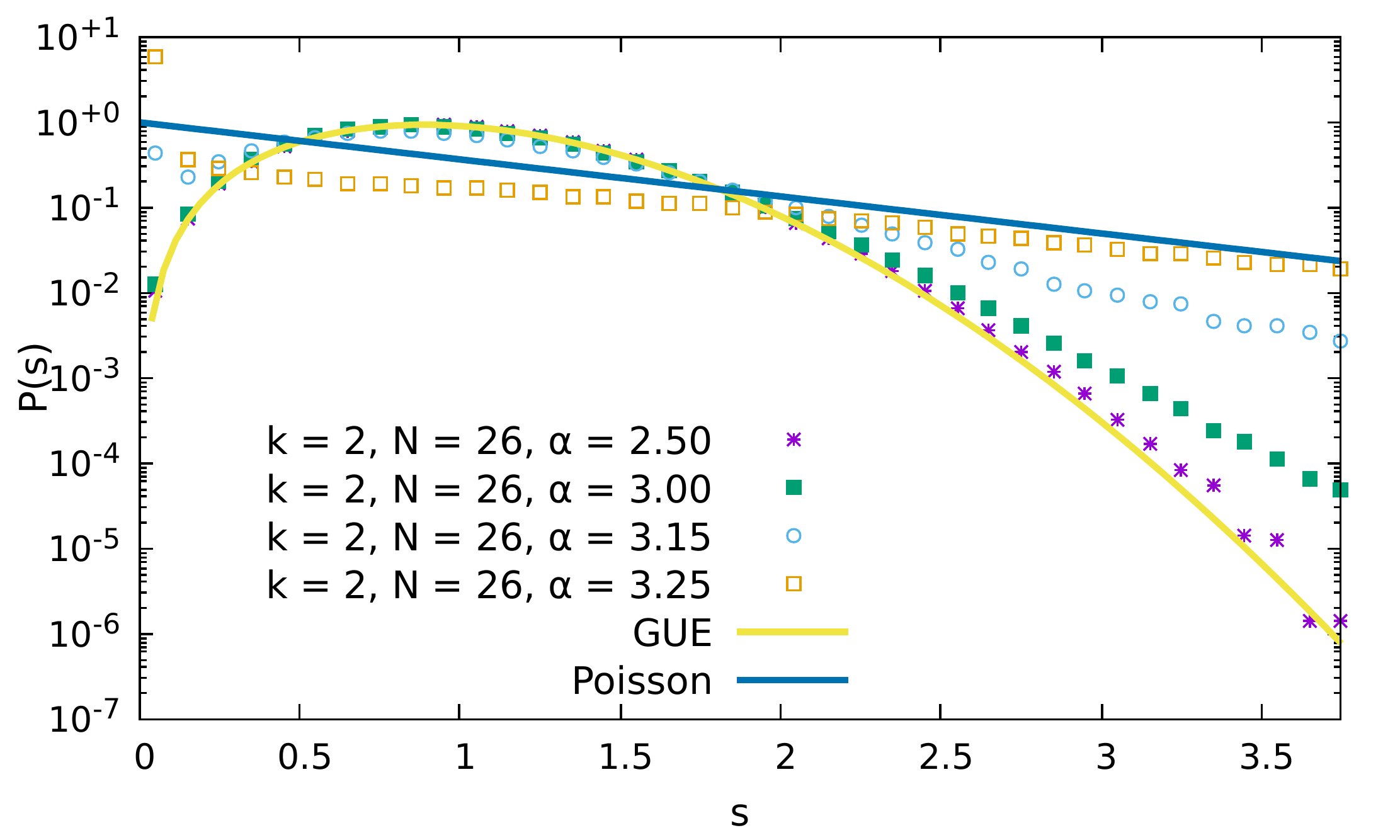}}
		\resizebox{0.49\textwidth}{!}{\includegraphics{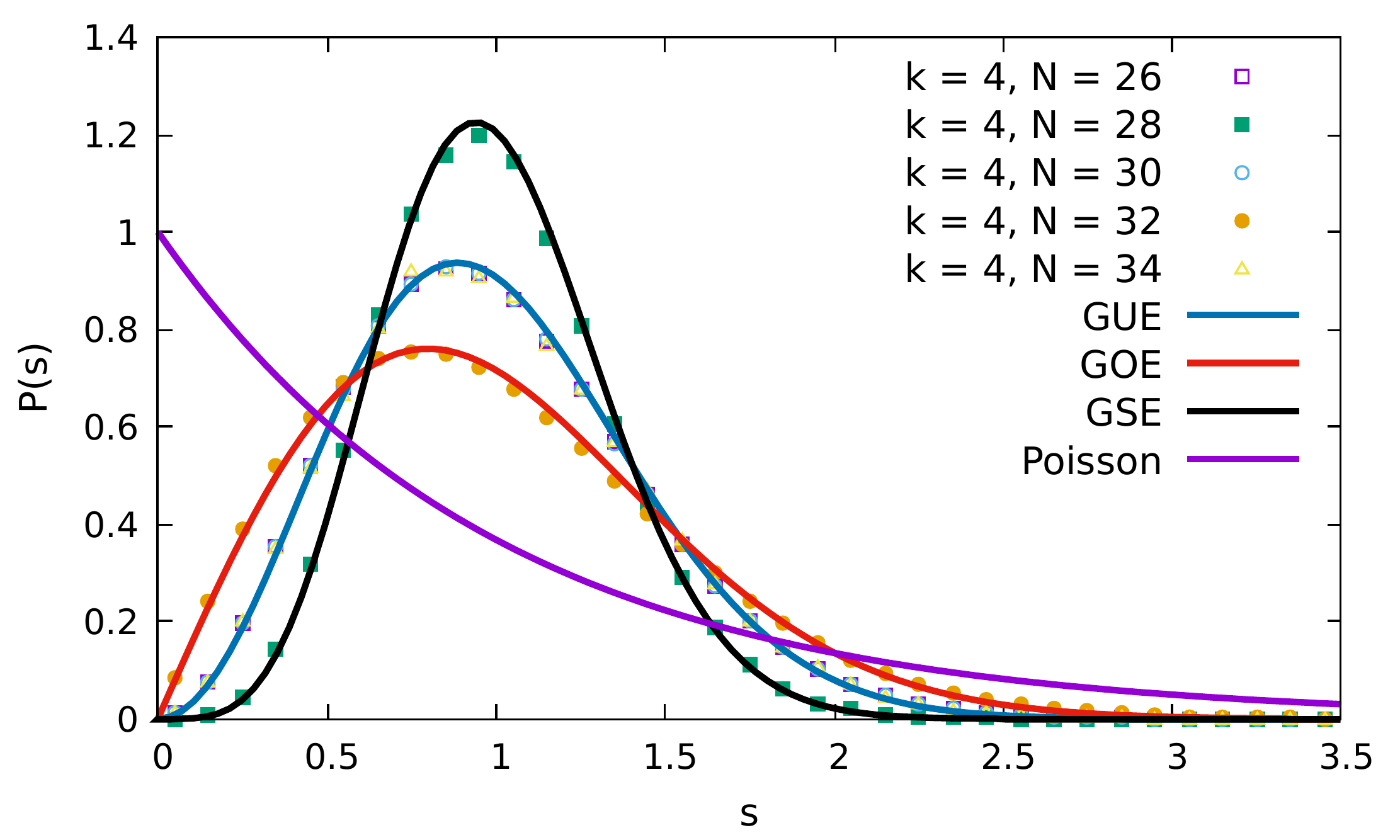}}
	\vspace{-4mm}
	\caption{Left: The nearest neighbor spacing distribution, $P(s)$, for $N = 26$ and different scalings $p  \propto 1/N^\alpha$ in the bulk of the spectrum. No regularity condition has been imposed. In agreement with the theoretical prediction, the critical scaling is at $\alpha = 3$. For $\alpha >3 $, corresponding to a more sparse Hamiltonian, we observe spectral quasi-degeneracy leading to an anomalous peak for small spacings and a gradual approach to Poisson statistics. Right: $P(s)$ for $\alpha = 3$, $k =4$ and different values of $N$. We observe an excellent agreement with the predictions of RMT for the different values of $N$ corresponding to different universality classes.}
	\label{psdifalphabulk}
\end{figure}

We define the bulk as the central part of the spectrum comprising $80\%$ of eigenvalues unless other percentage is explicitly stated.
Our first task is to determine the critical scaling $p {N \choose 4} = k N^{4-\alpha}$ for which the dynamics is quantum chaotic, namely, level statistics are well described by RMT. For that purpose, we first compute $P(s)$ defined above for $N = 26$, $k = 2$ and different scalings of the probability $p$ parameterized by $\alpha$.
The results depicted in Fig.~\ref{psdifalphabulk} strongly suggest that the maximum sparseness consistent with quantum chaotic dynamics is approximately $p \propto 1/N^3$, namely, $\alpha = 3$. This is in agreement with the prediction for Erdos-Renyi graphs adapted to random hypergraph  represented by the Hamiltonian Eq. (\ref{hami}).  We note that for $\alpha > 3$, not only the tail is exponential, as for Poisson statistics, but also there is a peak for small $s$ related to spectral degeneracies that we shall see soon are related to the presence of emergent global symmetries for sufficiently strong sparseness.

In order to see that $\alpha = 3$ corresponds to the maximum sparseness, called from now on the \textit{critical scaling}, we first study the level statistics for a larger $k = 4$ and different $N$'s. The global symmetries of the SYK model depend on $N$ \cite{you2016,garcia2016}, so a study of the $N$ dependence in the sparse case will also provide useful information about the robustness of these symmetries against the sparsing procedure.
We have found that the agreement with the RMT results corresponding to the different universality classes (GOE, GUE, GSE) is excellent, see Fig.~\ref{psdifalphabulk}. Moreover, the results for $N = 26$ and $N = 30$, both belonging to GUE, are almost indistinguishable. Both features provide convincing evidence
that in the region $k \gg 1$ and $\alpha =3$ the system is still fully quantum chaotic with not much difference with the dense case at least for short range correlations of few neighboring eigenvalues.

We note that this robustness of quantum chaos is remarkable. The dense SYK model has $\sim N^4$ non-zero different entries while the sparse one only $4N$. This is however the analytical prediction resulting from a heuristic extrapolation of the rigorous mathematical results \cite{erdos2012,huang2015}, and numerical simulations \cite{evangelou1992}, for random sparse graphs to hypergraphs such as the sparse SYK model: the dynamics is quantum chaotic and spectral correlation are described by RMT only for sufficiently large $k$.

We now turn  to the study of the dependence of spectral correlations on $k$ for this critical scaling ($p \sim k/N^3$) to determine the minimum $k = k_c$ for which this agreement to RMT persists. The theoretical expectation for random graphs \cite{evangelou1992} is that $k_c \gtrsim 1$. In the previous investigation of level statistics for $\alpha > 3$, we have noticed the emergence of level degeneracies at least for some disorder realizations. Qualitatively, the reason  is that the quantum dynamics is very sensitive to the overall connectivity of the hypergraph. Therefore, for sufficiently small $k$, or large $\alpha$, level statistics strongly depend on the connectivity of the disorder realization. As an example, for sufficiently small $k \leq 2$, in some cases, we observe double degeneracy while in others realizations, the spectrum has a chiral symmetry $E \to -E$. For some disorder realizations,
both a double degeneracy and a chiral symmetry occur at the same time. We will study this phenomenon in more detail in later sections. For the moment, we remark that large sparseness allows extra symmetries and chiral symmetries to emerge. Without the regularity condition, this effect of sparseness is more pronounced because disconnected hypergraphs can be present; the regularity condition eliminates  the
disconnectedness and mitigates the complication of emergent symmetries, however symmetries can still emerge once sparseness is further increased.

 \begin{figure}[t!]
	\centering
	\resizebox{.49\textwidth}{!}{\includegraphics{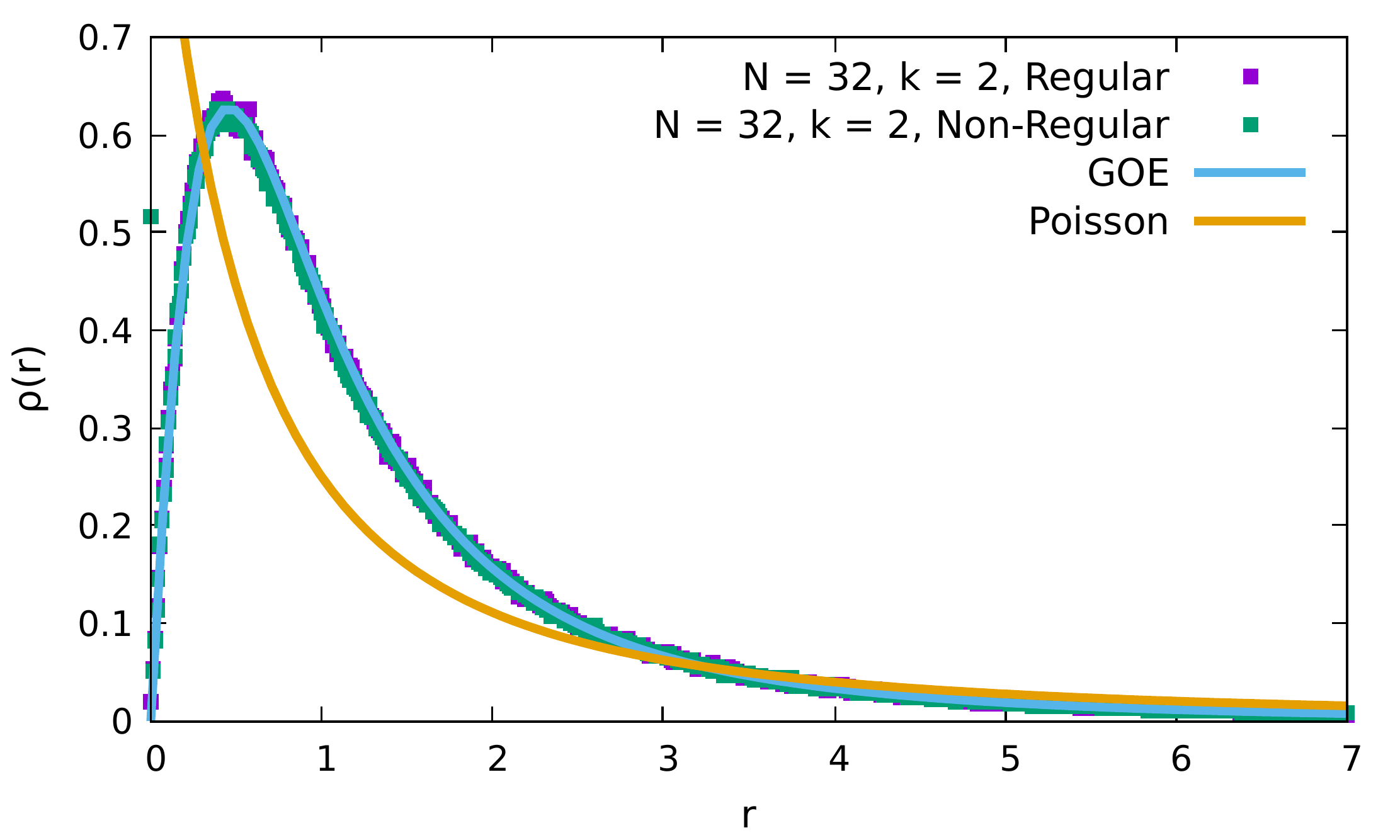}}
   \resizebox{.49\textwidth}{!}{\includegraphics{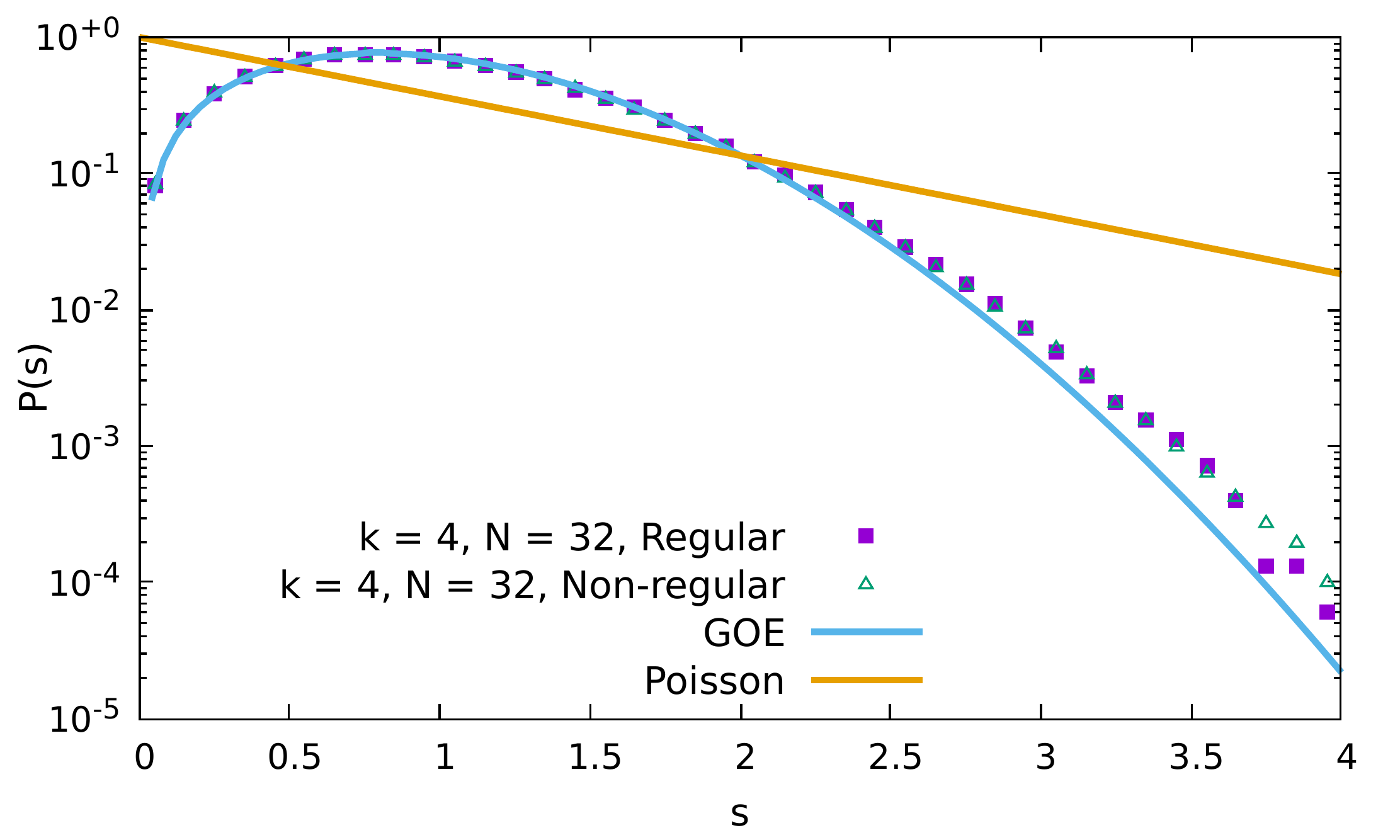}}
	\vspace{-4mm}
	\caption{Left: Distribution function of the adjacent gap ratio $\rho(r)$ for $k = 2$ with and without imposing the regularity condition. In the latter, we observe a peak in the $r \approx 0$ region while in the regular case the agreement with RMT prediction is excellent for all values of $r$. Right: $P(s)$ for $k = 4$ with and without regularity condition. We do not observe any difference even in the tail of the distribution.
	}
	\label{pradjcomp}
\end{figure}
 As an indication of the effect of the regularity condition, in Fig.~\ref{pradjcomp}, we compare results for the distribution of gap ratios $\rho(r)$ and $P(s)$ with and without the regularity condition for the $N=32$ sparse SYK model. For $k = 4$ no difference is observed even in the tail of the distribution. For $k = 2$, the degeneracy only appears in some of the realizations of the non-regular case, which results in a large peak at the origin.
(see green square in Fig. \ref{pradjcomp}, left). Here the effect of the hypergraph disconnectedness is concretely at display: for $N=32, k=2$ without regularity condition, often enough a realization misses a fermion, say $\gamma_{32}$. Hence it is really an $N=31$ model in disguise, which is incidentally still in the GOE class \cite{sun2020}. This produces a 2-fold degeneracy because the extra symmetry $\gamma_{32}$ anticommutes with the  chirality operator $\gamma_c = \prod_{i=1}^{32}\gamma_{i}$. However in this case fixing the $\gamma_c$ chirality, which we always do,  is enough to eliminate the degeneracy.  What happens much less often, but still with a non-negligible probability, is that a realization can altogether miss two fermions, say $\gamma_{31}$ and $\gamma_{32}$. In this case we have an $N=30$ model in disguise, which is in the GUE class. This is a 4-fold degenerate situation: 2-fold from the extra symmetries $\gamma_{31}, \gamma_{32}$, and another 2-fold from the fact that the $N=30$ model (GUE) has a time-reversal operator that anticommutes with the $N=30$ chirality operator \cite{garcia2016}. Fixing the $\gamma_c$ chirality only eliminates the former 2-fold degeneracy and this explains the degenerate data point in the left figure of Fig. \ref{pradjcomp}.
Therefore, to reach any firm conclusion from the study of spectral statistics, we have to classify
 the realizations of the Hamiltonian with all emergent symmetries taken into account (see next section).
 Therefore, for the study of the critical $k$ for quantum chaos to occur, it is advantageous to rely on regular hypergraphs, which will also exhibit degeneracies and emergent symmetries but only for smaller $k$'s with respect to those in the non-regular case.

\begin{figure}
	\centering
	\resizebox{0.49\textwidth}{!}{\includegraphics{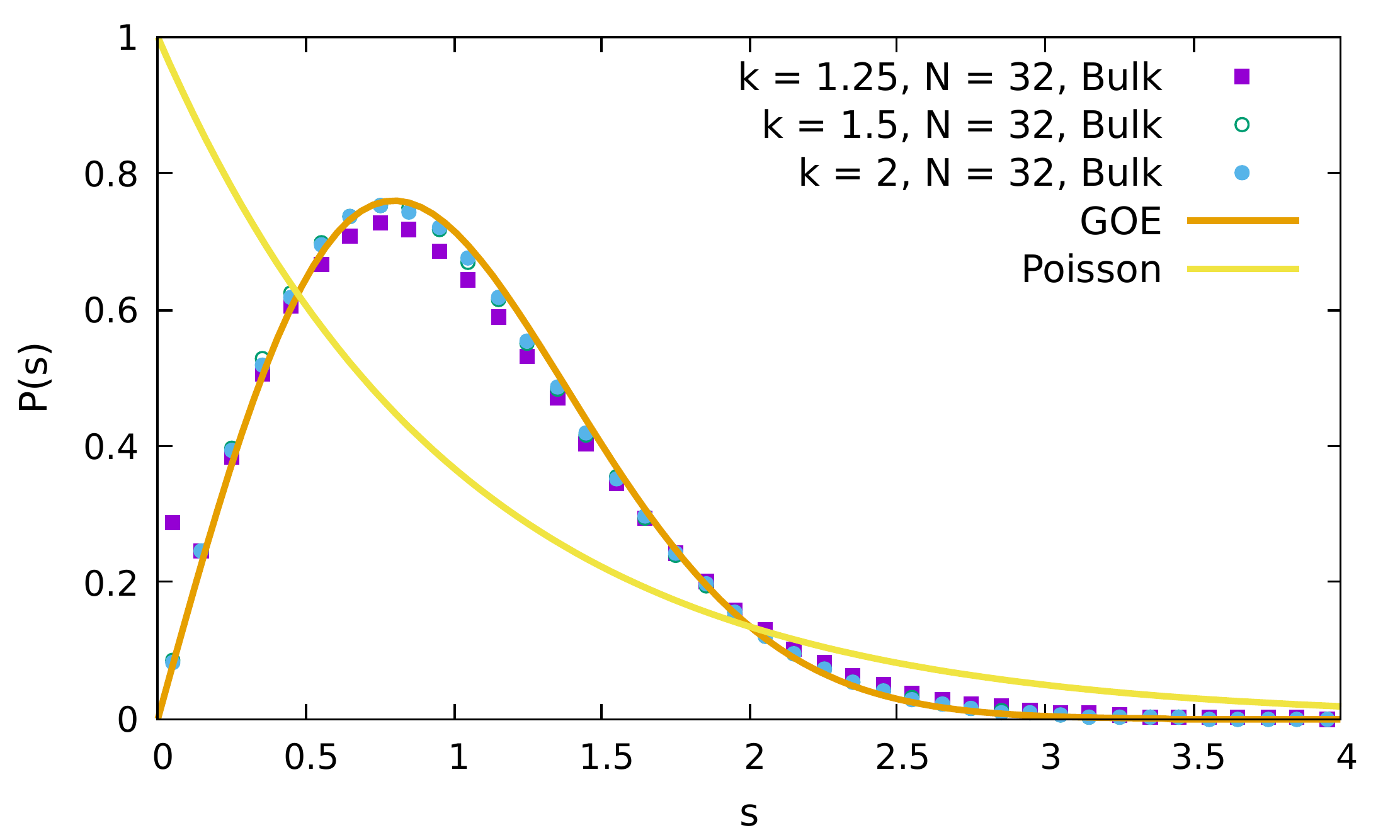}}
	\resizebox{0.49\textwidth}{!}{\includegraphics{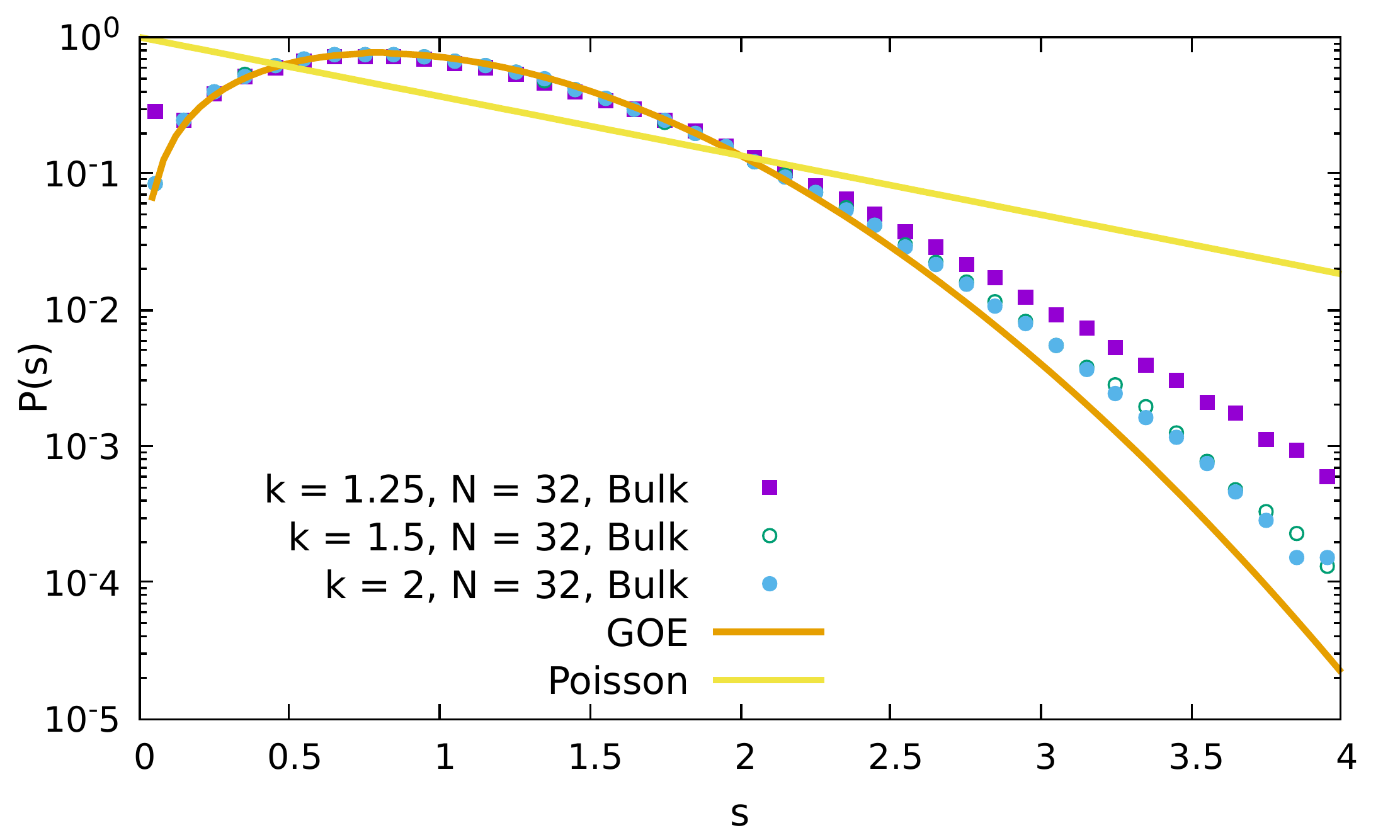}}
	\vspace{-4mm}
	\caption{The nearest neighbor spacing distribution, $P(s)$, for $N = 32$ and different $k$ with $p {N \choose 4} = kN$ in linear (left) and logarithmic scale (right). In agreement with the theoretical prediction, we find good agreement with RMT for sufficiently large $k$. As $k$ decreases, we observe a bump for small spacings which suggests that the spectrum starts to develop a twofold degeneracy. For $k = 1$, not shown, the degeneracy is exact for some realizations. We shall see that this is due to additional global symmetries induced by the increased sparseness. The regularity condition is imposed.}
	\label{psdifkbulk}
\end{figure}

In agreement with the theoretical expectation, see Fig.~\ref{psdifkbulk}, deviations from RMT become more evident as $k$ decreases. The tail becomes gradually exponential and more importantly, for $k = 1.25$, we again observe a peak in $P(s)$ for very small $s$ instead of the expected level repulsion $P(s) \to 0$ as $s \to 0$.
 By direct inspection of the spectrum, we have found that, even after the regularity condition is imposed, the peak is related to an emergent spectral degeneracy. As $k \to 1$, an almost exact two-fold eigenvalues degeneracy occurs for some disorder realizations. The peak becomes again very large which prevents a meaningful spectral analysis without further processing of the spectra.

 We postpone this analysis to later sections. For the moment,
 we just mention this degeneracy in the $k \to 1$ limit is related to the existence of additional global symmetries, represented by commuting and anti-commuting
 operators,  induced by the sparseness of the Hamiltonian. Once they are taken into account, the level spacing distribution still shows level repulsion but deviates markedly from the RMT prediction. The asymptotic decay is indeed exponential as for Poisson statistics.
 However, strictly speaking, it is unclear whether the nature of the quantum chaos transition is quantitatively similar to that of the Anderson metal-insulator transition or an chaos-integrable transition. The route to integrability  is highly non-universal. In many cases it is not properly a transition but rather
 a crossover at least from the point of view of spectral statistics.
 To be specific, harmonic oscillators are integrable and a rectangular billiard is also integrable but the spectral correlations are very different so the transition from chaos to integrability will depend on the integrable system. By contrast, an Anderson insulator has Poisson statistics and the transition can be typically characterized by critical exponents and the scale invariance of level statistics at the transition so it is largely universal. We will return to this point when we investigate the $k = 1$ case in more detail.

One disadvantage of $P(s)$ is that it requires unfolding of the spectrum. This does not pose any problem for large $k$, but as spectral degeneracies start to appear for smaller $k$, it is more challenging to carry out the unfolding procedure. In order to further characterize the deviations from RMT, we investigate the average adjacent gap ratio $\langle r \rangle$ which does not require any unfolding and also provides information on the nature of very short-range spectral correlations. Taking the $60\%$ of the eigenvalues around $E=0$, we have found that even for $k = 1.25$, the deviation from RMT are very small. For $N = 34$, $\langle r \rangle = 0.600195$ while $\langle r \rangle_{GUE} \approx 0.5996$. Similar results are obtained for other $N$'s or $k > 1.25$. If we consider the $90\%$ of the spectrum around $E = 0$, we observe small deviations, for $N = 32$,
$k = 1.25$, $\langle r \rangle = 0.5068$ while $\langle r \rangle_{GOE} \approx 0.5307$.

\begin{figure}[t!]
 	\centering
 		\resizebox{0.49\textwidth}{!}{\includegraphics{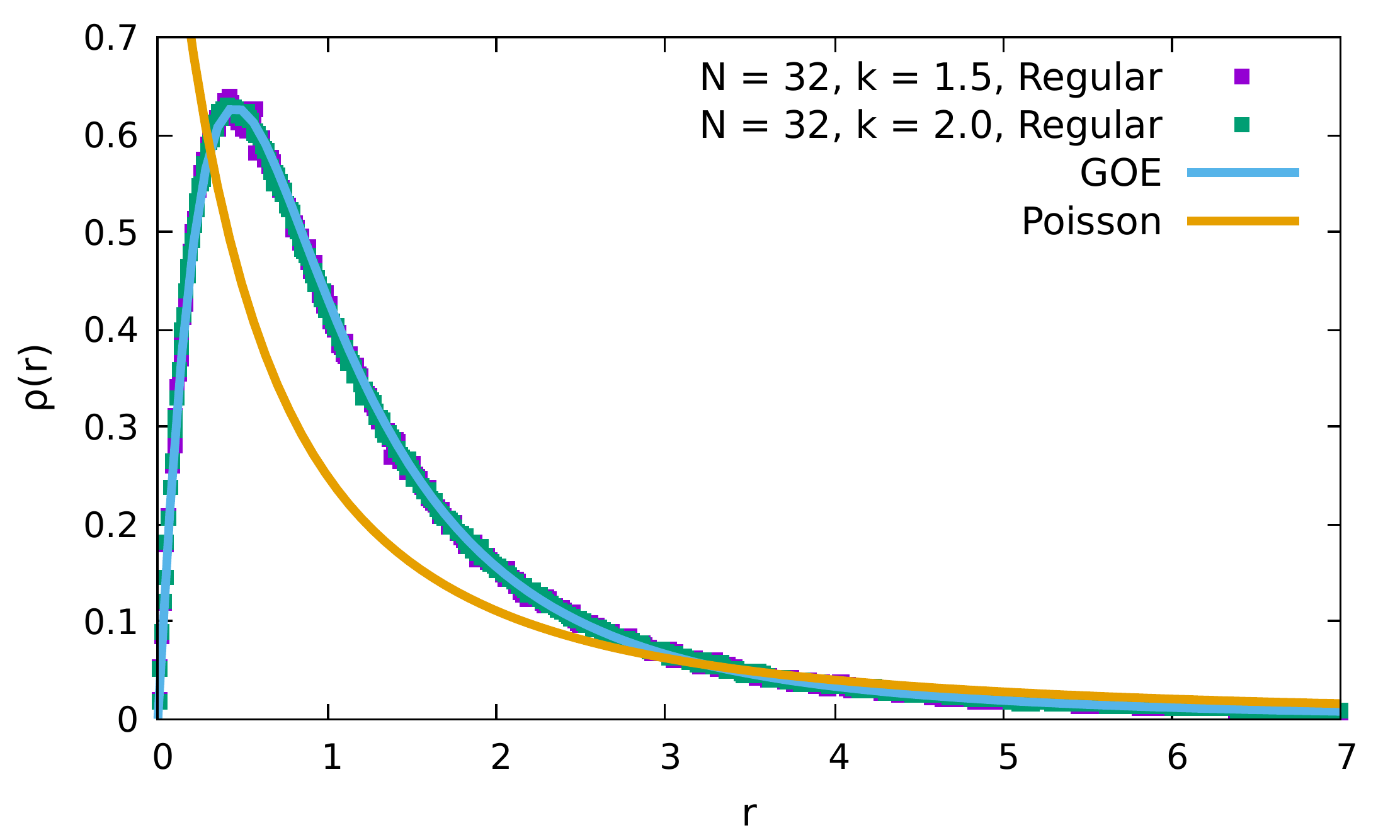}}
 	\resizebox{0.49\textwidth}{!}{\includegraphics{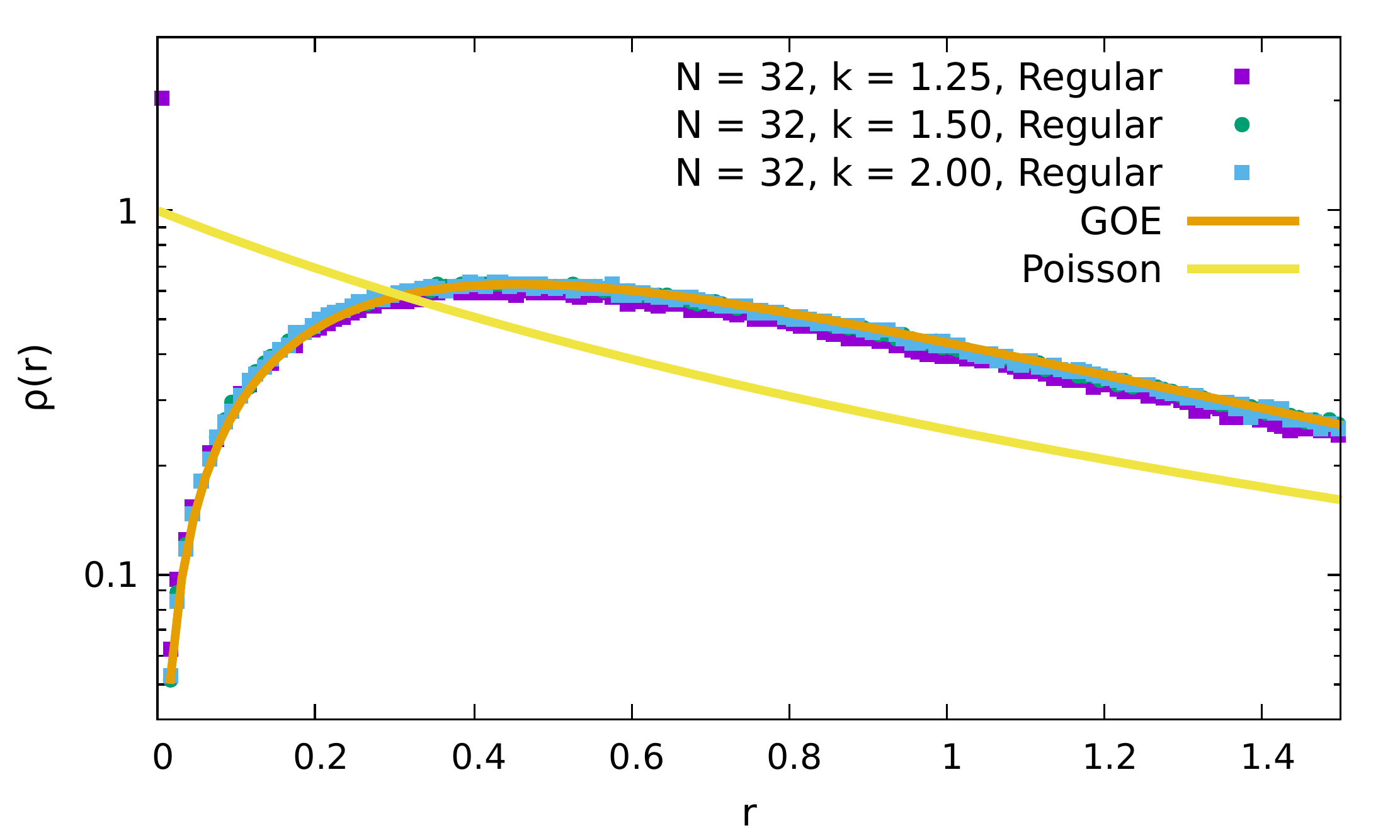}}
 	\vspace{-4mm}

 	\caption{Distribution function of the adjacent gap ratio $\rho(r)$ in linear (left) and log (right) scale. For $k > 1.25$, the agreement with the RMT prediction is excellent with no visible deviations. However, for $k = 1.25$, it has a large peak (right plot) at small $r$ which suggests that, even for $k > 1$, some disorder realization may have spectral degeneracies. The regularity condition is imposed.}
 	\label{prbulk}
 \end{figure}

Although these results are not inconsistent with those from
the level spacing distribution, it appears that deviations from RMT predictions are smaller for this correlator. A possible reason for this quantitative difference is that the gap ratio provides information of spectral correlations of shorter range than $P(s)$.
In order to confirm this prediction, we compute the full distribution of the adjacent gap ratio $\rho(r)$ using the $90\%$ of the eigenvalues.
Results depicted in Fig.~\ref{prbulk}, are consistent with those of $\langle r \rangle$.
Agreement with the RMT prediction is excellent except for $k = 1.25$. The main difference being the large enhancement of $\rho(r)$ for very small $r$ at $k = 1.25$.
By direct inspection of the spectrum, we associate this peak to an emergent degeneracy of the spectrum. Therefore, even considering only regular hypergraphs, it is not enough to remove these spectral degeneracies related to new global symmetries of the system. The regularity condition only shifts its appearance to even smaller values of $k \approx 1.25$.

 In summary, we have found that a sparse SYK model with $N$ Majoranas is still quantum chaotic for sufficiently high energies, or temperatures, provided that the probability $p \propto 1/N^\alpha$ with $\alpha < 3$. For $\alpha = 3$, spectral correlations are still well described by random matrix theory for $k \gg 1$. For $k \sim 1$, we gradually notice deviations from this prediction. In the $k \to 1$ limit spectral degeneracies are frequently observed which makes the spectral analysis difficult even if the regularity condition is taken into account.

 However, the spectral region related to the possible existence of a gravity dual is the edge corresponding to the lowest eigenvalues, and not the bulk of the spectrum. We now move to the study of this region.

\begin{figure}[t!]
	\centering
	\resizebox{0.49\textwidth}{!}{\includegraphics{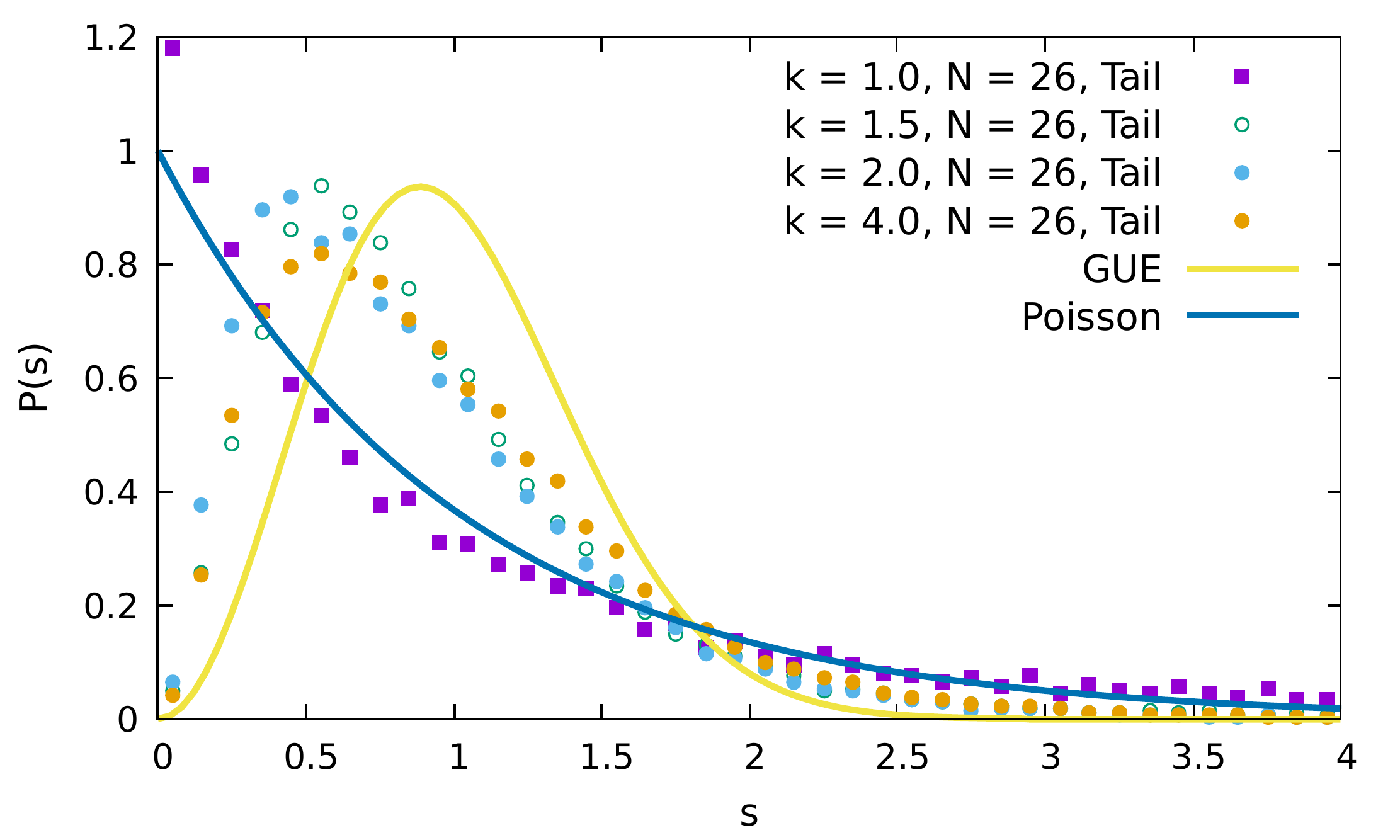}}
	\resizebox{0.49\textwidth}{!}{\includegraphics{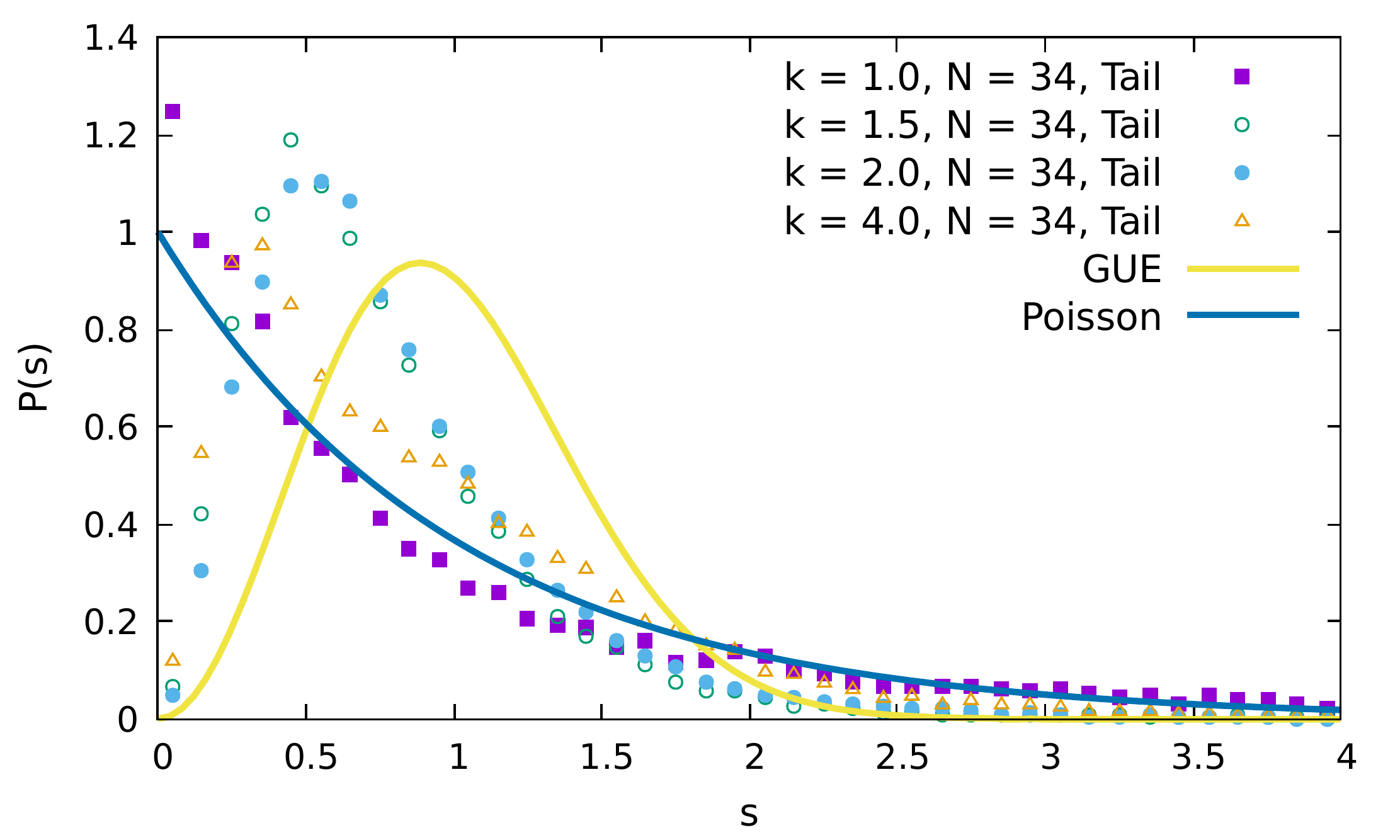}}
	\vspace{-4mm}
	\caption{Left: The nearest neighbor spacing distribution, $P(s)$, $p {N\choose 4} = kN$, $N = 26$ and different values of $k$. Right: Same for $N = 34$. Even for $k \gg 1$ spectral correlations deviate strongly from the RMT prediction (GUE). Results for different values of $N$ are qualitatively similar which reinforces the idea that the quantum chaos transition occurs at $k \gtrsim 1$. For $k = 1$, we have noticed spectral degeneracies in some of the disorder realizations which we have removed for the calculation of $P(s)$. We do not fully understand the reason why $P(s)$ for $N = 34$ and $k = 4$ deviates from the RMT prediction more strongly than for smaller $k$.}
	\label{pstaildifk}
\end{figure}

\subsection{The edge}
In this section we study the spectral correlations of the lowest eigenvalues relevant for the time evolution of the system in the low temperature limit. This is the only region that may be related to a gravity dual.
Technically, it is more challenging to reach firm conclusions because the small spectral window close to the ground state limits substantially the use of
spectral averaging to diminish statistical fluctuations. Moreover, the rigorous mathematical results for sparse random graphs are less sharp for the edge of the spectrum as compared to the bulk. As was mentioned earlier, for a random graph, RMT spectral correlations at the edge of the spectrum and a semi-circular spectral density law occur for $p \gg 1/L^{2/3}$ where $L$ is the matrix size. A naive translation of these result to the sparse SYK would lead to a critical scaling $p \gg 1/N^{8/3}$. However, we stress that the results for graphs are not necessarily applicable here and that, even if they are applicable, the bound on $p$ to observe RMT correlation does not have to be optimal, namely, it may be that an even stronger sparsing, such as $1/N^3$, may still lead to RMT correlations at the edge of the spectrum.

In order to proceed with the spectral analysis,
we obtain only the $2N$ lowest eigenvalues by exact diagonalization using special techniques for sparse matrices based on the Lanczos algorithm which allows us to reach $N = 42$. For a given set of parameters, we carry out ensemble average until we have at least $10^4$ eigenvalues.

 \begin{figure}[t!]
 	\centering
 	\resizebox{0.49\textwidth}{!}{\includegraphics{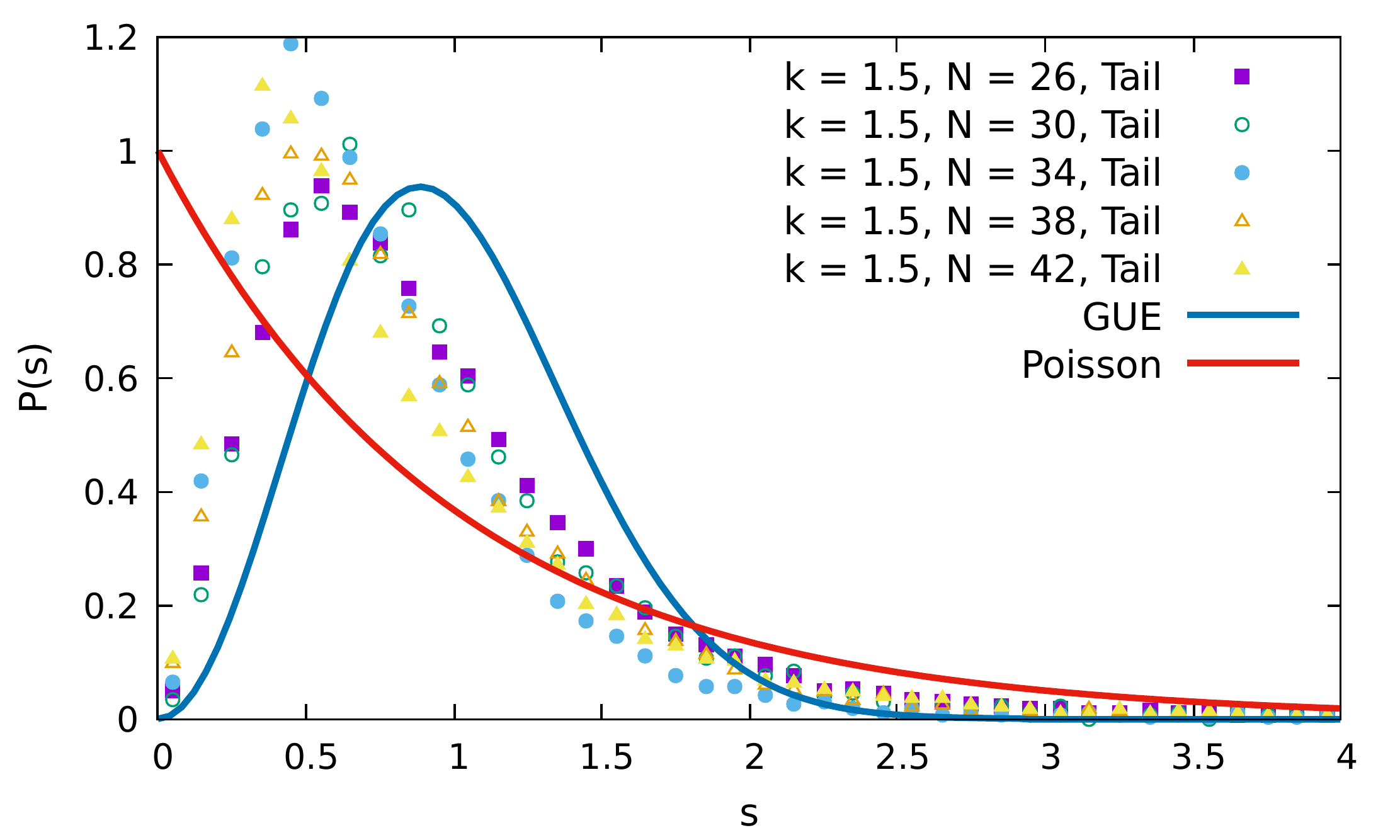}}
 	\resizebox{0.49\textwidth}{!}{\includegraphics{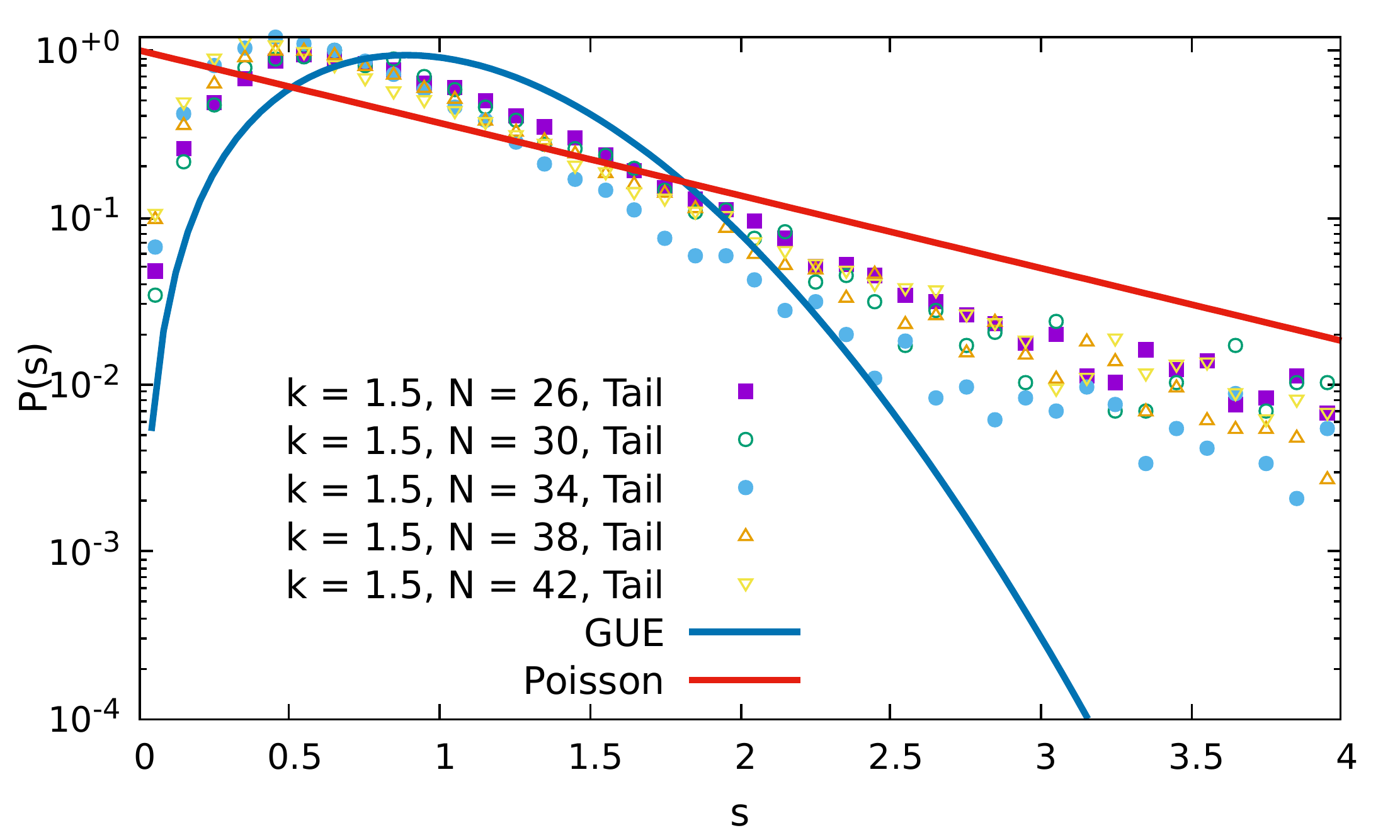}}
 	\vspace{-4mm}
 	\caption{Left: The nearest neighbor spacing distribution, $P(s)$, $p {N \choose 4} = kN$, $k = 3/2$ and different values of $N$. Right: Log-scale. In this critical region, spectral correlations show deviations from the RMT prediction though level repulsion is still clearly observed. The $N$ dependence is relatively weak. These features are qualitatively similar to those of a critical system \cite{altshuler1988,shapiro1993} approaching a quantum chaos transition.}
 	\label{pstaildifN}
 \end{figure}

\begin{figure}[b!]
	\centering
	\resizebox{0.6\textwidth}{!}{\includegraphics{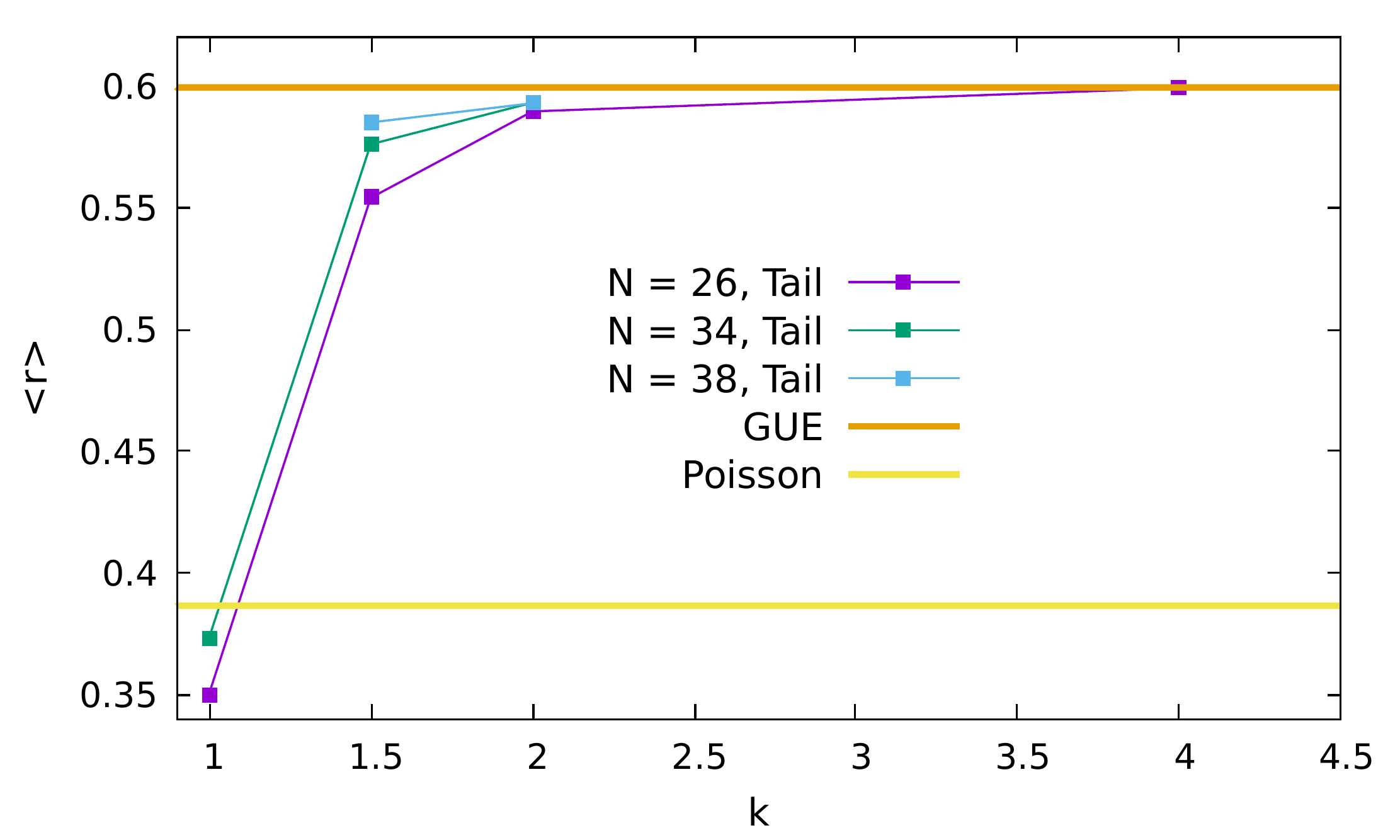}}
	\vspace{-4mm}
	\caption{Adjacent gap ratio $\langle r \rangle$, Eq.~(\ref{eq:agr})
		for $p {N \choose 4} = kN$ and different $k$'s corresponding to the lowest $2N$ eigenvalues of the Hamiltonian Eq.~(\ref{hami}). As $k \to 1$, the adjacent gap ratio decreases and approaches the Poisson limit. We note that, for $k = 1$, the spectrum of some disorder realizations, especially for larger values of $N$, shows a twofold degeneracy which was removed before the calculation of $\langle r \rangle$. Emergent global symmetries are the origin of the spectral degeneracy. See text for more detail.}
	\label{adjtail}
\end{figure}

We first investigate the critical scaling $p {N \choose 4}= kN$ identified in the bulk of the spectrum. We study the dependence of level statistics on $k$ with the goal to clarify whether spectral correlations are still quantum chaotic and, if so, to identify the approximate critical $k = k_c$. Results, depicted in Fig.~\ref{pstaildifk}, show a gradual weakening of quantum chaotic features as $k$ is reduced.
An exception to this trend is $P(s)$ for $k = 4$ and $N = 34$ which is closer to Poisson than that of $k = 2$. Presently, we do not have a clear understanding of this anomalous deviation. Results for the adjacent gap ratio, see Fig.~\ref{adjtail}, indicates that the spectrum, at least for very short range correlations, is quantum chaotic in the large $k$ limit. 
 
 In the $k \approx 1$ region no level repulsion is observed which indicates that the Hamiltonian is too sparse to sustain quantum chaotic features. It is important to note that, also in the tail of the spectrum, we observe degeneracies of the spectrum for $k \to 1$ though not in all disorder realizations. For the analysis of the spectral correlations, we have removed them ad hoc. This will be justified in the next section by the existence of global symmetries that cause the spectral degeneracies.
 
A feature of criticality \cite{altshuler1988,shapiro1993} is the weak or no dependence of spectral correlations on the system size $N$. Results depicted in Fig.~\ref{pstaildifN} show a weak $N$ dependence in the $k \sim 1$ region. Furthermore, spectral correlations deviate strongly from the RMT prediction. This is consistent with the idea that $k \approx 1$ is the maximum sparseness consistent with quantum chaotic features.
\begin{figure}[t!]
	\centering
	\resizebox{0.6\textwidth}{!}{\includegraphics{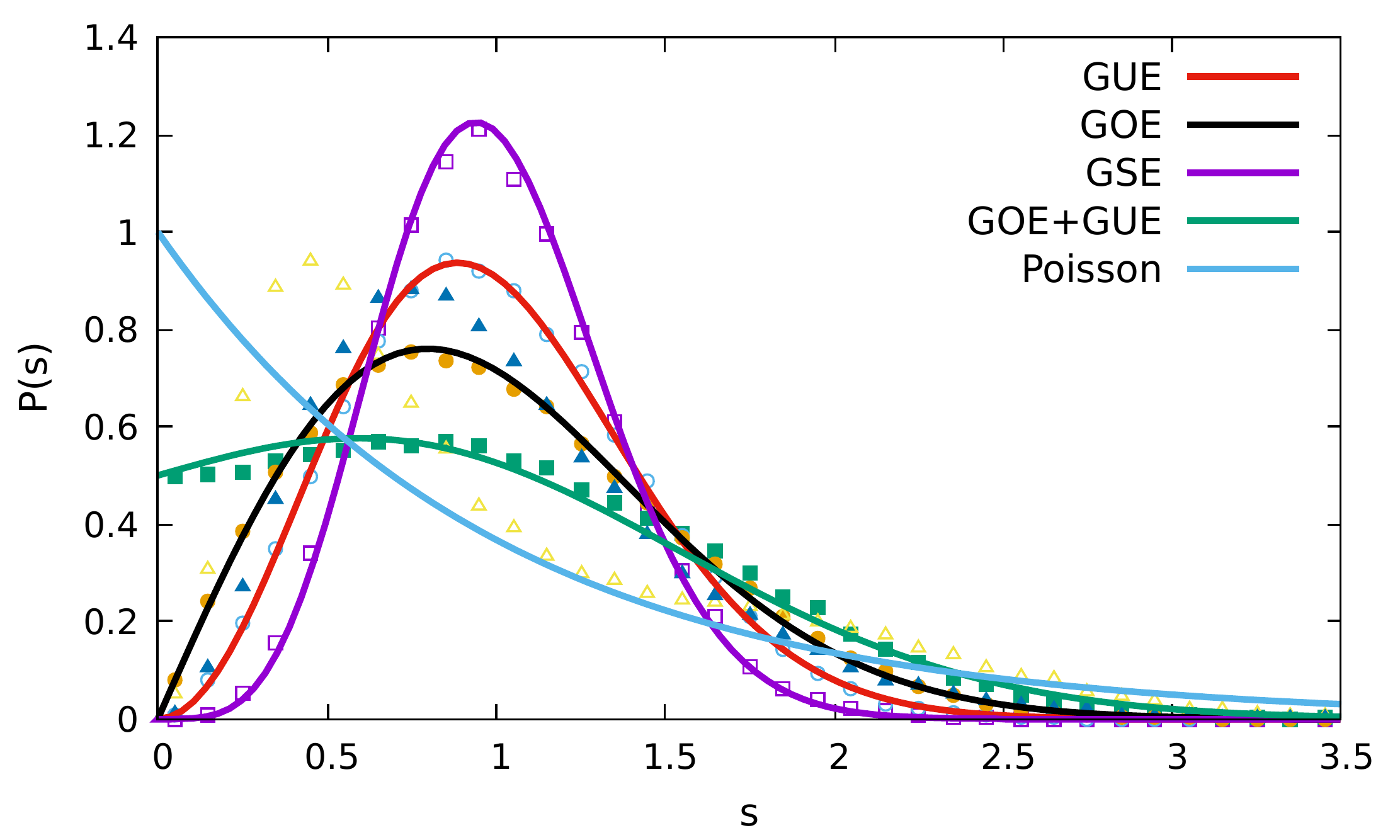}}
	\vspace{-4mm}
	\caption{$P(s)$ for $N = 34$ and $k =1$. The $90\%$ of the eigenvalues around the center of the spectrum are considered. Each curve represents results for different disorder realizations, namely, without any ensemble average. For comparison, we also include the random matrix prediction for different universality classes and Poisson statistics.}
	\label{bulktailfixxone}
\end{figure}

The calculation of the adjacent gap ratio $\langle r \rangle$, see Fig.~\ref{adjtail}, confirms that the maximum degree of sparseness consistent with quantum chaotic features $k = k_c \gtrsim 1$. For smaller $k$, the gap ratio deviates strongly from the random matrix prediction and approaches the Poisson limit. Our main aim here is identify the region of parameters for which quantum chaos occurs rather than the description and nature of the transition. 
Although a transition to Poisson statistics and a critical region with an approximately size invariant spectral correlations are typical of metal-insulator transitions induced by disorder, further investigations , beyond the scope of the paper, would be necessary to reach a firm conclusion.

\section{Emergent Symmetries and Quantum Chaos}\label{sec:emersym}

Having identified the critical sparseness $p {N \choose 4} = kN$ with $k \gtrsim 1$ to observe quantum chaotic features,
we now focus on the limiting case $k = 1$.
Depending on the value of $N \mod 8$, the SYK model for even $q$ is in one of the three Wigner-Dyson universality classes,
while the SYK model for odd $q$ is in one
of the three chiral random matrix classes. In this section, we show that,
for small $k = 1$, at least six of the ten RMT universality classes emerge from
a SYK model for a single value of $N$ in the GUE class. Since the joint spectral density
of the superconducting ensembles \cite{Altland:1997zz} is of the same general form as
the chiral ensembles, our observables cannot distinguish the two.
The study of emergent symmetries  requires a large ensemble and, although
we show some results for $N=34$, our main analysis focuses on $N=26$ where we
can easily generate a large number of disorder realizations with and without the regularity condition.

\begin{figure}[t!]
  \centering
	\resizebox{0.6\textwidth}{!}{\includegraphics{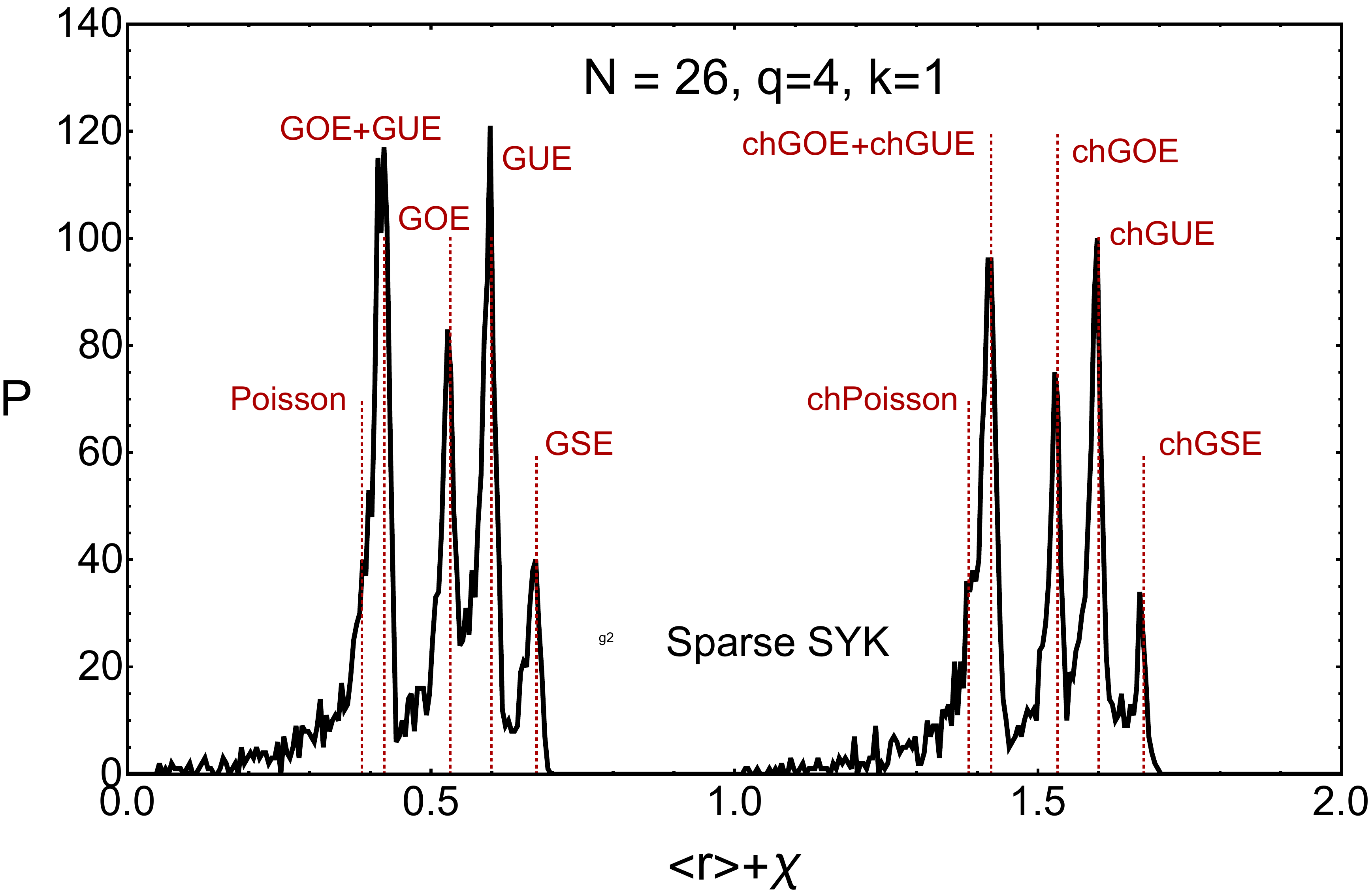}}
  \caption{Distribution of the adjacent gap ratios for an ensemble of $5000$ disorder realizations of the sparse SYK Hamiltonian with $N=26$, $q=4$, $k=1$ and no regularity condition. The red line shows the analytical
  value of the adjacent gap ratio of the corresponding ensemble. We note that for convenience we set $\chi = 1 (0)$ for realizations with (without) chiral symmetry. }  \label{fig:emer}

  \end{figure}

Even after the regularity condition is imposed, for some disorder realizations, we observe what
appears to be an exact two-fold degeneracy while for other realizations there is no degeneracy or only a quasi-degeneracy. In addition,
the spectrum for some disorder realizations has a chiral symmetry $E \to -E$ while for others there are both chiral and two-fold degeneracy. As an example, in Fig.~\ref{bulktailfixxone}, we depict results for different disorder realizations for $N=34$ and $k=1$ where only a spectral average is carried out in
the central spectral window comprising $90\%$ of the total number of eigenvalues $\sim 65000$.
Surprisingly, despite the fact that the symmetry for $N = 34$ is GUE, we observe for some disorder realizations GOE and GSE symmetry. For others disorder realizations, a spacing distribution resembling that of the superposition of two random matrix ensembles is observed.

For a more systematic study, we turn to $N=26$ and $k=1$ where more disorder realizations can be generated. We start our analysis with an ensemble of 5000 configurations without imposing
the regularity condition. To determine if a spectrum has chiral symmetry,
we compute\footnote{This is the $\eta$ index (not to be confused
  with the parameter $\eta$ used throughout this paper) \cite{Atiyah:1973a}
  $
  \eta(s) =
\sum_i \frac {{\rm sign}(E_i)}{|E_i|^s}
$
for $s\to 1$.
}
 \be
\sum_{E_i \neq 0} \frac {{\rm sign}(E_i)}{|E_i|}.
\ee
 For a finite spectrum, it vanishes if the eigenvalues occur in pairs, while
it is of order $2^{N/2}$ for realizations of the sparse SYK model without chiral symmetry. Disorder realizations are labeled
by the index $\chi=0$, when there is no chiral symmetry, and $\chi=1$, when there is chiral symmetry.
  In Fig. \ref{fig:emer}, we show a histogram of $\langle r \rangle +\chi$ for
  this ensemble. The red dotted lines
  denote the values of the adjacent ratio for the various ensembles. Since
  the adjacent ratio is averaged over the spectrum (with the exclusion of 100
  eigenvalues in both tails), up to  corrections that vanish for large $N$,
  the chiral ensembles and the Wigner-Dyson ensembles have the same values
  depending on the Dyson index. It is clear from this figure that it does not
  make sense to calculate spectral correlation by averaging over the full
  ensemble. Rather, it is necessary to partition the ensemble into sub-ensembles
  corresponding to the peaks in Fig. \ref{fig:emer}.

\begin{figure}
	\centering
	\resizebox{0.6\textwidth}{!}{\includegraphics{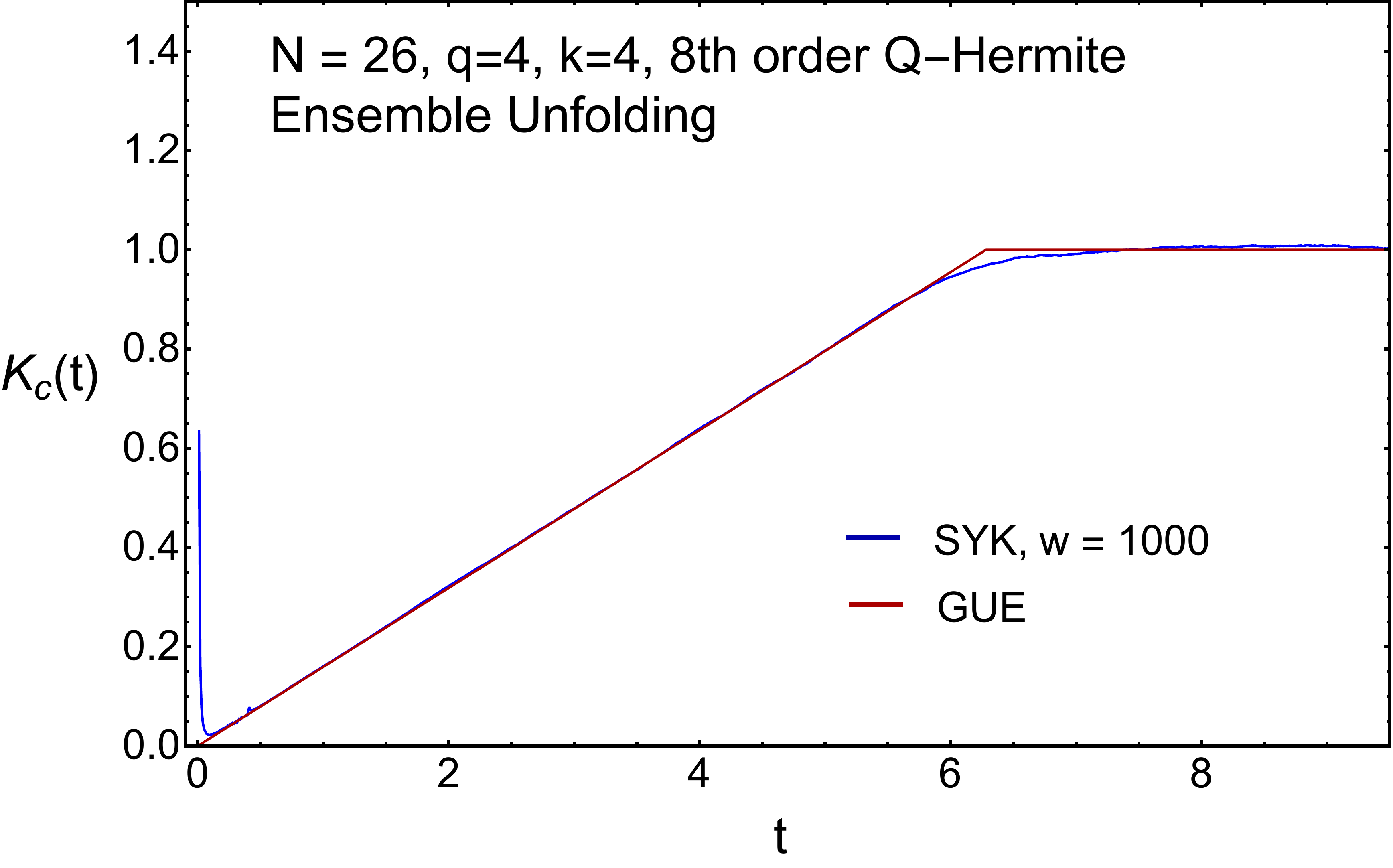}}
	\vspace{-4mm}
	\caption{Connected spectral form factor $K_c(t)$ in units of the Heisenberg time. Unfolding was carried out by the Q-Hermite method \cite{Jia:2019orl}. For $k=4$, we find a large ramp and saturation in excellent agreement with the random matrix prediction. The peak for short time is a well known finite size effect \cite{Jia:2019orl}.
	}
	\label{sfff}
\end{figure}

In order to investigate the scale to which  quantum chaotic features persists,
we turn to the connected spectral form factor
  \cite{Delon-1991,alhassid1992,borgonovi2016,Torres-Herrera2017,cotler2016} for  the unfolded spectrum,
\be\label{eq:sff}
K_c(t) = \frac{\langle Z^*(t)Z(t)\rangle}{\langle Z(0)\rangle^2}-\frac{\langle Z^*(t)\rangle\langle Z(t)\rangle}{\langle Z(0)\rangle^2}
\ee
with $Z=\sum_ie^{i\lambda_it-\beta\lambda_i}$ with $\lambda_i$ the unfolded eigenvalues and $\beta$ the inverse temperature (only the
$\beta =0$ case will be considered). We have removed the disconnected
part related to the one-point function. In order to reduce finite size effects, the sum of $\lambda_k$ is cut-off by  a Gaussian factor
\be
e^{-\frac {\lambda_i^2}{2W^2}}
\ee
with a width $W$ determined such that a significant fraction of the  eigenvalues is included in the calculation. For example, in the case of $N=26$ with 4096 eigenvalues, we choose $W=500$ or $W=1000$.
In agreement with previous spectral analysis \cite{cotler2016,Jia:2019orl},
we have observed, see Fig.~\ref{sfff}, for $k \gg 1$, an excellent agreement with RMT predictions even for relatively short times.
The smearing of the peak at $t = 2\pi$ (the Heisenberg time) is a well documented finite size effect.

\begin{figure}[t!]
\centering
    \includegraphics[width=0.45\textwidth]{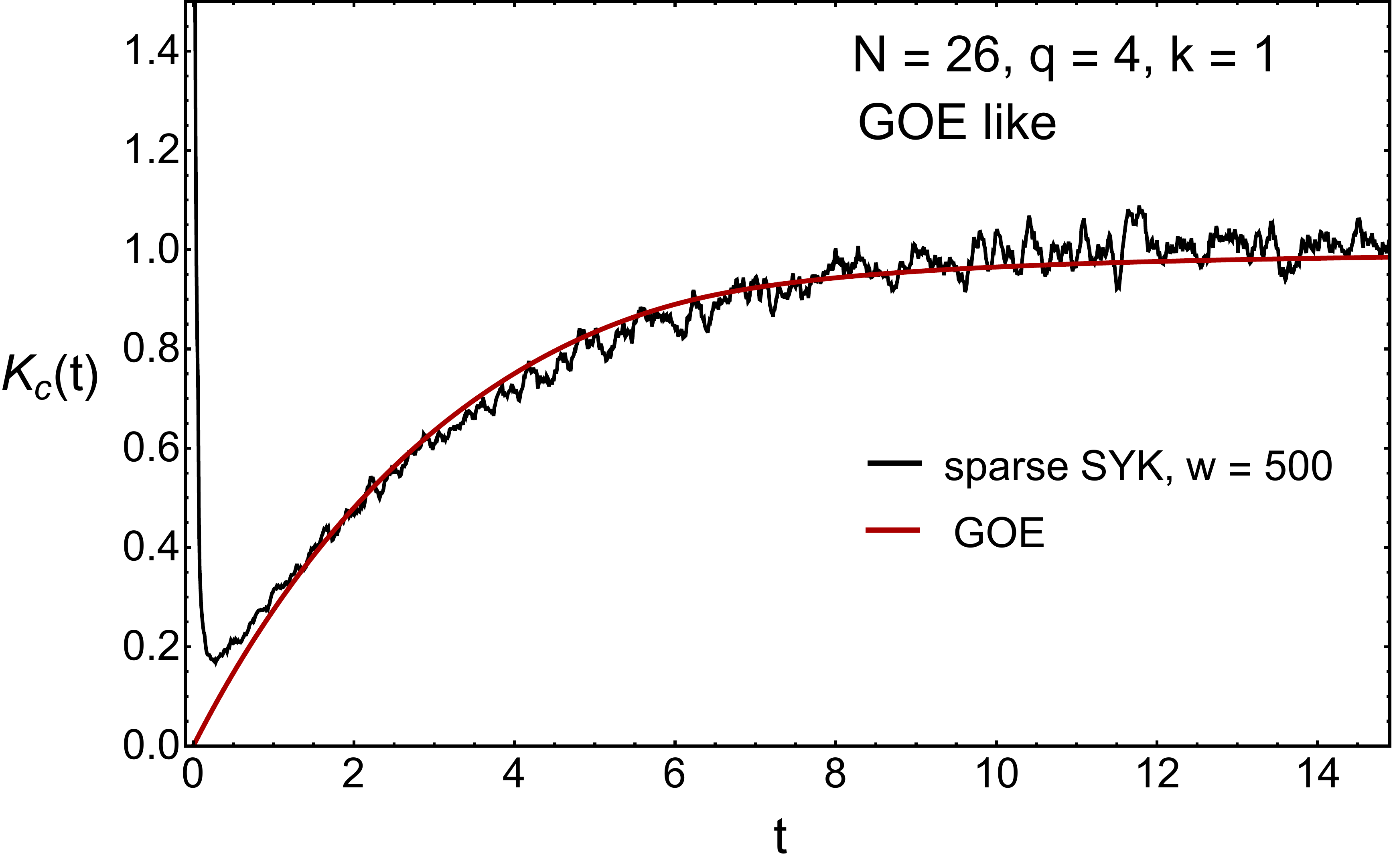}
    \includegraphics[width=0.45\textwidth]{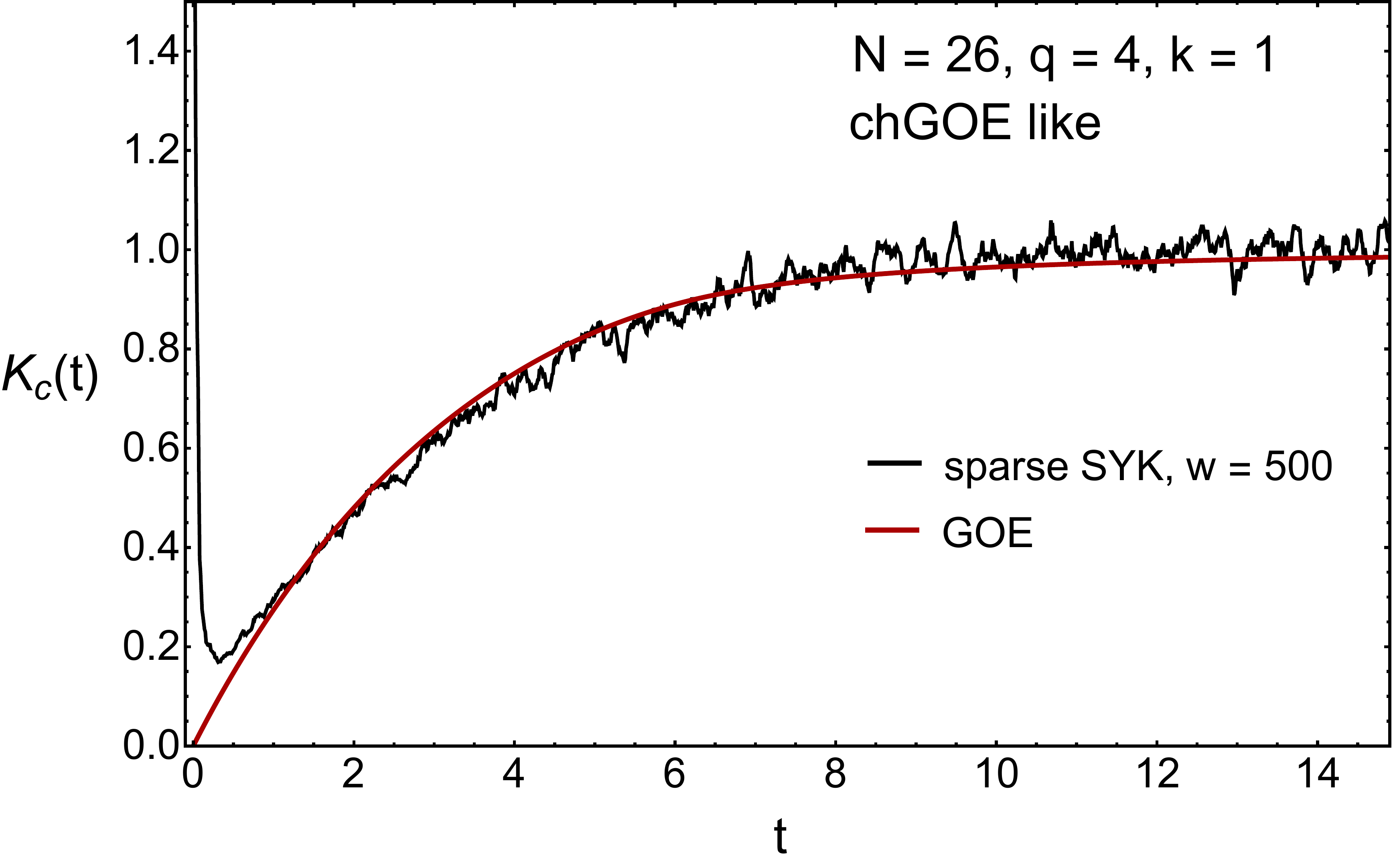}\\
    \includegraphics[width=0.45\textwidth]{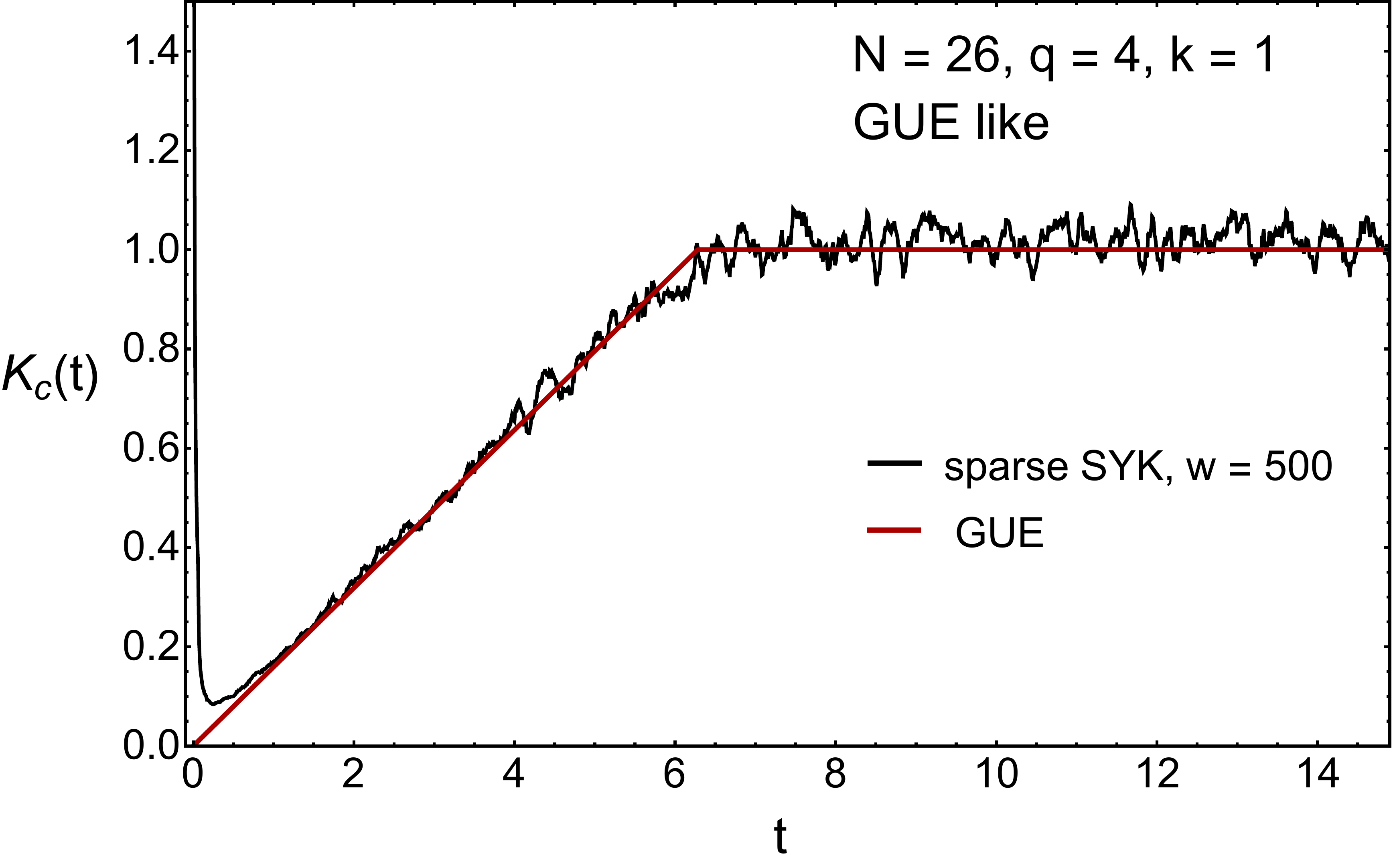}
\includegraphics[width=0.45\textwidth]{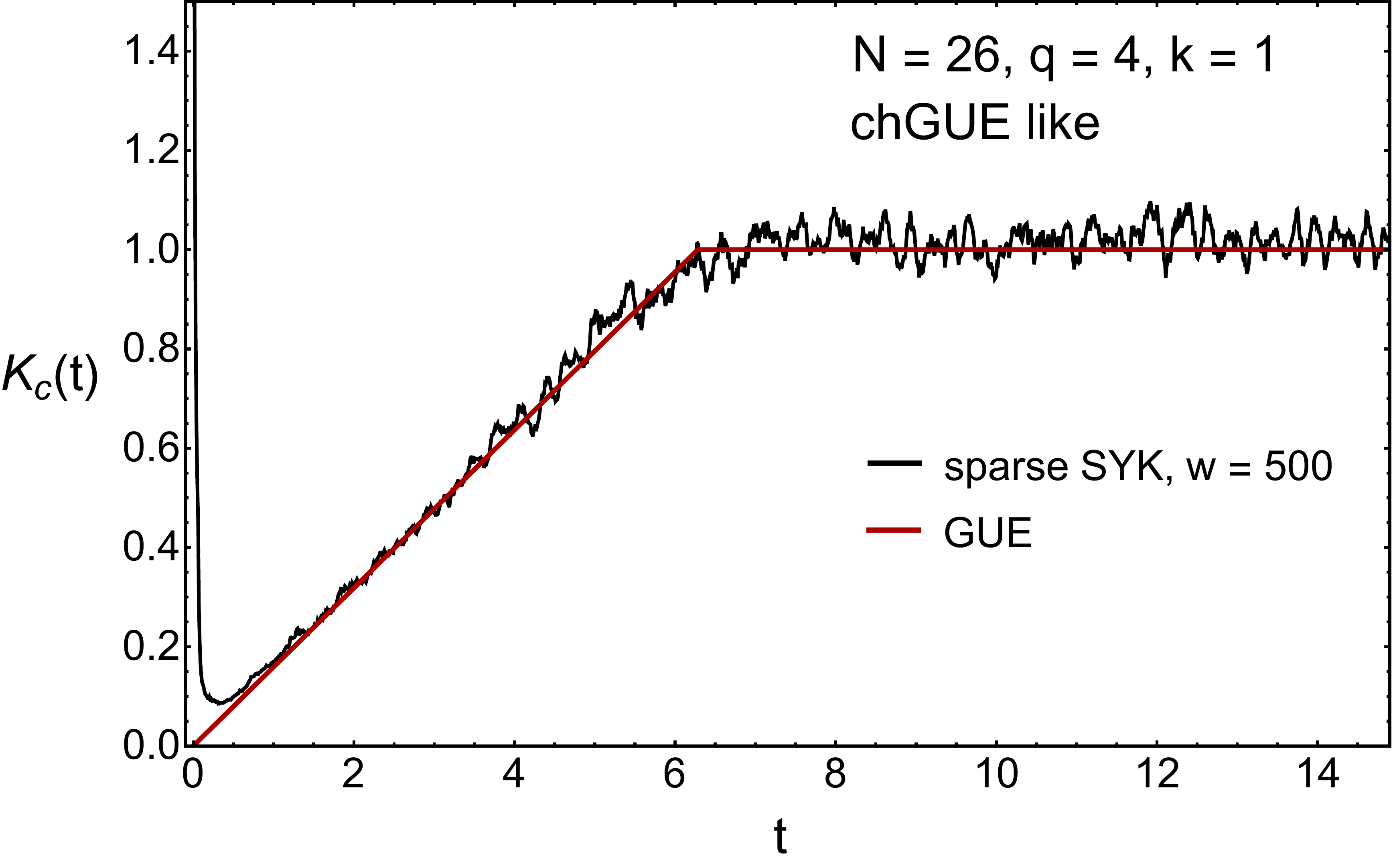}\\
 \includegraphics[width=0.45\textwidth]{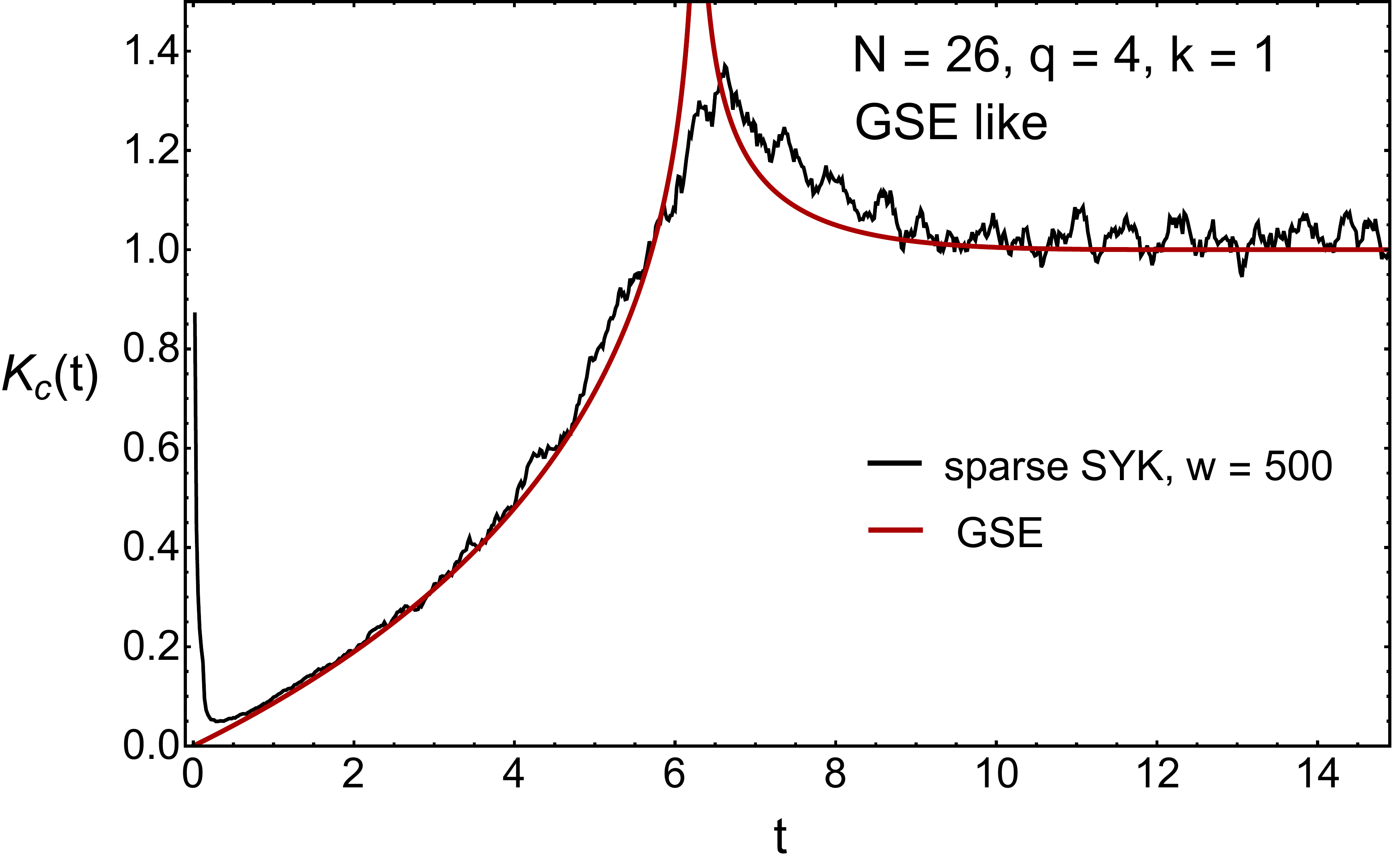}
    \includegraphics[width=0.45\textwidth]{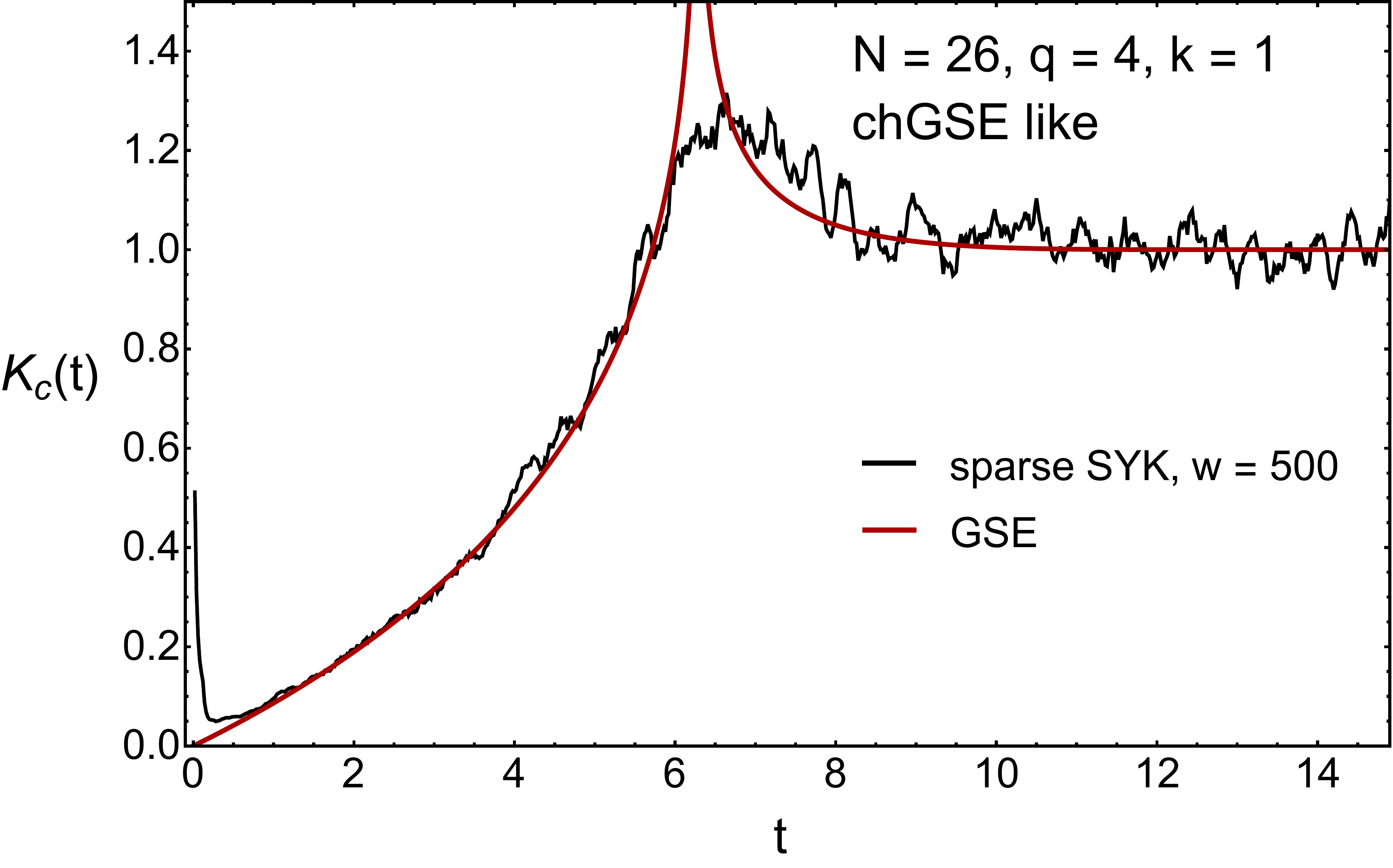}
    \caption{
      Connected form factor for the unfolded spectrum where we have grouped disorder realizations with the same global symmetries. No regularity condition is imposed. Surprisingly, despite  the large degree of sparseness, $k = 1$, the numerical results follows rather closely the predictions of random matrix theory. For each of
      the figures the value of  $\langle r\rangle +\chi $ is within 0.01 equal
      to the corresponding random matrix theory.
}\label{fig:form33}
    \end{figure}

  In Fig. \ref{fig:form33} we show the connected form factor calculated from
  the unfolded eigenvalues for a subensemble with $\langle r\rangle+\chi$ within 0.01
  from the value of the random matrix theory in the legend of the figure.
  The eigenvalues have been unfolded by fitting
  the spectral density of the Q-Hermite polynomials
  corrected by $1 +a_2 H^Q_2(x)+a_4 H^Q_4(x)+a_6 H^Q_6(x)+a_8 H^Q_8(x)$.
  In addition, the unfolded eigenvalues of each realization have been rescaled
  to have an average spacing equal to 1.
  Despite the limitations on the ensemble average to reduce statistical fluctuations due to the different universality classes, we observe very good agreement with universal random matrix results.
  Deviations from RMT occur at $t < 0.5$ where we observe a large peak which should not be confused with the
  peak due to the disconnected part of the form factor. The width of the peak is of the order $1/W$, but its
  area, which is responsible for deviations of the number variance from the RMT results,
  does not depend on $W$.

\begin{figure}[t!]
  \centering
    \includegraphics[width=0.45\textwidth]{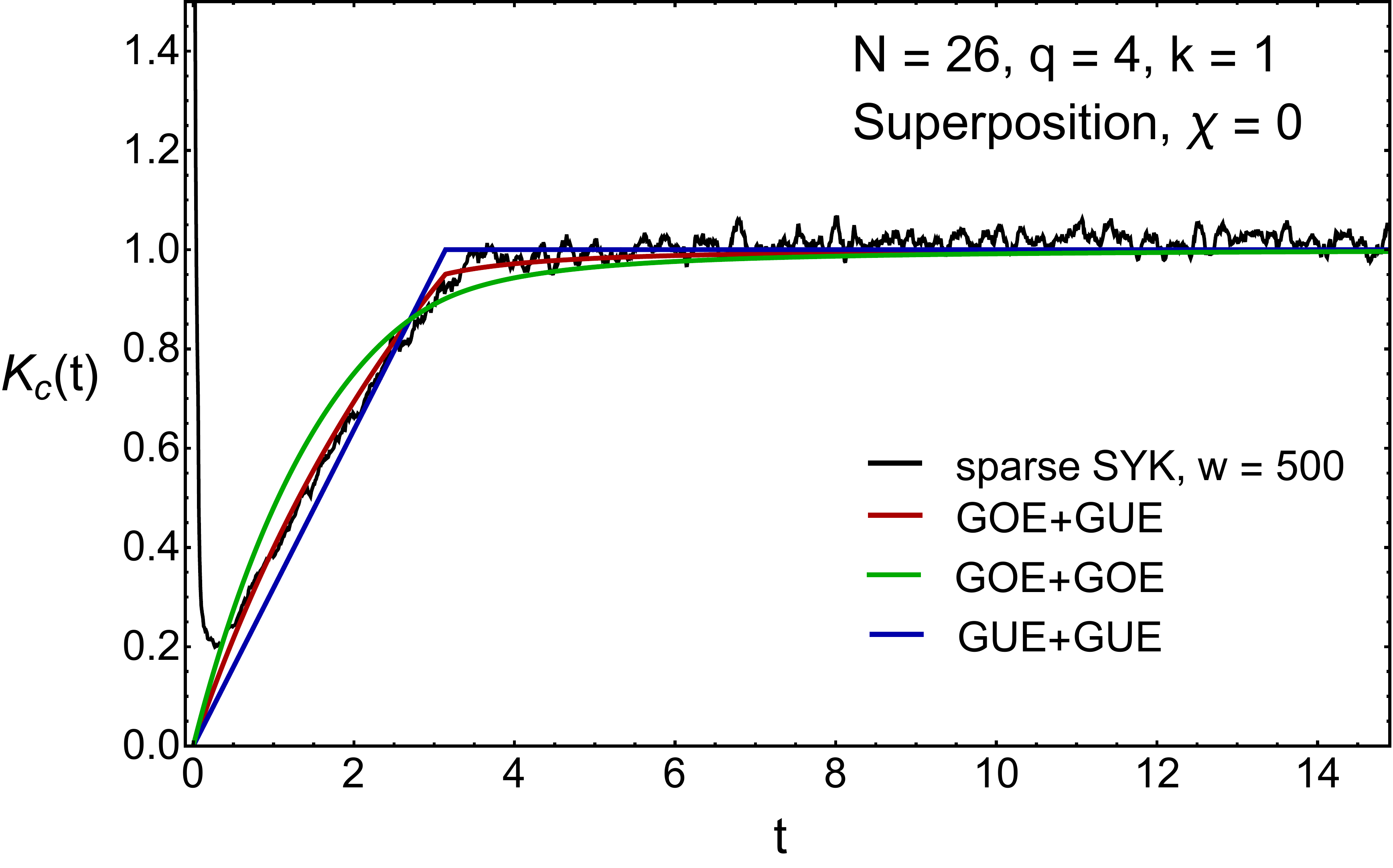}
    \includegraphics[width=0.45\textwidth]{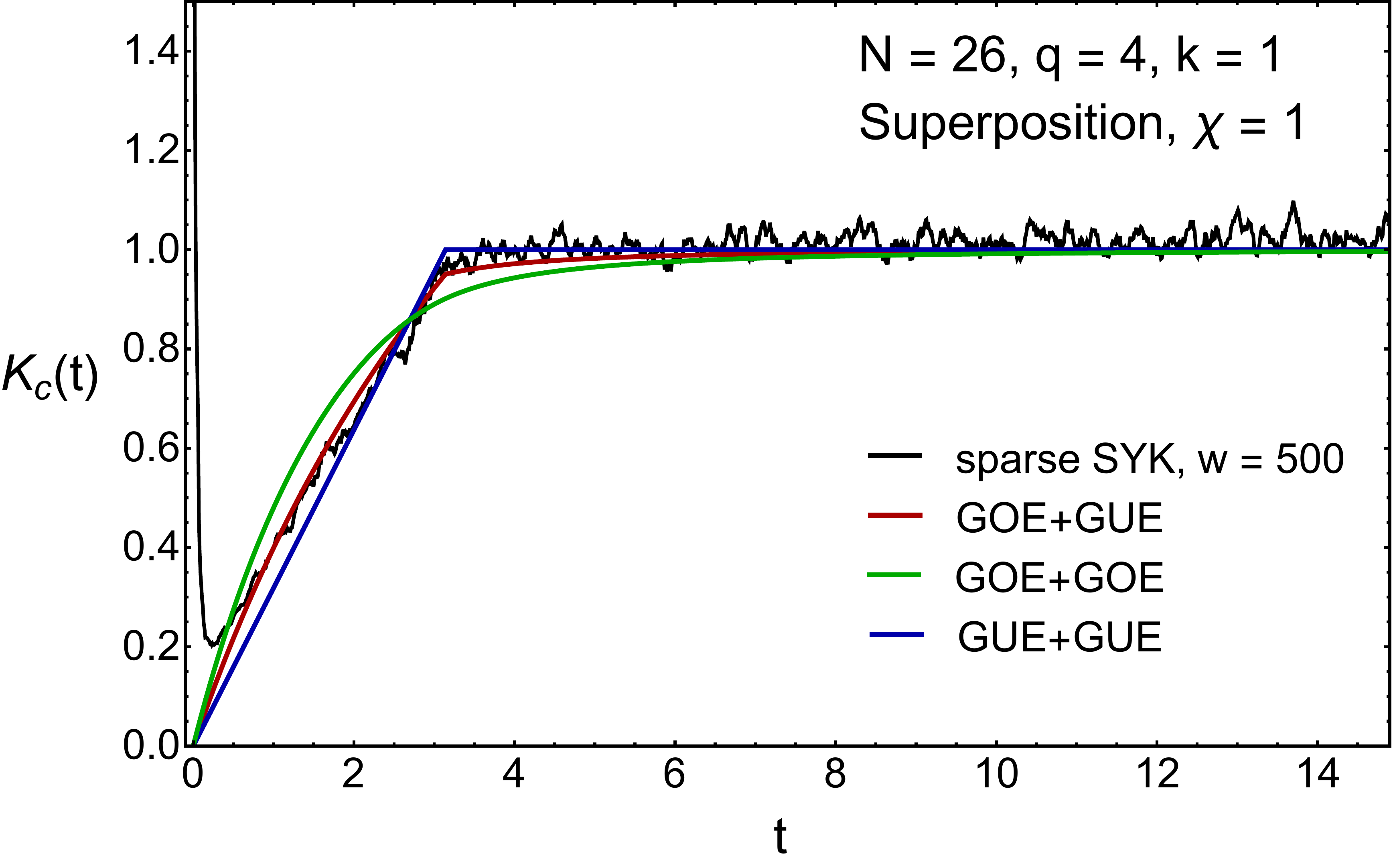}
      \caption{The spectral form factor of the sparse SYK model with
        $N=26$, $q=4$ and $k=1$ for realizations with adjacent ratio in the
        interval $[0.41,0.43]$ with no chiral symmetry in the left panel and
        with chiral symmetry in the right panel. Also shown are the analytical results
         for the superposition of GOE+GOE, GOE+GUE and GUE+GUE, see Eq.~(\ref{eq:superposition}) with $\alpha = 1/2$. No regularity condition is imposed.
  }      \label{fig:mix}
    \end{figure}

  Next we consider realizations with a adjacent ratio of about 0.42 which show a pronounced peak
  in Fig. \ref{fig:emer} both with $\chi =1$ and without chiral symmetry $\chi =0$. The value of
  this ratio corresponds to the superposition of two GOEs, two GUEs or a GOE and
  and GUE with ratios equal to 0.421, 0.423 and 0.423, in this order. The analytical result of the form
  factor for the superposition of two ensembles with an equal total number of eigenvalues follows from the superposition rule for the point correlator of
  unfolded eigenvalues
  \cite{guhr1998}. It  is simply given by
    \be
		\label{eq:superposition}
  K_c(t) = \alpha K_{c,1}(t/\alpha) + (1-\alpha)K_{c,2}(t/(1-\alpha))
  \ee
  with $\alpha$, not to be confused with the scaling of probability introduced earlier, the fraction of the realizations in class 1, and the rest, $1-\alpha$, in class 2.
  In Fig.
  \ref{fig:mix}, we show the spectral form factor of the realizations with adjacent ratio
  in the interval $[0.41,0.43]$ (left) and  in the interval $[1.41,1.43]$. The solid curves represent
  the analytical results \eref{eq:superposition} for $\alpha = \frac 12$. We find good agreement with
  the result of the superposition of a GOE and a GUE.

\begin{figure}[t!]
\centering
    \includegraphics[width=0.45\textwidth]{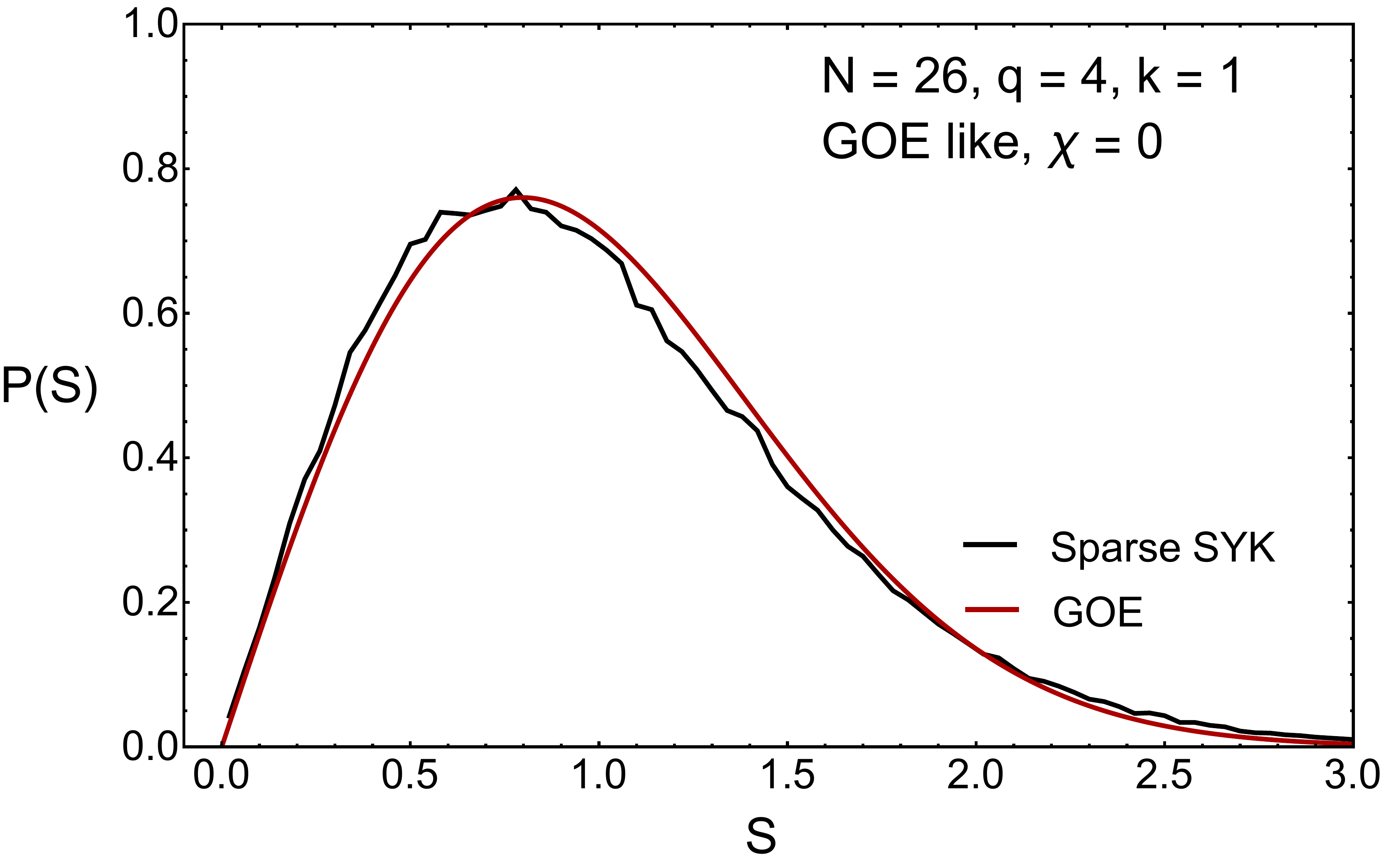}
    \includegraphics[width=0.45\textwidth]{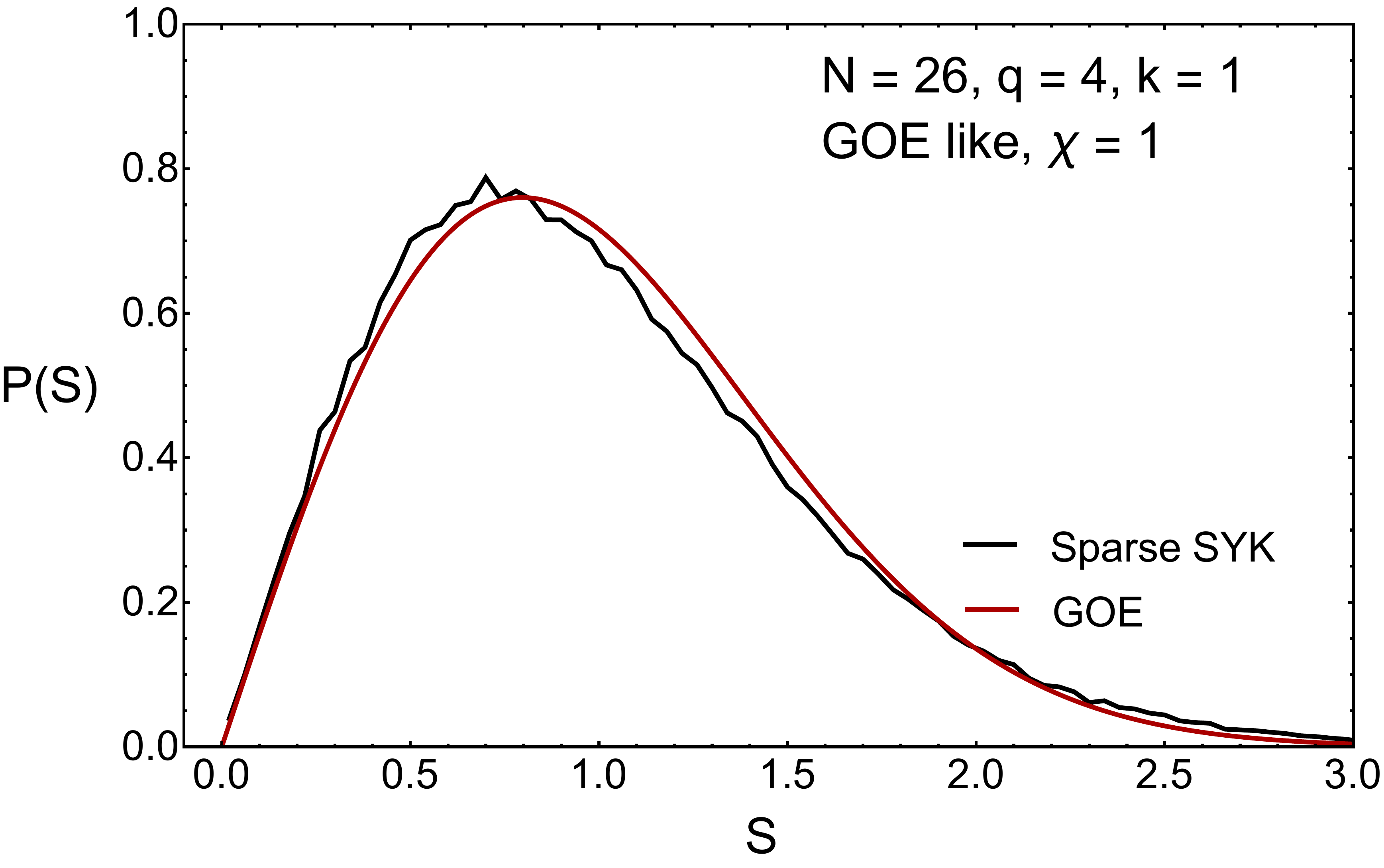}\\
    \includegraphics[width=0.45\textwidth]{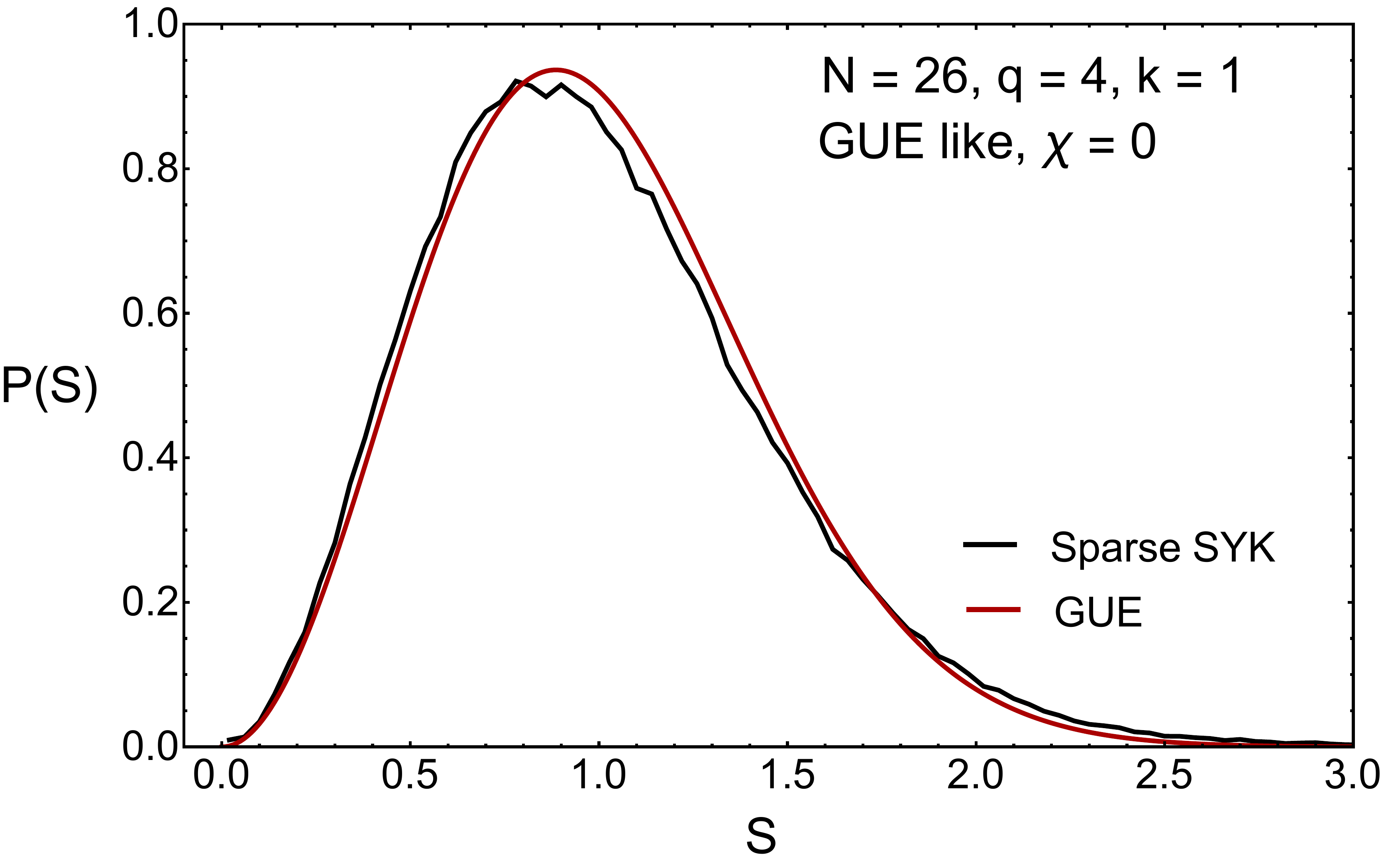}
\includegraphics[width=0.45\textwidth]{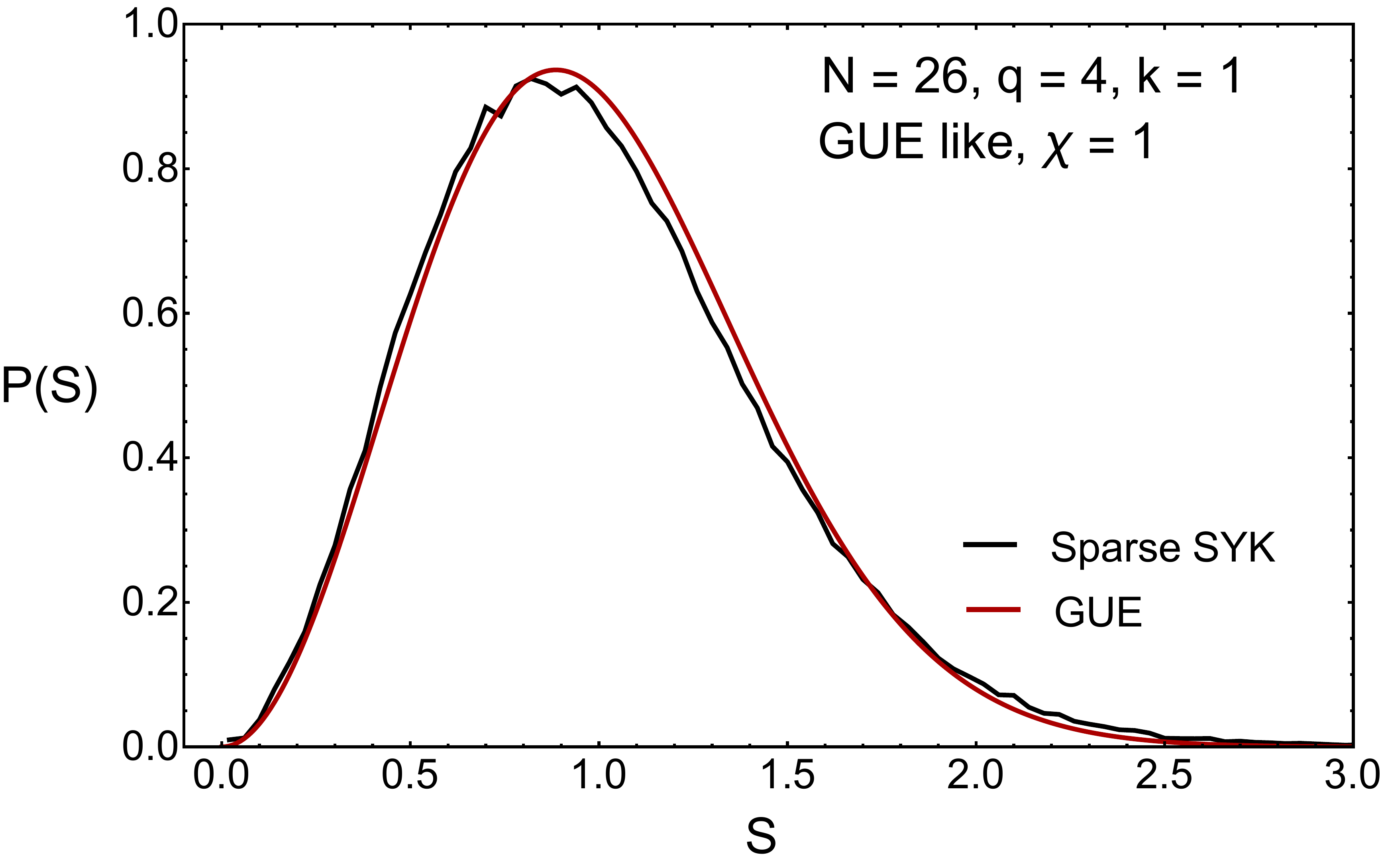}]\\
      \includegraphics[width=0.45\textwidth]{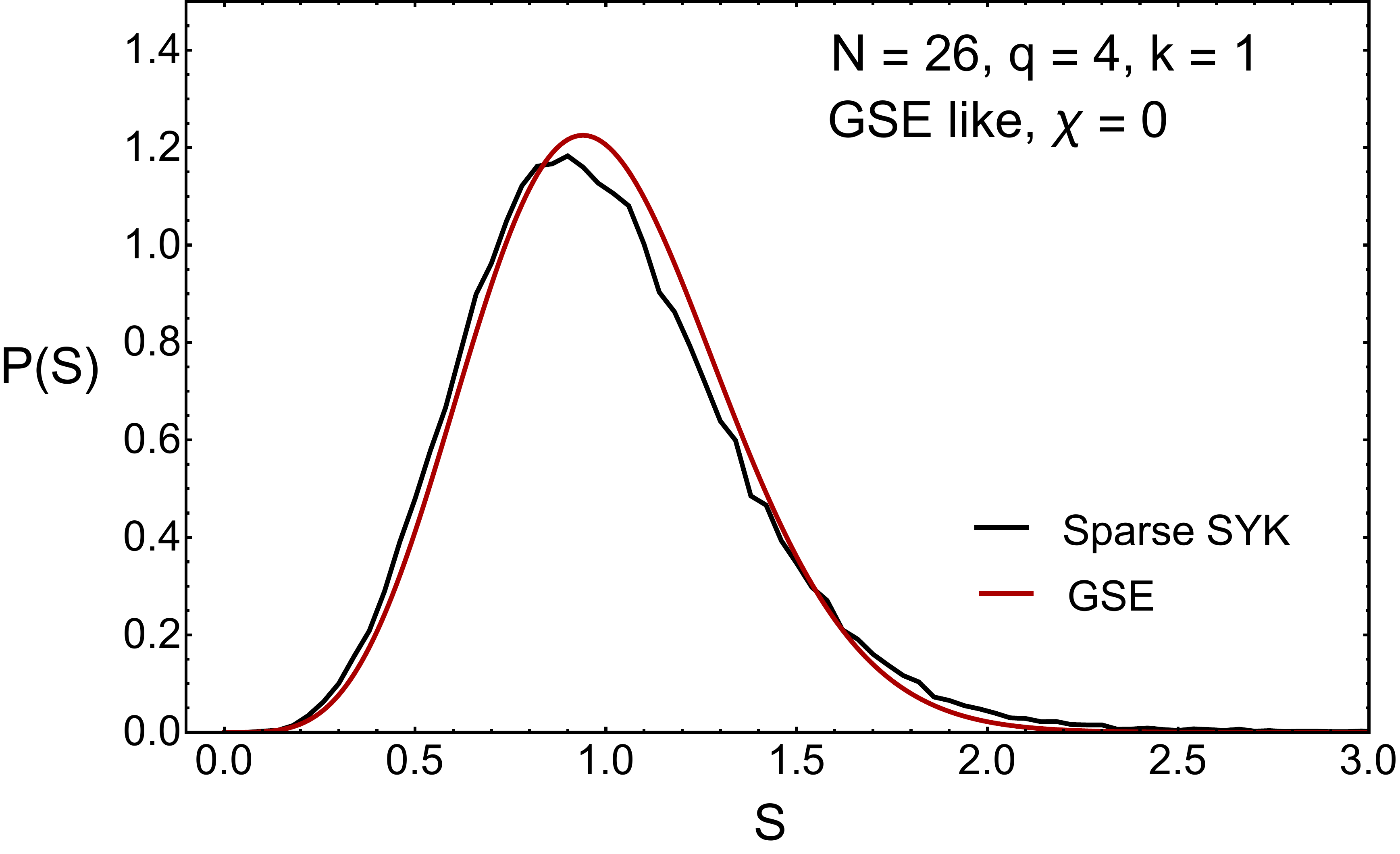}
    \includegraphics[width=0.45\textwidth]{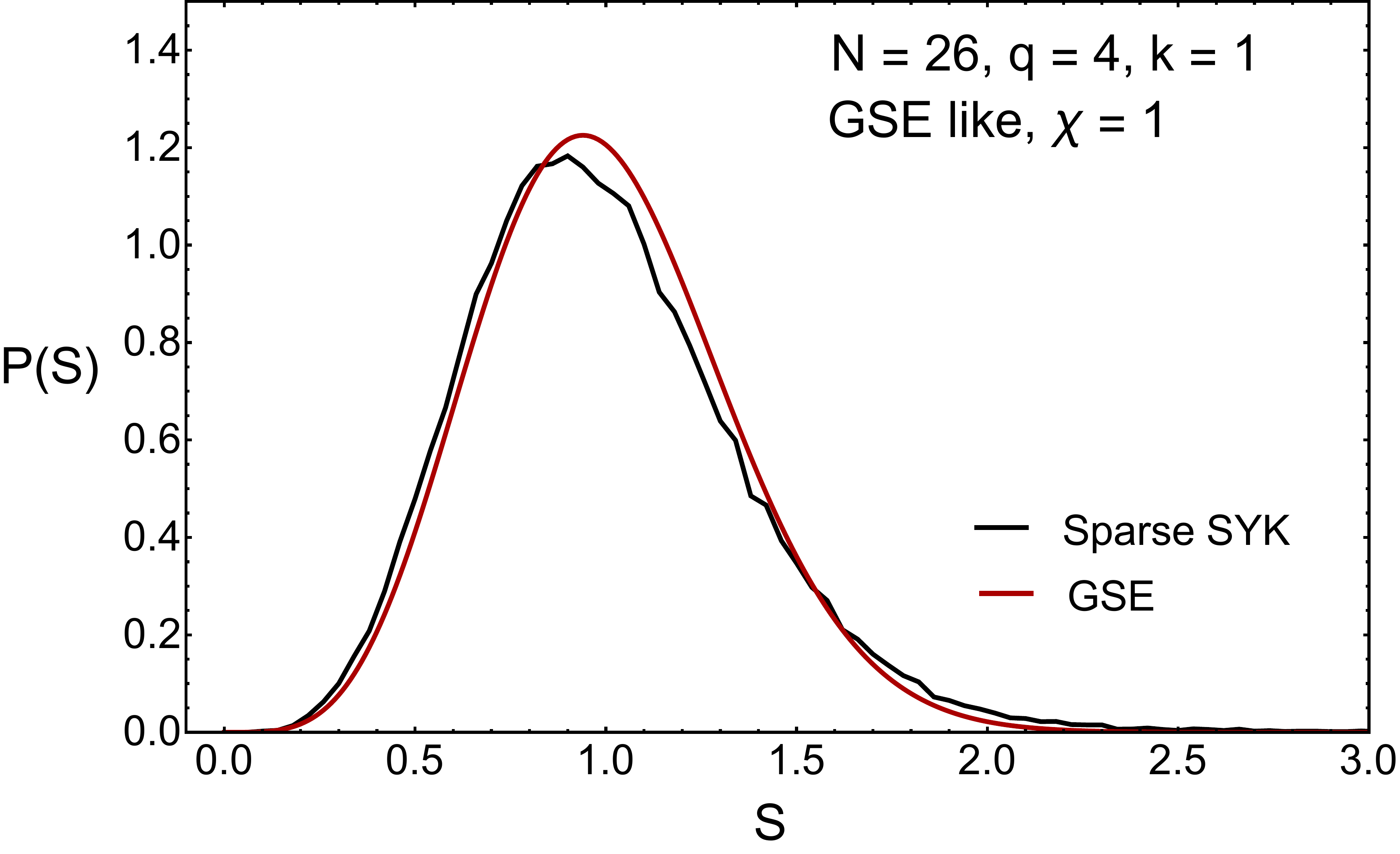}
    \caption{Level spacing distribution $P(s)$ for the same spectrum as Fig.~\ref{fig:form33}. We observe both a good agreement with the RMT prediction and, depending on the disorder realization, results corresponding to different universality classes including the chiral random matrix ensembles ($\chi =1$).}
\label{fig:ps33}
    \end{figure}

Since the spectral form factor agrees well with the universal random matrix results, we expect that also the
nearest neighbor spacing distribution is given by RMT. In Fig. \ref{fig:ps33}, we show the spacing distribution
corresponding to the eigenvalues of Fig. \ref{fig:form33}.
We have excluded realizations with spacings $> 5$ which actually  occur quite frequently. We even have observed spacings of
order 100 times the average spacing. Including these realizations would shift the peak somewhat to the left, but otherwise
the agreement with RMT is again good.

We now study the same parameters $k=1$, $N=26$ but imposing the regularity condition. In this case, about half of the realization have
chiral symmetry, and about a quarter are doubly degenerate, but we did not observe
higher degeneracies in our ensemble of 5000.
For smaller values of $k$, below  $k=1$, the number of emergent symmetries rapidly
increases. In an ensemble of $5000$ disorder realizations, for $N=26$ and $k= \frac 34$,  with  no regularity condition, the maximum degeneracy is $2^9$-fold. Degeneracies always
appear in powers of $2$ which points to the presence of discrete symmetries that
square to one or zero, or to symmetries of the Hamiltonian that contain
both  commuting and  anti-commuting combinations.
Already for $k = \frac 34$, in particular with the regularity condition, almost
all realizations have chiral symmetry 
and a large number of degeneracies appear in the  spectrum.
In Fig. \ref{fig:degen}, we show a histogram of the 2-logarithm of the degeneracy of the spectrum for $N=26$ and $k=0.75$. With regularity condition (left) the spectrum of almost all configurations is either 8 fold or 16 fold degenerate. Without regularity condition, we find
a wider distribution of the degeneracies up to a 512 fold degeneracy.

\subsection{Origin of the Emergent Symmetries} \label{sec:EmergentSymOrigin}

We now investigate the origin of the emergent symmetries in the sparse limit.  For the model without regularity condition, we can imagine an extreme sparseness $k \sim 1/N$. Then a realization of the Hamiltonian typically involves only one product of four Dirac matrices, namely
\begin{equation}\label{eqn:verySparseirregular}
H= J_{i_1 i_2 i_3 i_4} \gamma_{i_1}\gamma_{i_2}\gamma_{i_3} \gamma_{i_4}
\end{equation}
with no Einstein summation convention. Such a Hamiltonian has two energy levels with energies $\pm J_{i_1 i_2 i_3 i_4}$, each level having a $2^{N/2 -1}$ degeneracy.  This Hamiltonian has a large number of symmetries including chiral symmetries and symmetries responsible for the observed spectral degeneracies. These symmetries are represented by a product of Majoranas
\begin{equation}\label{eqn:mBodyOp}
i^{l(l-1)\over 2}\prod_{m=1}^l \gamma_{j_m}
\end{equation}
that commute with the Hamiltonian if $\{i_1, i_2, i_3, i_4\}$ and $\{j_1, j_2,\ldots, j_l\}$ have an even number of common elements  or anti-commutes with the Hamiltonian if $\{i_1, i_2, i_3, i_4\}$ and $\{j_1, j_2,\ldots, j_l\}$ have an odd number of common elements. The former operators form a large nonabelian symmetry group which explains the observed large degeneracy; the latter kind of operators are the chiral symmetries which explain why the energies come in pairs $\pm E$.

The same story holds when the model becomes slightly less sparse, when the Hamiltonian is a sum several products of Dirac matrices: an $l$-body operator defined in Eq. \eqref{eqn:mBodyOp} is a symmetry (chiral symmetry) if $\{j_1, j_2,\ldots, j_l\}$  have even (odd) number of common elements with the set of subscripts of every term in the Hamiltonian. A simple example is the following: for $N=10,\  q=4$ and $k=0.5$ with regularity condition, we can for example obtain a Hamiltonian of the form
\begin{equation}
H= J_{1357} \gamma_1\gamma_3\gamma_5\gamma_7+ J_{25610} \gamma_2\gamma_5\gamma_6\gamma_{10}+ J_{34610} \gamma_3\gamma_4\gamma_6\gamma_{10}+ J_{2789} \gamma_2\gamma_7\gamma_8\gamma_9+ J_{1489} \gamma_1\gamma_4\gamma_8\gamma_9.
\end{equation}

\begin{figure}[t!]
\centering
    \includegraphics[width=0.45\textwidth]{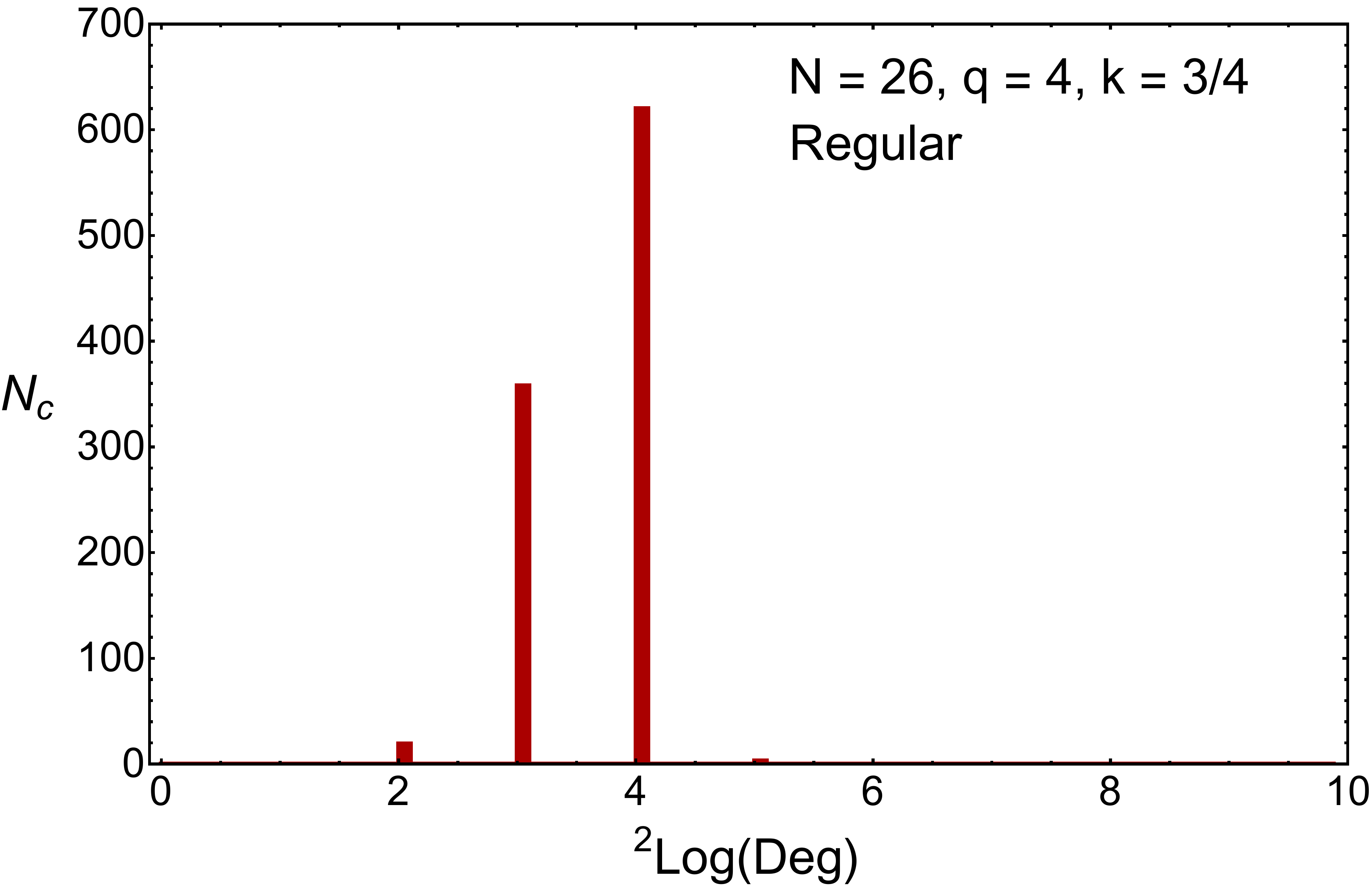}]
    \includegraphics[width=0.45\textwidth]{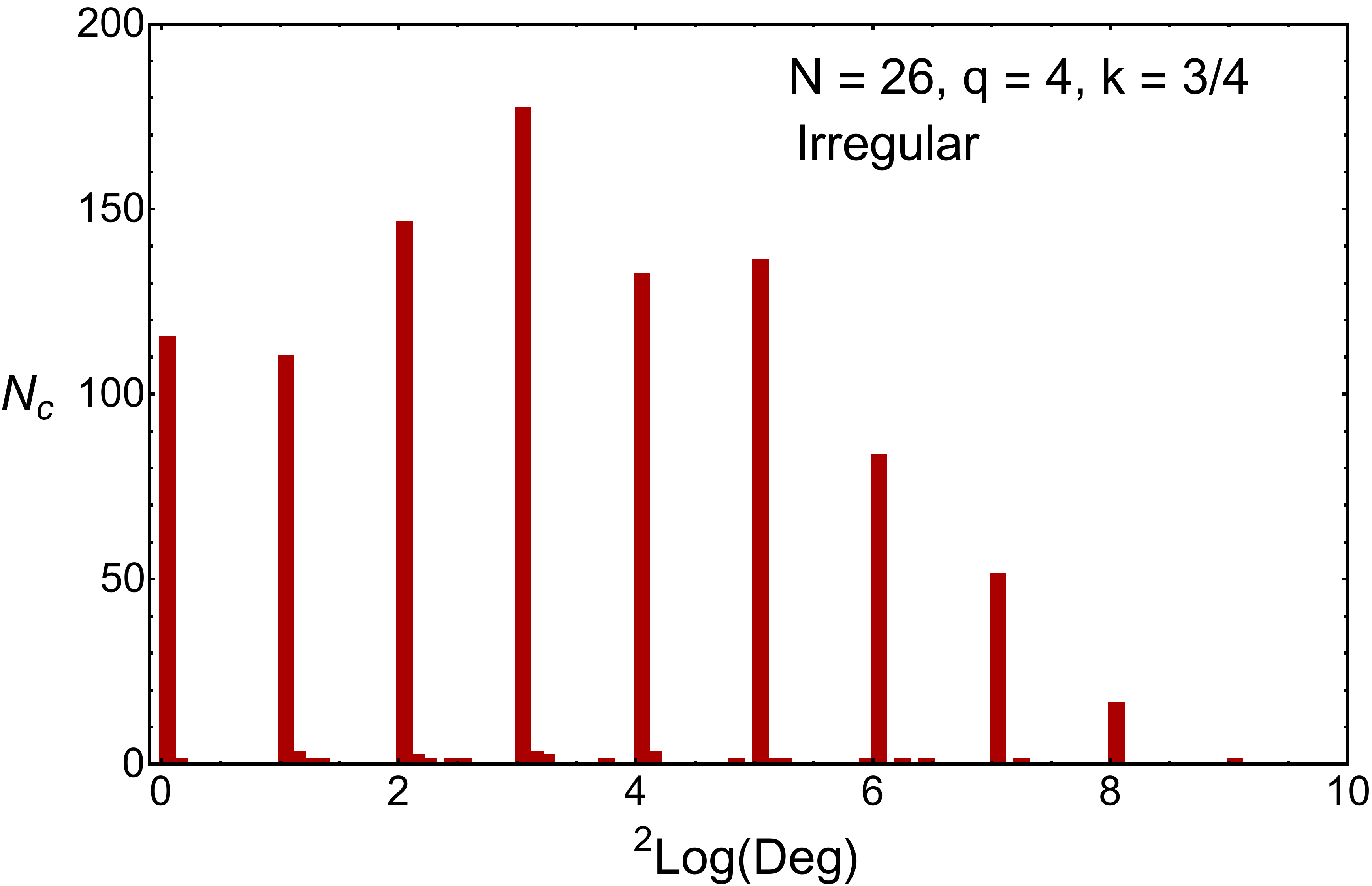}]\\
      \caption{Histogram of the 2-logarithm of the degeneracy of the spectrum for $N=26$ and $k=0.75$ both with
      regularity condition (left) and without regularity condition (right). The total number of disorder configurations is $1000$.}      \label{fig:degen}

    \end{figure}

The symmetries are,
\begin{align}
A_1 & = \gamma_2\gamma_4\gamma_6\gamma_8,\
A_2  = \gamma_3\gamma_4\gamma_7\gamma_8,\ \
A_3  = \gamma_c = i \prod_{l=1}^{10}\gamma_l,\\
B_1 & = \gamma_1\gamma_2\gamma_5\gamma_9,\
B_2  = \gamma_6\gamma_8\gamma_9\gamma_{10},\
B_3 = \gamma_1\gamma_2\gamma_4\gamma_7\gamma_{10},
\end{align}
and all the operators generated by the above six operators. There is no chiral symmetry for this Hamiltonian.  Note that $A_1, A_2, A_3$ commute with each other and
\begin{align}
\{A_1, B_1\} &=0,\ [A_2,B_1]=0,\ [A_3,B_1]=0,\nn\\
\{A_2, B_2\} &=0,\ [A_1,B_2]=0,\ [A_3,B_2]=0,\\
\{A_3, B_3\} &=0,\ [A_1,B_3]=0,\ [A_2,B_3]=0.\nn
\end{align}
Hence $(H, A_1, A_2, A_3)$ gives a complete set of quantum numbers of the form $(E,\pm 1, \pm 1, \pm 1)$ and $B_1, B_2, B_3$ respectively flip the quantum numbers of $A_1, A_2,A_3$ without changing the energy. Therefore, for such a system, each eigenvalue is $2^3=8$ fold degenerate. This explains the numerical degeneracies depicted in Fig. \ref{fig:degen}. We stress that the spectral degeneracy is directly related to the non-abelian nature of the symmetry group. Symmetry in itself only implies simultaneous diagonalization with the Hamiltonian but not degeneracy.
We can also find examples where there are chiral symmetries but no other symmetries leading to degeneracies,
and examples where  both  are present.

 Given the above discussion, it becomes interesting to ask statistically how the number of symmetries and chiral symmetries scales with respect to $N$ at different values of $k$. A precise study of this question is beyond the scope of the current paper, but we mention on the fly our preliminary numerical observations for spectra obtained imposing the regularity condition in the generation of the Hamiltonian:
 \begin{enumerate}
 \item If $k<1$, the number of emergent symmetries grows quickly as $N$ grows.
\item If $k=1$, the number of emergent symmetries stays more or less constant (or grows very slowly) as N grows.
\item If $k>1$, emergent symmetries only rarely occur and with a frequency that decreases rapidly as $k$ increases.
\end{enumerate}

\begin{figure}[t!]
	\centering
	\resizebox{0.49\textwidth}{!}{\includegraphics{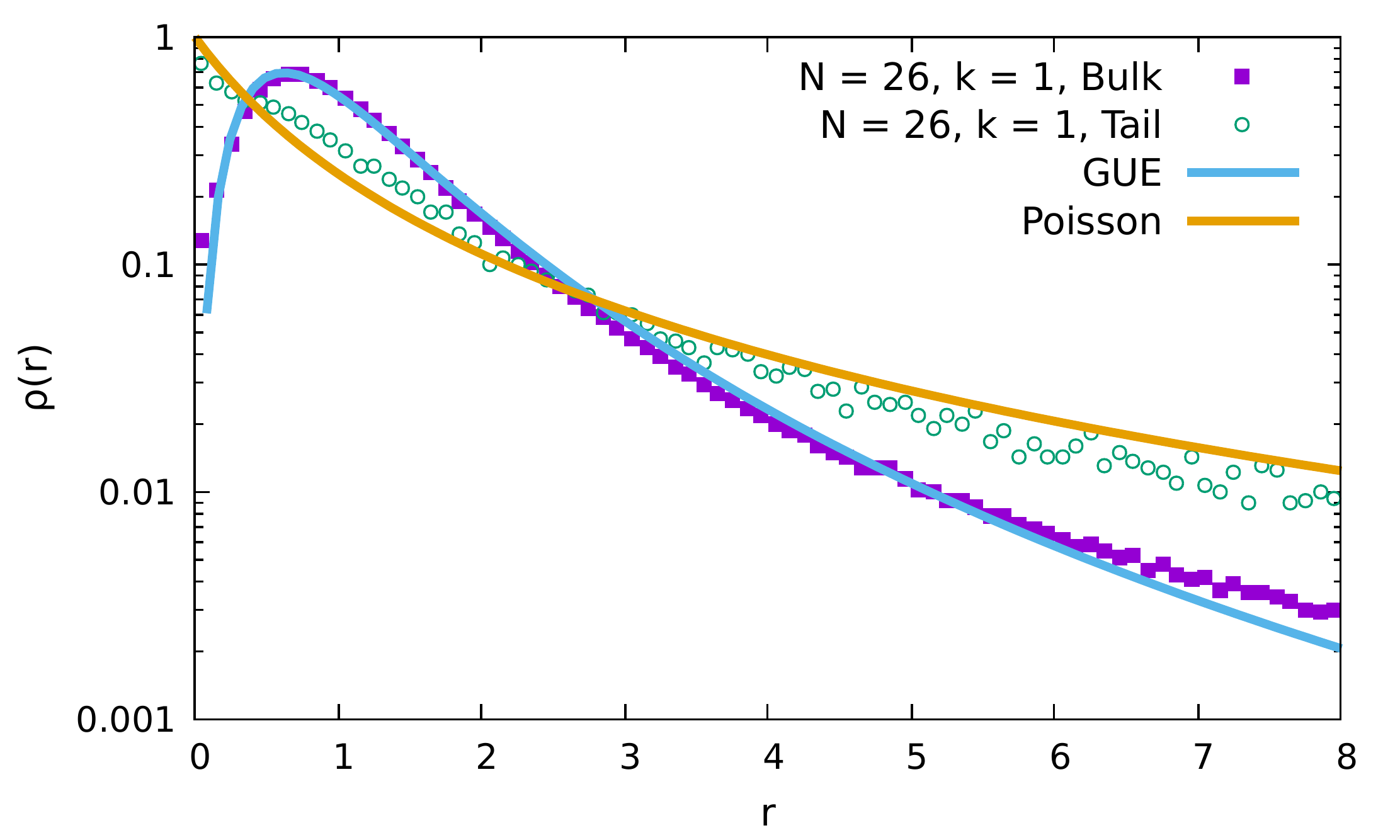}}
	\resizebox{0.49\textwidth}{!}{\includegraphics{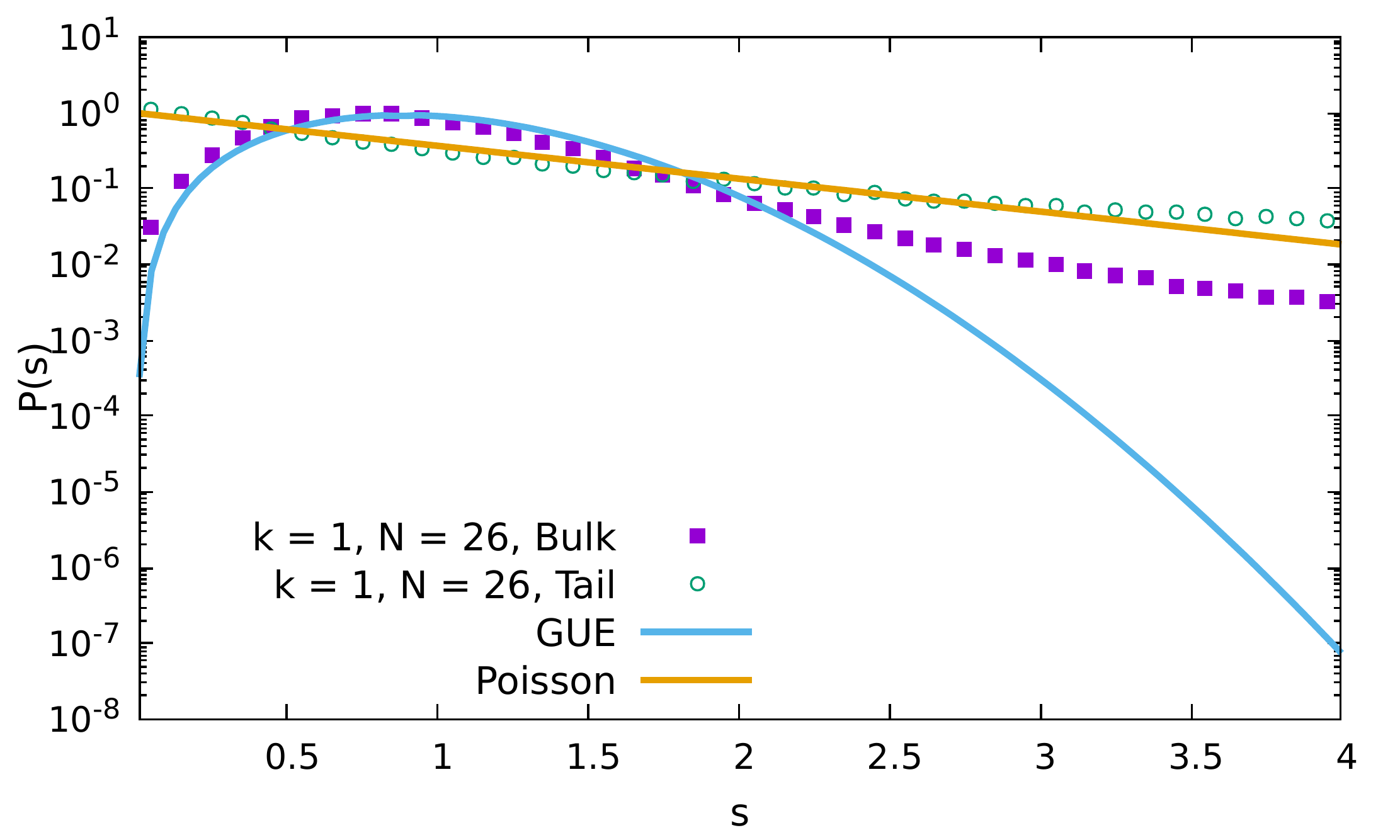}}
	\vspace{-4mm}
	\caption{Left: Adjacent gap ratio distribution $\rho(r)$ for $N = 26$ and $k = 1$, $1000$ disorder realizations, for both the bulk and the tail ($2N$ lowest eigenvalues) of the spectrum. The regularity condition is imposed. Right: The same for $P(s)$. We have fixed the non-zero $x_{ijkl}$ so that the system has a global symmetry that leads to a double degeneracy of the spectrum for all disorder realizations. This degeneracy is removed in the calculation of $P(s)$ and $\rho(r)$.}
	\label{bulktailfixx}
\end{figure}

The presence of emergent symmetries or chiral symmetries can alter the RMT symmetry class naively expected from the corresponding dense SYK model. For example the $q=4$ dense SYK model does not have any chiral symmetry and hence always falls into one of the three Wigner-Dyson ensembles, whereas in the very sparse regime of the $q=4$ sparse SYK model we see how chiral symmetry can emerge, and hence chiral ensembles can appear. The emergent symmetries can also alter the symmetry class in more subtle ways. We see that $N=26, q=4$ (regular) sparse SYK model, whose dense counterpart always lies in the GUE class, can have realizations in the GOE and GSE classes in the very sparse regime $k=1$ (Fig. \ref{fig:ps33}). To explain this we first briefly recap why the dense $N=26, q=4$ SYK model is always GUE. For any even $N$, there are exactly two independent symmetries for the dense SYK model: a unitary symmetry
\begin{equation}
\gamma_c:= i^{\frac{N(N-1)}{2}} \prod_{l=1}^N \gamma_l
\end{equation}
and an anti-unitary symmetry
\begin{equation}
T:= C K
\end{equation}
where $K$ is the complex conjugation and $C$ is the charge conjugation matrix such that $C\gamma_i C^{-1} = \pm \gamma_i^T$. Since we always look at the eigenvalue statistics in a fixed quantum number sector of the unitary symmetry, for the anti-unitary symmetry to play a role it must commute with the unitary symmetry. Hence we have
\begin{align}
[T,\gamma_c]\neq 0 &\implies \text{GUE}, \nn\\
[T,\gamma_c]= 0,\ T^2=1 &\implies \text{GOE},\\
[T,\gamma_c]= 0,\ T^2=-1 &\implies \text{GSE}.\nn
\end{align}
For $N=26$, $T$ and $\gamma_c$ do not commute (in fact they anti-commute) and hence we have GUE for the dense model.   However in the case of $N=26, q=4, k=1$ (regular) sparse model, it could happen that we have an emergent unitary symmetry $A$ such that
\begin{equation}
\{A, \gamma_c\}=0,\ [A, T] =0.
\end{equation}
Then we can define a new anti-unitary symmetry
\begin{equation}
T' = A T,
\end{equation}
which commutes with $\gamma_c$, and $T'^2 = 1$ or $-1$ depending on which Dirac matrices $A$ contains. The former case gives us GOE and the latter gives us GSE. In Appendix \ref{app:N26GOEGSE} we give two concrete examples of this phenomenon. There also can be scenarios where the emergent symmetries give rise to degeneracies but do not change the symmetry class, such as the example shown in Fig. \ref{bulktailfixx}.

Once these degeneracies are taken into account, so that we fix $x_{ijkl}$ and carry out disorder average over $J_{ijkl}$ only, we observe the following (see Fig.~\ref{bulktailfixx}): $P(s)$ in the bulk of the spectrum is well described by RMT but only for $s <1$, for larger $s$,
the agreement with Poisson statistics is excellent. By contrast, $P(s)$ in the tail of the spectrum comprising the lowest $2N$ eigenvalues,
shows excellent agreement overall with Poisson statistics. Results for the distribution of the gap ratio $\rho(r)$ are qualitatively similar,
the bulk of the spectrum is well described by RMT while the tail by Poisson statistics. There is no discrepancy with the level spacing results because the gap ratio provides spectral information of the shortest-range scale, a region where $P(s)$ still agrees with GUE. The tail of the spectrum is close to the prediction for Poisson statistics though we observe a peak at small $r$ likely related to some other emergent symmetry that we have failed to identify.

In summary, once the symmetries are factored out, it seems that even for $k =1$, it remains some degree of level repulsion in the bulk of th spectrum that may indicate some remaining quantum chaotic features though deviations from the RMT prediction are very strong. By contrast, in the tail of the spectrum, the results are consistent with Poisson statistics. The latter suggests that the system may have a mobility edge at finite energy. It would be interesting to further characterize the exact nature of the transition though our main motivation here is only to determine the maximum sparseness for which quantum chaos is observed.

\section{Conclusions and outlook}\label{sec:conclusions}
We have investigated the spectral density and spectral correlations of
a sparse SYK model as a function of the degree of sparseness. We have identified the maximum sparseness strength consistent with a Schwarzian spectral density, once collective fluctuations are factored out, and quantum chaotic level statistics.
These are features of
a field theory with a quantum gravity dual. 
We have carried out explicit analytical calculations of the spectral density moments that have revealed a striking relation between the leading correction due to the sparsity of the SYK Hamiltonian, $\sim 1/(kN)$, and the leading large $d$ correction  of the Parisi's model, a $U(1)$ gauge theory on a $d$-dimensional hypercubic lattice. As the critical sparseness for quantum chaos is approached, we have noticed the emergence of novel global symmetries that not only induce spectral degeneracies but result in an ensemble that, for a single value of $N$, contains
disorder realizations with level statistics well described by any of the three Wigner-Dyson symmetry classes, and the three chiral random matrix ensembles.

 Our results raise some interesting questions: effectively, the sparse SYK Hamiltonian is represented by a sparse random matrix. Can the matrix defined in this way be relevant for matrix models describing quantum JT gravity? Is the critical sparseness to observe quantum chaos of relevance in the description of realistic interacting quantum dots \cite{altshuler1997}. Is there some explicit relation between Fock-space geometry and space-time so that these sparse SYK models have a natural gravity dual? Is it possible to characterize more generally the connectivity and regularity of a hypergraph so that we can establish the condition for quantum chaos and the existence of a gravity dual in terms of these parameters? About this last point, it would be interesting to study how the sparsity of the random hypergraph affects the early time diagnostics of quantum chaos: the OTOCs and the related diagnostics of operator growth.
 In particular, it would be interesting to clarify whether the high degree of sparsity has sharp effects on the growth of local operators built out of products of Majorana fermions \cite{carrega2020unveiling}.
 It would also be interesting to push further the relation between the sparse SYK model and the Parisi's model to, among other things, to identify the role of the latter in the context of holography. We expect to address some of these problems and questions in the near future.
\acknowledgments
A.M.G.G acknowledges financial support from a Shanghai talent program and from the National Natural Science
Foundation of China (NSFC) (Grant number 11874259).
DR is supported by a KIAS Individual Grant PG059602 at Korea Institute for Advanced Study. Some numerical computations were done thanks to the computing resources provided by the KIAS Center for Advanced Computation (Abacus System).
Y.J and J.J.M.V. acknowledge partial support by  U.S. DOE Grant No. DE-FAG-88FR40388. Y.J thanks S.H.Chan for a discussion of the connectivity of random regular hypergraphs. We also acknowledge the Simons Center for Geometry and Physics, where the talk by Brian Swingle for the conference ``Random
	Matrix Theory to Many-Body Physics'' initiated this work.
Some of the numerical results have been obtained by making use of the Wolfram Mathematica package \verb|QuantumManyBody|, freely available on \href{https://github.com/Dario-Rosa85/QuantumManyBody}{GitHub}.

\appendix

\section{Details on the algorithm to build regular hypergraphs}
\label{app:regularity}

In this appendix we provide some additional details on the algorithm we used to implement the $kq$-regularity condition on the sparse SYK Hamiltonians.

For us, the $kq$-regularity condition simply means that each fermion, $\gamma_i$, must be included in \textit{exactly} $kq$ non-vanishing  independent couplings, and not just on average. So let us see how we can implement this requirement in practice.

The fact that each fermions must appear in exactly $kq$ non-vanishing couplings, implies that in total the non-vanishing couplings must be extracted from a list, which we call $L$ , including each fermionic indices $kq$ times. For example, for $k = 1$ and $q = 4$, we have the list of indices $L \equiv \left( 1, 1, 1, 1,  \dots , N, N, N, N \right)$.

Hence, to construct a $kq$-regular hypergraph, what we have to do is just to sample from this list of indices sub-groups of exactly $q$ indices, such that the following two conditions are met:
\begin{itemize}
\item[a)] each group does not include repeated indices,
\item[b)] there are no repeated groups.
\end{itemize}

If we group $L$ into subgroups of $q$ indices such that  both the conditions a) and b) are met, we have a regular hypergraph.
In this case, the non-vanishing components of the $x$-couplings, \textit{i.e.} the values for which we have $x_{ijkl} = 1$, are then given by the groups of four indices just created.

In practice, we found that the following algorithm, inspired by the so-called \textit{pairing model} for regular graphs \cite{wormald_1999}, is efficient in building random regular hypergraphs:

\begin{itemize}
\item[1.] We create the list, made of two sub-lists
\begin{equation}
x_\mathrm{try} \equiv \left(\{ \} , L \right) \ ,
\end{equation}
where the first sub-list is empty and the second sub-list is the full list $L$ already introduced.

\item[2.] We randomly select a group of $q$ indices from the second sub-list and we check whether the first sub-list continues to meet the criteria a) and b) if the new group of $q$ indices is added to the first sub-list.
In the affirmative case, we move the selected indices from the second to the first sub-list.
Otherwise, we do nothing.

\item[3.] We iterate the procedure for $2N$ times (or more). In the end we check if the second sub-list in $x_\mathrm{try}$ is empty or not.
In the affirmative case, the first sub-list in $x_\mathrm{try}$ defines a $kq$-regular hypergraph (and correspondingly, the $x$-couplings $x_{ijkl}$).
In the negative case, we start again from the first point of the iteration.

\end{itemize}

\section{Examples of GOE and GSE for $N=26$}\label{app:N26GOEGSE}
In section \ref{sec:EmergentSymOrigin}
we discussed how emergent symmetries can make some of the realizations of the $N=26, q=4$ very sparse SYK (with or without regularity condition) fall into the GOE and GSE classes. In this appendix we show some explicit examples for $k=1$ with regularity condition.
\subsection{GOE}
We choose the Dirac matrices with the following subscripts to appear in the Hamiltonian
\begin{equation}
\begin{split}
\{&\{4, 7, 10, 17\}, \{8, 11, 13, 20\}, \{4, 5, 9, 22\}, \{11, 14, 18, 25\}, \{7,11, 19, 22\}, \{1, 5, 16, 20\}, \\
&\{2, 7, 8, 26\}, \{3, 15, 21, 22\}, \{5, 6,16, 19\}, \{1, 3, 7, 17\}, \{6, 19, 23, 26\}, \{2, 9, 12, 19\}, \{2, 16,24, 26\}, \\
&\{3, 12, 15, 23\}, \{2, 8, 10, 14\}, \{9, 10, 11, 12\}, \{6, 13,15, 21\}, \{10, 22, 23, 25\}, \{13, 14, 18, 20\},\\
& \{1, 9, 13, 17\}, \{3, 8,14, 21\}, \{15, 18, 23, 24\}, \{4, 6, 12, 24\}, \{1, 5, 17, 24\}, \{20, 21,25, 26\}, \{4, 16, 18, 25\}\}.
\end{split}
\end{equation}
That is, the Hamiltonian is
\begin{equation}
H= J_1 \gamma_4\gamma_{7}\gamma_{10}\gamma_{17}+J_2 \gamma_{8}\gamma_{11}\gamma_{13}\gamma_{20}+\cdots+ J_{26} \gamma_4\gamma_{16}\gamma_{18}\gamma_{25},
\end{equation}
where $\{J_1,\ldots,J_{26}\}$ are the random couplings. This Hamiltonian has an emergent 9-body symmetry
\begin{equation}
A= \gamma_{5}\gamma_{6}\gamma_{9}\gamma_{11}\gamma_{13}\gamma_{16}\gamma_{18}\gamma_{19}\gamma_{24}.
\end{equation}
This emergent symmetry $A$ makes the Hamiltonian belong to GOE through the mechanism described in section \ref{sec:EmergentSymOrigin}.
\subsection{GSE}
We choose the Dirac matrices with the following subscripts to appear in the Hamiltonian
\begin{equation}
\begin{split}
\{ &\{4, 15, 20, 22\}, \{10, 17, 18, 20\}, \{5, 10, 19, 21\}, \{2, 12, 19, 22\}, \{1, 5, 16, 25\}, \{4, 17, 25, 26\}, \\
 &\{3, 11, 18, 22\}, \{8, 18, 20, 21\}, \{4, 15, 23, 24\}, \{3, 7, 8, 14\}, \{12, 17, 23, 25\}, \{6, 7, 9, 15\}, \{2, 5, 16, 24\},\\
 & \{1, 10, 11, 26\}, \{8, 15, 18, 25\}, \{7, 13, 14, 19\}, \{2, 5, 17, 24\}, \{2, 3, 11, 16\}, \{6, 7, 16, 21\}, \{6, 8, 11,14\},\\
 & \{1, 12, 13, 21\}, \{9, 13, 19, 23\}, \{9, 20, 24, 26\},\{4, 10, 13, 14\}, \{9, 12, 22, 23\}, \{1, 3, 6, 26\}\}.
\end{split}
\end{equation}
That is, the Hamiltonian is
\begin{equation}
H= J_1 \gamma_4\gamma_{15}\gamma_{20}\gamma_{22}+J_2 \gamma_{10}\gamma_{17}\gamma_{18}\gamma_{20}+\cdots+ J_{26} \gamma_1\gamma_{3}\gamma_{6}\gamma_{26},
\end{equation}
where $\{J_1,\ldots,J_{26}\}$ are the random couplings. This Hamiltonian has an emergent 7-body symmetry
\begin{equation}
A= i \gamma_1\gamma_{12}\gamma_{18}\gamma_{20}\gamma_{22}\gamma_{25}\gamma_{26}.
\end{equation}
This emergent symmetry $A$ makes the Hamiltonian belong to GSE through the mechanism described in section \ref{sec:EmergentSymOrigin}.
\bibliographystyle{apsrev4-1}
\bibliography{library}

\end{document}